\pdfoutput=1
\documentclass[11pt]{article}
\usepackage[pdftex]{graphicx,color} 
\usepackage{jheppub}
\usepackage{amsmath}
\usepackage{amssymb}
\usepackage{ytableau}
\usepackage[multiple]{footmisc}

\usepackage{multirow}
\usepackage{mathtools}
\usepackage{dsdshorthand}

\renewcommand{\a}{\alpha}
\renewcommand{\b}{\beta}

\newcommand{\bali}{\begin{align}}
\newcommand{\eali}{\end{align}}
\newcommand{\bea}{\begin{equation}\begin{aligned}}
\newcommand{\eea}[1]{\label{#1}\end{aligned}\end{equation}}
\newcommand{\beg}{\begin{equation}\begin{gathered}}
\newcommand{\eeg}[1]{\label{#1}\end{gathered}\end{equation}}

\graphicspath{ {./figures/} }

\newcommand{\beq}{\begin{equation}}
\newcommand{\eeq}{\end{equation}}

\newcommand{\D}{\Delta}

\newcommand{\calO}{\mathcal{O}}

\newcommand{\calD}{\mathcal{D}}
\newcommand{\calL}{\mathcal{L}}
\newcommand{\calV}{\mathcal{V}}

\newcommand{\barz}{{\bar z}}
\newlength\Colsep
\usepackage[vcentermath]{youngtab}

\usepackage{tikz}
\usetikzlibrary{trees}
\usetikzlibrary{decorations.pathmorphing}
\usetikzlibrary{decorations.markings}
\usetikzlibrary{shapes}

\tikzset{
  threept/.style={
    circle,
    draw,
    inner sep=2pt,
  },
  twopt/.style={
    circle,
    draw,
    fill=black,
    inner sep=1pt,
    minimum size=1pt
  },
  cross/.style={
    cross out,
    draw=black, 
    minimum size=7pt, 
    inner sep=0pt,
    outer sep=0pt
  },
  bulkprop/.style={
    diamond,
    draw,
    fill=black,
    inner sep=2pt,
    minimum size=2pt
  },
  scalar/.style={
    thick,
    dashed,
    postaction={
      decorate,
      decoration={
        markings,
        mark=at position 0.5 with {\arrow{>}}
      }
    }
  },
  spinning/.style={
    thick,
    postaction={
      decorate,
      decoration={
        markings,
        mark=at position 0.5 with {\arrow{>}}
      }
    }
  },
  scalar no arrow/.style={
    thick,
    dashed,
  },
  spinning no arrow/.style={
    thick,
  },
  finite with arrow/.style={
    decoration={
      snake,
      amplitude=1pt,
      segment length=6pt,
      post length=2pt
    },
    decorate,
    thick,->
  },
  finite/.style={
    decoration={
      snake,
      amplitude=1pt,
      segment length=6pt,
    },
    decorate,
    thick
  },
  scalar bulk/.style={
    semithick,
    double,
    dashed,
    postaction={
      decorate,
      decoration={
        markings,
        mark=at position 0.5 with {\arrow{>}}
      }
    }
  },
  scalar bulk no arrow/.style={
    semithick,
    double,
    dashed,
  },
  spinning bulk/.style={
    double,
    semithick,
    postaction={
      decorate,
      decoration={
        markings,
        mark=at position 0.5 with {\arrow{>}}
      }
    }
  }
}

\def\be#1\ee{\begin{align}#1\end{align}}
\newcommand\cN{\mathcal{N}}
\newcommand\cS{\mathcal{S}}
\newcommand\nn\nonumber
\newcommand*{\uniq}{\raisebox{-0.7ex}{\scalebox{1.8}{$\cdot$}}}

\newcommand{\diagramEnvelope}[1]{#1}

\title{AdS Weight Shifting Operators}
\author{Miguel S. Costa$ ^{\dagger}$,}
\author{Tobias Hansen$ ^{\sharp}$}
\affiliation{$ ^{\dagger}$Centro de F\'\i sica do Porto,
Departamento de F\'\i sica e Astronomia\\
Faculdade de Ci\^encias da Universidade do Porto\\
Rua do Campo Alegre 687,
4169--007 Porto, Portugal}
\affiliation{$ ^{\sharp}$Department of Physics and Astronomy,
	Uppsala University,\\
	Box 516,
	SE-751 20 Uppsala,
	Sweden}
\emailAdd{miguelc@fc.up.pt, tobias.hansen@physics.uu.se}

\abstract{We construct a new class of  differential operators that naturally act on AdS harmonic functions. 
These are weight shifting operators that change the spin and dimension of AdS representations. 
Together with CFT weight shifting operators, the new operators obey crossing equations that relate distinct representations of the conformal group.
We apply our findings to the computation of Witten diagrams, focusing on the particular case of cubic interactions and on massive, symmetric and traceless fields. In particular we show that 
tree level 4-point Witten diagrams with arbitrary  spins, both in the external fields and in the exchanged field, can be reduced to the action of weight shifting operators on similar 4-point Witten diagrams where all fields are scalars. 
We also show how to obtain the conformal partial wave expansion of these diagrams using the new set of operators.
In the case of 1-loop diagrams with cubic couplings we show how to reduce them to similar 1-loop diagrams with scalar fields except for a single external spinning field (which must be a scalar in the case of a two-point diagram).
As a bonus, we provide new CFT and AdS weight shifting operators for mixed-symmetry tensors.}

\begin{document}
\maketitle

\section{Introduction}

Spinning fields are an essential ingredient in the AdS/CFT duality \cite{Maldacena:1997re}. 
Unless we consider a very specific situation where the spectrum becomes
sparse with a parametrically large gap, the duality requires infinitely many higher spin fields. Well known examples are type IIB
string theory in a stringy AdS space which is dual to weakly coupled ${\cal N}=4$ Super Yang-Mills, and higher spin gauge  theories \cite{Fradkin:1987ks,Vasiliev:1990en} which are dual to vector models \cite{Klebanov:2002ja} (see \cite{Didenko:2014dwa,Giombi:2016ejx} for reviews).

To compute CFT correlation functions holographically, one must compute Witten diagrams \cite{Witten:1998qj} which in general contain spinning fields. Witten diagrams are notoriously difficult to compute \cite{Muck:1998rr,Freedman:1998tz,Liu:1998ty,Freedman:1998bj,Liu:1998th, DHoker:1998ecp, DHoker:1999kzh,DHoker:1999mic,DHoker:1999mqo,Hoffmann:2000mx,Arutyunov:2002fh,Costa:2014kfa,Bekaert:2014cea,Hijano:2015zsa,Sleight:2017fpc,Nishida:2016vds,Castro:2017hpx,Dyer:2017zef,Chen:2017yia}, so it would be very helpful if we could compute such diagrams starting from a small set of known diagrams involving mostly scalar fields. The computation of Witten diagrams becomes even more difficult if one considers loop diagrams, for which 
it is fair to say our understanding  is still rather incomplete \cite{Cornalba:2007zb,Penedones:2010ue,Fitzpatrick:2011hu,Fitzpatrick:2011dm,Aharony:2016dwx,Alday:2017xua,Cardona:2017tsw,Giombi:2017hpr,Bertan:2018khc}. In fact, most loop diagrams in AdS typically consider simpler observables such as the partition function \cite{Mansfield:2000zw,Giombi:2008vd,Gopakumar:2011qs,Giombi:2013fka,Ardehali:2013xya,Beccaria:2014xda,Bae:2016rgm,Bae:2016hfy,Bae:2017spv,Skvortsov:2017ldz}.

A related development, that ignited this work, was the discovery of CFT weight shifting
operators \cite{Karateev:2017jgd}. 
With this new technology, it is possible to  compute conformal blocks for  a correlation
function with operators in any representation, in terms of scalar conformal blocks. This is a significant technical advance in the
finding of such functions, which play a central role
in setting up the CFT bootstrap equations. We will show in this paper that 
field theory in AdS also has such operators, which can be used to simplify the computation of Witten diagrams.
This is expected in the light of the AdS/CFT duality.

Representations of the conformal group can be realized either on the space of homogeneous functions on the conformal boundary or, equivalently, on the space of (homogeneous) harmonic functions in the bulk of AdS. We will make the relation between these two realizations very concrete and show that the weight shifting operators for both sets of functions are related by simple transformations.
Let us illustrate this fact for the simple case of two-point functions of scalars with conformal dimension $\De$. We encode points by vectors in embedding space $P, X \in \mathbb{R}^{d+1,1}$, making it easy to construct conformally invariant quantities.
Points on the boundary of AdS are encoded by null vectors $P$, so that the conformal two-point function $G_{\De}(P,P')$ obeys the restrictions
\beq
P^2 = 0 \,, \qquad \left( P \cdot \partial_P + \De \right)G_{\De}(P,P') = 0\,.
\label{eq:constr_boundary}
\eeq
AdS
harmonic functions  $\Omega_{\De}(X,X')$ are instead constrained by
\beq
\partial_X^2 \Omega_{\De}(X,X') = 0 \,, \qquad
\left( X \cdot \partial_X + \De \right)\Omega_{\De}(X,X') = 0
 \,.
\label{eq:constr_bulk}
\eeq
The weight shifting operators of \cite{Karateev:2017jgd} and the new bulk weight shifting operators derived in this paper obey crossing equations involving 
two- and three-point functions, for example in the simple case of scalar two-point functions we have
\bea
\calD_{+}^a(P) \,G_{\De}(P,P') &\propto
\calD_{-}^a(P') \,G_{\De + 1}(P,P')\,,\\
\calL_{+}^a(X) \,\O_{\De}(X,X') &\propto
\calL_{-}^a(X') \,\O_{\De + 1}(X,X')\,.
\eea{eq:crossing_intro}
These equations will be generalised to other representations, also including  shifts in the spin.
A crucial observation that we make in this paper,  also valid for more general representations, is that  the close relation between the constraints in \eqref{eq:constr_boundary} and \eqref{eq:constr_bulk} can be exploited to find simple relations between weight shifting operators for bulk and boundary functions. Concretely, for the operators (\ref{eq:crossing_intro}) one finds the relation
\beq
\overleftarrow{\calL}_{\pm}^a(X) = \calD_{\mp}^a(P) \big|_{P \to \partial_X, \partial_P \to X}\,,
\eeq
where the arrow on $\calL$ indicates that derivatives in this expression should act to the left. Reordering terms to remove the arrow, the operators read
\bea
\calD_{-}^a(P) &= P^a \,, &\qquad&
&\calD_{+}^a(P) &= \left( d+2 P \cdot \partial_P \right) \partial_P^a - P^a \partial_P^2 \,, \\
\calL_{+}^a(X) &= \partial_X^a \,, &\qquad&
&\calL_{-}^a(X) &= X^a \left( d+2 X \cdot \partial_X \right) - X^2\partial_X^a \,.
\eea{eq:wso_scalars}

A special role is played by the bulk-to-boundary propagator $\Pi_{\De}(X,P)$, the object transforming as a harmonic function in one argument and as a 
conformal function in the other. In the present simple case it obeys the crossing relation
\beq
\calL_{\pm}^a(X)\, \Pi_{\De}(X,P) \propto
\calD_{\mp}^a(P)\, \Pi_{\De \pm 1}(X,P)\,.
\label{eq:bulk-boundary-crossing-intro}
\eeq
One can actually see this relation as defining the bulk weight shifting operators given the boundary ones.
Using this equation, the bulk-to-boundary propagator can be used to translate between relations for boundary and bulk functions. For example, crossing relations for conformal three-point structures can be translated into crossing relations for local cubic couplings in AdS. Virtually all relations of \cite{Karateev:2017jgd} have a corresponding bulk counterpart.

These crossing relations for bulk functions can be used to make dramatic simplifications in the computation of tree-level, as well as loop Witten diagrams.
Making use of the  spectral decomposition of the bulk-to-bulk propagator \cite{Costa:2014kfa} in terms of harmonic functions,
we will illustrate this simplification in the case of Witten diagrams with cubic interactions involving massive spinning fields described by symmetric, transverse and traceless tensors. For tree level exchange diagrams with arbitrary spinning fields, both on the internal and external legs, we will be able to reduce the result to a number of weight shifting operators acting on an all-scalar Witten diagram. For loop diagrams we will also be able to make a dramatic simplification. More concretely, the two-point correlation function of a spinning field, computed at one-loop in the bulk with the exchange of spinning fields, can again be reduced to an all-scalar one-loop Witten diagram. For higher correlation functions at one-loop, a similar simplifications occurs but with a single external spinning field.

We shall start in section \ref{sec:Embedding formalism}
with a revision of the embedding space formalism for AdS fields. We point out that AdS harmonic functions can be extended to its embedding space in such a way that the constrains obeyed by these functions are a sort of Fourier transform of the constrains for CFT correlation functions. This fact is used in section \ref{sec:3} to construct the new AdS weight shifting operators, from the CFT ones.
This includes weight shifting operators for mixed-symmetry tensors, for which even the CFT version was previously unknown.
Then we extend the diagrammatic language introduced in \cite{Karateev:2017jgd} to the case of AdS harmonic functions and bulk-to-boundary propagators.
We shall see in section \ref{sec:crossing} that 
this gives an economic way of writing equations involving these objects, such as crossing equations for harmonic functions and for AdS cubic couplings. 
In section \ref{sec:WittenDiagrams} we apply these ideas to the computation of Witten diagrams, showing how to reduce diagrams involving spinning fields to simpler diagrams with mostly scalar fields. We also reproduce, using weight shifting operators,
the known conformal partial wave expansion of a scalar Witten diagram describing the exchange of a field with arbitrary spin. 
Finally, we point out that the computation of loop diagrams is equivalent to the computation of $6j$ symbols for infinite dimensional representations of the conformal group. We conclude in section \ref{sec:conclusion}. 
In appendix \ref{sec:projectors} we explain how the AdS weight shifting operators are related to
weight shifting operators for projectors to mixed-symmetry tensors.
Appendix \ref{app:crossing} contains some of the more detailed computations.

\section{Embedding formalism}
\label{sec:Embedding formalism}

We are interested in tensor fields defined in Euclidean (d + 1)-dimensional Anti de-Sitter space. As it is well known,
extending these fields from  $AdS_{d+1}$ to its embedding space $\mathbb{R}^{d+1,1}$  significantly simplifies the computations.
This approach is sometimes called ambient space formalism \cite{Bekaert:2010hk} (see \cite{Sleight:2017krf} for a review).
In this section we introduce notation and  develop the  formalism, for further details see \cite{Costa:2011mg,Costa:2014kfa}.

A point in AdS is defined by choosing a vector $X$ in embedding space with the additional restrictions
\beq
X^2 = -1 \,, \qquad X^0 \geq 0\,.
\eeq
Our approach will be \emph{not} to impose these restrictions. Instead, we will work in embedding space with homogeneous functions of fixed 
degree in $X$.  The unique map between AdS functions and homogeneous functions in
embedding space is determined by the dimension of the fields (or of their dual operators, to be more precise).
We shall see that  working with homogeneous functions will facilitate the use of weight shifting operators, which relate fields of different
conformal dimension and spin.

\subsection{Harmonic functions}
\label{sec:Harmonic}
Let us start with the simplest example of our approach, the scalar harmonic function in AdS. As a function in AdS, it satisfies the equation
\beq
\Big(\nabla_{X}^{2}-\D(\D-d)\Big)\Omega_{\D}(X,Y) = 0\,,
\eeq
where $\nabla_{X}^{2}$ is the AdS Laplacian acting on $X$.
If instead we work in embedding space, with  $\Omega_{\D}(X,Y)$ a homogeneous function of degree $-\D$ in $X$ and $Y$,
the above equation simply becomes
\beq
\eta^{ab} \frac{\partial\ }{\partial X^a}  \frac{\partial\ }{\partial X^b} \,\Omega_{\D}(X,Y) \equiv
\partial_{X}^2 \,\Omega_{\D}(X,Y) = 0\,,
\label{eq:laplace_scalar_embedding}
\eeq
where Latin indices $a,b,\dots$ denote embedding space coordinates.
To see this, one computes covariant derivatives by taking partial derivatives and then contracting all indices with the 
induced AdS metric which acts as a projector
\beq
P^a_b = \delta^a_b - \frac{X^a X_b}{X^2}\,,
\label{eq:ads_projector}
\eeq
giving
\bea
&\nabla_{X}^2 \Omega_{\D} (X,Y)
=  P^a_b  P^d_a  \frac{\partial\ }{\partial X^d}  P^b_c  \frac{\partial}{\partial X_c} \, \Omega_{\D} (X,Y)=
\\
&
\Big( \partial_{X}^2 - X^{-2} 
\left( (X \cdot \partial_X)^2 + d (X \cdot \partial_X) \right)
\!\Big)\Omega_{\D} (X,Y)
= 
\Big( \partial_X^2 + \D (\D-d) 
\Big)\Omega_{\D} (X,Y)\,,
\eea{eq:LaplacianOnOmegaScalar}
where in the last step we set $X^2=-1$. This computation 
justifies the assumption that the homogeneity of $\Omega_{\D} (X,Y)$ is $-\D$ (a second possible choice would be $(\D - d)$).

Without the covariant derivative and knowing that the function is homogeneous, equation \eqref{eq:laplace_scalar_embedding}
immediately looks easier to solve.
In fact, if $\D$ were a negative integer $-l$, the solution is given by the projectors to $SO(d+2)$  traceless symmetric tensors contracted with two vectors
\bea
\pi_{(l)}^{d+2} (X,Y) = {} &X^{a_1} \ldots X^{a_l}
\pi_{(l)}^{d+2}{\,\!}_{a_1 \ldots a_l , b_1 \ldots b_l}
Y^{b_1} \ldots Y^{b_l}
\\
={}&
 \frac{l!}{2^l \left(\frac{d}{2}\right)_l} \left(X^2\right)^{\frac{l}{2}} \left(Y^2\right)^{\frac{l}{2}}
C_l^{(\frac{d}{2})}  \!\left( \frac{X \cdot Y}{\sqrt{X^2} \sqrt{Y^2}} \right)
\,,
\eea{eq:proj_gegenbauer}
usually expressed in terms of a Gegenbauer polynomial.
This equation can be analytically continued to arbitrary values of $\D$ by writing it in terms of a hypergeometric function
\beq
\pi_{(l)}^{d+2} (X,Y) =
\frac{(d)_l}{2^l \left(\frac{d}{2}\right)_l} \left(X^2\right)^{\frac{l}{2}} \left(Y^2\right)^{\frac{l}{2}}\,
{}_2F_1  \! \left(-l, d+l, \frac{d+1}{2},-\frac{u}{2}\right) ,
\label{eq:projector_Gegenbauer}
\eeq
where $u$ is the chordal distance
\beq
u =-1 - \frac{X \cdot Y}{\sqrt{-X^2} \sqrt{-Y^2}}\,.
\eeq
Indeed, replacing $l=-\D$ we  obtain, up to a relative normalization constant,
the AdS scalar harmonic function  defined in  \cite{Cornalba:2007fs}
\beq
\O_{\D}(X,Y) = 
\frac{(-2)^{-\D-3}(d-2\D)\Gamma(\D)}{\pi^{\frac{d+2}{2}} \Gamma\!\left(1-\frac{d}{2}+\D\right)} \,\pi_{(-\D)}^{d+2} (X,Y)\,.
\eeq

Let us now discuss spinning AdS harmonic functions $\Omega_{\D,J}(X_1,X_2;W_1,W_2)$  \cite{Costa:2014kfa}, where
the $W_i \in \mathbb{R}^{d+1,1}$ are polarizations encoding the spacetime indices of the functions.
These AdS harmonic functions satisfy the following conditions
\begin{align}
\left(\nabla_{X_1}^{2}-\D(\D-d)+J\right)\Omega_{\D,J}(X_1,X_2;W_1,W_2) & =0\,,
\label{eq:LaplacianOmega}\\
\nabla_{X_1}\cdot \partial_{W_1}\,\Omega_{\D,J}(X_1,X_2;W_1,W_2) & =0	\,,
\label{eq:divergenceOmega}\\
\partial_{W_1}^2 \,\Omega_{\D,J}(X_1,X_2;W_1,W_2) & =0 \label{eq:dZ1squaredOnOmega}	\,,\\
X_{1}\cdot \partial_{W_1}\,\Omega_{\D,J}(X_1,X_2;W_1,W_2) & =0	\label{eq:X1dZ1OnOmega}\,,
\end{align}
i.e. they are respectively eigenfunctions of the Laplace operator, divergence free, traceless and transverse.\footnote{
In  \cite{Costa:2014kfa} we imposed that $X_i\cdot W_i=0$ and $W_i^2=0$, while here we keep  $W_i$
general but impose  conditions (\ref{eq:dZ1squaredOnOmega}) and (\ref{eq:X1dZ1OnOmega}).}
The bulk-to-bulk propagator $\Pi_{\D,J}$ satisfies an inhomogeneous version of the same equations, with \eqref{eq:LaplacianOmega} and \eqref{eq:divergenceOmega} for $\Pi_{\D,J}$ having terms that involve delta functions and derivatives of delta functions on the right hand side. As we shall see, 
these contributions lead to contact terms in the propagator that are not generated directly by the use of weight shifting operators. 
Because of this fact we first focus on the harmonic functions and later make use of the spectral representation of the propagator in terms of harmonic functions.

As in the case of the scalar harmonic function, we extend to homogeneous functions in $X_i$. The constraints \eqref{eq:LaplacianOmega} and \eqref{eq:divergenceOmega} become
\begin{align}
\partial_{X_1}^2 \, \Omega_{\D,J}(X_1,X_2;W_1,W_2) & =0\,,\label{eq:dX1squaredOnOmega}\\
\partial_{X_1}\cdot \partial_{W_1}\,\Omega_{\D,J}(X_1,X_2; W_1,W_2) & =0 \label{eq:dX1dZ1OnOmega}\,.
\end{align}
To show this we start with equation \eqref{eq:divergenceOmega}. Let us show explicitly only one set of indices and,   for simplicity, suppress the dependence on 
$X_2$ and $W_2$,
\beq
 \Omega_{\D,J}(X,X_2; W,W_2) \equiv
 W_{a_1} \ldots W_{a_J} \Omega_{\D,J}^{a_1 \ldots a_J} (X) \,.
\eeq
The covariant derivative is computed using the projector \eqref{eq:ads_projector},
thus \eqref{eq:divergenceOmega} becomes
\bea
\nabla_{X a_i } \Omega_{\D,J}^{a_1 \ldots a_J}(X)
=  P_{b_1}^{a_1} \ldots P_{b_J}^{a_J}   P_{a_i}^c \partial_{X c} \,\Omega_{\D,J}^{b_1 \ldots b_J}(X)
= \partial_{X a_i} \Omega_{\D,J}^{a_1 \ldots a_J}(X)\,,
\eea{eq:divergenceOmega_homo}
where \eqref{eq:dZ1squaredOnOmega} and \eqref{eq:X1dZ1OnOmega} where used for simplifications.
The action of the Laplacian can be computed in a similar way
\bea
&\nabla_{X}^2 \Omega_{\D,J}^{a_1 \ldots a_J} (X) 
=  P_{b_0}^{a_0} P_{b_1}^{a_1} \ldots P_{b_J}^{a_J} \partial_{X a_0}
P_{c_0}^{b_0} P_{c_1}^{b_1} \ldots P_{c_J}^{b_J} \partial_X^{c_0}
\Omega_{\D,J}^{c_1 \ldots c_J}(X)\\
={}& P_{b_1}^{a_1} \ldots P_{b_J}^{a_J}
\Big( \partial_{X}^2 - X^{-2} 
\left( (X \cdot \partial_{X})^2 + d (X \cdot \partial_X) - J \right) \!\Big)\,\Omega_{\D,J}^{b_1 \ldots b_J} (X)\,.
\eea{eq:LaplacianOnOmega}
Assuming $\Omega_{\D,J}$ is homogeneous of degree $-\D$ in $X$, we obtain that
\beq
\Big(\nabla_{X}^2 + X^{-2} \big( \D(\D-d)-J \big) \Big)\,  \Omega_{\D,J}^{a_1 \ldots a_J} (X)
=P_{b_1}^{a_1} \ldots P_{b_J}^{a_J}
 \partial_{X}^2\Omega_{\D,J}^{b_1 \ldots b_J} (X)\,,
\eeq
so \eqref{eq:dX1squaredOnOmega} for a homogeneous $\Omega_{\D,J}$ implies \eqref{eq:LaplacianOmega} when setting $X^2=-1$.

Together the four conditions (\ref{eq:dZ1squaredOnOmega}-\ref{eq:dX1dZ1OnOmega}) have the same form as the ones satisfied by the projectors to
$SO(d+2)$ traceless mixed-symmetry tensors whose Young Tableaux have two rows  with $X$ and  $W$ serving as polarization vectors. This is 
spelled out in appendix \ref{sec:projectors}. Thus, the relation between such projectors and AdS harmonic functions,  observed above for scalar functions, 
continues to hold for the case with spin. This fact will become clear from the construction of these functions by means of weight shifting operators.

\subsection{Bulk-to-bulk propagator}

The AdS propagator  for a spin $J$ field solves an inhomogeneous version of equations (\ref{eq:LaplacianOmega}-\ref{eq:X1dZ1OnOmega}). More concretely, (\ref{eq:LaplacianOmega}) and (\ref{eq:divergenceOmega}) will contain delta functions and their derivatives
in the right hand side, in particular (\ref{eq:LaplacianOmega}) becomes
\beq
\Big(\nabla_{X_1}^{2}-\D(\D-d)+J\Big)\Pi_{\D,J}(X_1,X_2;W_1,W_2)  = -\, 
\delta(X_1,X_2) \,  \left( W_{1} \circ W_2\right)^J + \ldots\,,
\label{eq:LaplacianPi}
\eeq 
where the $\circ$ product  indicates a traceless and transverse contraction of the vectors denoted here by
curly brackets 
\bea
\left( W_{1} \circ W_2\right)^J = {}&W_{1 \{a_1} \ldots W_{1 a_J\}}  W_2^{\{a_1}\ldots  W_2^{a_J\}} \\
={}&
W_{1}^{a_1} \ldots W_{1}^{a_J}
P_{a_1}^{b_1} \ldots  P_{a_J}^{b_J}
\pi^{d+1}_{(J) \, b_1 \ldots b_J, c_1 \ldots c_J}
P_{d_1}^{c_1} \ldots  P_{d_J}^{c_J}
W_{2}^{d_1} \ldots W_{2}^{d_J}\,,
\eea{eq:tt_contraction}
with $\pi^{d+1}_{(J)}$  the projector to symmetric traceless rank $J$ tensors in $d+1$ dimensions.
The dots in (\ref{eq:LaplacianPi}) represent local source terms that change the propagator by contact terms.
The traceless part of the propagator was computed in \cite{Costa:2014kfa} as a spectral integral over (derivatives of) harmonic functions.
It is of the form \cite{Costa:2014kfa}\footnote{\label{fn:covariant_derivative}In \cite{Costa:2014kfa} this expression is written with covariant derivatives instead of the $\partial_{X_i}$. However when doing the covariant derivatives as  defined in \eqref{eq:divergenceOmega_homo} one finds that any terms containing a $X^{a_i}$ or $\eta^{a_i a_j}$ are projected out by the curly brackets and only the partial derivatives remain.}
\begin{align}
&\Pi_{\Delta,J}(X_1,X_2;W_1,W_2)=
\sum_{l=0}^J \frac{1}{(l!)^2} \int d\nu \,a_{J,l}(\nu)W_{1 \{a_1} \ldots W_{1 a_{J}\}} \partial_{X_1}^{\{a_1} \ldots \partial_{X_1}^{a_{J-l}}
\partial_{W_1}^{a_{J-l+1}} \ldots \partial_{W_1}^{a_{J}\}}
\nonumber\\
& \times W_{2 \{b_1} \ldots W_{2 b_{J}\}} \partial_{X_2}^{\{b_1}\ldots \partial_{X_2}^{b_{J-l}}
\partial_{W_2}^{b_{J-l+1}} \ldots \partial_{W_2}^{b_{J}\}}
 \, \Omega_{\D(\nu),l}(X_1,X_2;W_1,W_2)\,,
\label{eq:SlipRepStart}
\end{align}
where $ \D(\nu) \equiv  h + i \nu$ and we set as usual $h=d/2$. For the maximal spin $l=J$ the spectral function has the simple form
\beq
a_{J,J}(\nu)=\frac{1}{\nu^2 + (\D-h)^2}\,.
\eeq
This is the universal part of the propagator that describes the propagating degrees of freedom. 
The other terms with $l<J$ were discussed in \cite{Costa:2014kfa}. 
They are important to cancel spurious poles.
The trace part of the propagator depends on the full equation \eqref{eq:LaplacianPi}, hence we use only the traceless part
and results are up to contact interactions.
In this paper
we will use the spectral representation \eqref{eq:SlipRepStart} as an input, that is, some of our results are written in terms of the  spectral 
functions $a_{J,l}(\nu)$.

\subsection{Bulk-to-boundary propagator}

The bulk-to-boundary propagator
is obtained from the bulk-to-bulk propagator by first setting $X_i^2=-1$ and then scaling the $X_i$ to infinity,
\begin{equation}
\Pi_{\Delta,J}(X,P;W,Z) = \lim_{\lambda\rightarrow\infty}\lambda^{\Delta}\Pi_{\Delta,J}(X,\lambda P+O(\lambda^{-1});W,Z)
\,.
\end{equation}
Obviously one first needs to set $X_i^2=-1$ because otherwise the whole homogeneous function would be just rescaled. 
The result of this computation is \cite{Costa:2014kfa}
\begin{equation}
\Pi_{\Delta,J}(X,P;W,Z)=\mathcal{C}_{\Delta,J}\,\frac{\big( 2(W\cdot P)(Z\cdot X) - 2 (P\cdot X)(W\cdot Z)\big)^{J}}{(-2P\cdot X)^{\Delta+J}}\,,
\label{eq:eq propagador boundary}
\end{equation}
with
\begin{equation}
{\cal C}_{\Delta,J}=\frac{ \left(J+\Delta-1\right) \Gamma(\Delta)}{2 \pi^{d/2}\left(\Delta-1\right)\Gamma(\Delta+1-h)}\,.
\label{eq:normalizationboundary}
\end{equation}
Note that we used the letters $X, W$ to label bulk points and polarizations, and $P, Z$ to describe points 
and polarizations on the boundary of AdS which satisfy as usual \cite{Costa:2011mg}
\beq
P^2 = P \cdot Z = Z^2 = 0\,.
\label{eq:cft_constraint_1}
\eeq
This convention is used throughout the paper (allowing to distinguish between bulk-to-bulk and bulk-to-boundary propagators).

\subsection{Conformal correlators}

We will also frequently work with conformal correlators. In the embedding space formalism for CFTs  \cite{Costa:2011mg}  
a correlator involving the operator $\cO_{\D,J} (P,Z)$ can be constructed by imposing that it is homogeneous of degree 
$-\De$ in $P$ and degree $J$ in $Z$, and satisfies
\beq
P \cdot \partial_Z \< \cO_{\D,J} (P,Z) \ldots \> =0\,.
\label{eq:cft_constraint_2}
\eeq
The two-point function of spin $J$ operators is given by
\beq
\< \cO_{\D,J} (P_1,Z_1) \cO_{\D,J} (P_2,Z_2) \>
=
\frac{\big(2 (Z_1 \cdot P_2)( Z_2 \cdot P_1) - 2 (Z_1 \cdot Z_2)( P_1 \cdot P_2) \big)^J}{P_{12}^{\D+J}}\,,
\eeq
where $P_{ij} = -2 P_i \cdot P_j$.
The single three-point tensor structure that appears in a three-point function of two scalars and one spin $J$ operator is
\beq
\< \cO_{\D_1} (P_1) \cO_{\D_2} (P_2) \cO_{\D,J} (P_3,Z_3)\>
=
\frac{\big((Z_3 \cdot P_1)P_{23} - (Z_3 \cdot P_2)P_{13} \big)^J}{P_{12}^{\frac{\D_1+\D_2-\D+J}{2}} P_{23}^{\frac{\D_2+\D-\D_1+J}{2}} P_{31}^{\frac{\D+\D_1-\D_2+J}{2}}}\,.
\label{eq:3pt_correlator}
\eeq

\section{Weight shifting operators}
\label{sec:3}

Weight shifting operators for CFTs have  been recently constructed in  \cite{Karateev:2017jgd}. These
operators can be used to relate correlation functions of operators in different
representations of the conformal group. They are a major technical advance in the construction of 
conformal blocks of arbitrary representations in terms of more basic scalar conformal blocks.
Our goal in this section is to extend this construction to the case of weight shifting operators that 
relate AdS propagators and harmonic functions of different $(\D, J)$. 

\subsection{Construction}

To construct the AdS weight shifting operators it is useful to consider first their CFT counterparts introduced in  \cite{Karateev:2017jgd}.
Let us consider the action of the CFT weight shifting operators $\calD$ on an operator placed at point $P$ and with polarization vector $Z$. 
These operators  act on $R/(R \cap I)$, where $R$ is the ring of functions of $P$ and $Z$ that are killed by $P \cdot \partial_{Z}$, and $I$ is the ideal generated by $\{P^2, P \cdot Z, Z^2\}$. These operators satisfy
\begin{align}
\calD R &\subseteq R,\label{eq:WS_in_R}\\
\calD (R \cap I) &\subseteq (R \cap I).
\label{eq:WS_in_RI}
\end{align}
We would like to find AdS weight shifting operators $\cL$ that map harmonic functions to harmonic functions, i.e.\ that preserve the properties (\ref{eq:dZ1squaredOnOmega}-\ref{eq:dX1dZ1OnOmega}).
If $R'$ is the ring of functions of $X$ and $W$ that are killed by $X \cdot \partial_{W}$ and $R''$ the ring of functions killed by $\{\partial_X^2, \partial_X \cdot \partial_W, \partial_W^2\}$ these weight shifting operators have to satisfy\footnote{Strictly speaking \eqref{eq:WS_in_RpRpp} is sufficient to relate harmonic functions. However \eqref{eq:WS_in_Rp} will hold as a result of the construction.}
\begin{align}
\calL R' &\subseteq R',\label{eq:WS_in_Rp}\\
\calL (R' \cap R'') &\subseteq (R' \cap R'').\label{eq:WS_in_RpRpp}
\end{align}
We will show that the operators $\calL$ can be obtained from the operators $\calD$.
Let us first rewrite \eqref{eq:WS_in_R} and \eqref{eq:WS_in_RI} in a more concrete way.
Equation \eqref{eq:WS_in_R} can be rephrased by demanding $\calD$ to satisfy
for any function $r\in R$
\beq
\left[ P \cdot \partial_{Z}, \calD \right] r = 0\,.
\label{eq:D_transverse_commutator}
\eeq
The condition \eqref{eq:WS_in_RI} means that one can define maps to $R \cap I$
by first multiplying by any $h \in R \cap I$ and then acting with $\calD$
\beq
\calD h : R \to R \cap I\,.
\eeq
This equation implies that for any $h \in R \cap I$ and $r\in R$ we have
\beq
\calD h r =P^2 \ldots + P \cdot Z \ldots + Z^2 \ldots ,
\eeq
where the right hand side simply means that any term of an element of $R \cap I$ is proportional to at least one of the generators of $I$.  
For some particular choices of $h$ we have
\bea
\calD P^2 r &=P^2 \ldots + P \cdot Z \ldots + Z^2 \ldots ,\\
\calD \left((P \cdot Z)( S\cdot P) - P^2 (S \cdot Z)\right) r &= P^2 \ldots + P \cdot Z \ldots + Z^2 \ldots,\\
\calD \left(Z^2 (S\cdot P)^2 - 2 (P \cdot Z)(S \cdot P)( S \cdot Z) + P^2 (S\cdot Z)^2\right) r &= P^2 \ldots + P \cdot Z \ldots + Z^2 \ldots ,
\eea{eq:Dg_map_examples}
where $S$ is an arbitrary vector.

We are interested in the case that the $\calD$ act on homogeneous functions $f_{\D,J}(P,Z) \in R$ of homogeneity $-\D$ in $P$ and $J$ in $Z$,
changing the homogeneities by integers $\de_\D$, $\de_J$
\bea
\calD_{\delta_\D, \delta_J}(P,Z): f_{\D, J}(P,Z) &\to f'_{\D + \delta_\D, J+\delta_J} (P,Z)\,.
\eea{eq:ws_operators_def}
This implies that $\calD_{\delta_\D, \delta_J}$ itself is homogeneous in $P$ and $Z$ (of degree $-\de_\D$ and $\de_J$) and depends on $\De$ and $J$. 
Consider the following object which has homogeneity zero in both $P$ and $Z$
\beq
g_{-J - \delta_J +1, -\D- \delta_\D +1}(\partial_{Z},\partial_{P})\big[ P \cdot \partial_{Z}, \calD_{\delta_\D, \delta_J} \big] f_{\D, J}(P,Z)\,,
\label{eq:D_transverse_commutator_gf}
\eeq
where $g_{\D,J}(P,Z)$ is a homogeneous function with homogeneities $(-\D,J)$ in $(P,Z)$. We shall consider 
 $\D$ to be a negative integer, so that $f$ and $g$ are polynomials in their arguments. In this case (\ref{eq:D_transverse_commutator_gf}) is just a combination of metrics, i.e.\ there is no dependence on $P$ or $Z$. The new weight shifting operators found below will then be defined for any $\D$ by analytic continuation, just as AdS harmonic functions can be defined from analytic continuation of projectors, as explained in  section
 \ref{sec:Harmonic}.

Next we do the following three operations in (\ref{eq:D_transverse_commutator_gf}) at once:
\begin{itemize}
 \item[$(i)$] Replace the boundary vectors $(P,Z)$ and corresponding derivatives by their bulk counterparts $(X,W)$ in the following way:\footnote{Note that this transformation relates the constraints (\ref{eq:cft_constraint_1},\ref{eq:cft_constraint_2}) for CFT correlators with the constraints (\ref{eq:dZ1squaredOnOmega}-\ref{eq:dX1dZ1OnOmega}) for AdS harmonic functions. A similar relation was observed in \cite{Bekaert:2012vt}.}
  \beq
   P \to \partial_W\,, \quad
   Z \to \partial_X\,, \quad
   \partial_P \to W\,, \quad
   \partial_Z \to X\,.
   \eeq
 \item [$(ii)$] Invert the whole equation (let derivatives act to the left instead of to the right).
 \item [$(iii)$] Shift $\D$ and $J$ to achieve $g_{-J - \delta_J +1, -\D- \delta_\D +1} \to g_{\D,J}$, that is
   \beq
   \D \to -J - \delta_\D +1\,, \qquad
   J \to -\D- \delta_J + 1\,.
   \eeq
\end{itemize}
For each operator $\calD_{\delta_\D, \delta_J} (P,Z) $ these three steps together define a new operator $\calL_{\delta_J, \delta_\D} (X,W)$ such that 
\beq
 \calL_{\delta_J, \delta_\D} (X,W): g_{\D, J}(X,W) \to g'_{\D + \delta_J, J+\delta_\D} (X,W)\,.
\eeq

Since the expression \eqref{eq:D_transverse_commutator_gf} is just a number (which is 0 due to \eqref{eq:D_transverse_commutator}), it is invariant under this transformation and can also be written as
\beq
f_{-J -\delta_\D+1, -\D-\delta_J+1}(\partial_{W},\partial_{X})\big[ \calL_{\delta_J, \delta_\D}, X \cdot \partial_{W}\big] 
g_{\D, J}(X,W)= 0\,,
\label{eq:D_transverse_commutator_inverted}
\eeq
implying that the operators $\calL_{\delta_J, \delta_\D}$ indeed satisfy \eqref{eq:WS_in_Rp}.
Moreover, equation \eqref{eq:Dg_map_examples} becomes after the transformation, when acting on any $r' \in R'$
\bea
r_{\text{tr}}\partial_{W}^2 \calL r'&= \left(\ldots \partial_{W}^2 + \ldots \partial_{X} \cdot \partial_{W} + \ldots \partial_{X}^2\right) r' \,,\\
r_{\text{tr}}\left(( S\cdot \partial_{W})(\partial_{X} \cdot \partial_{W}) -  (S \cdot \partial_{X}) \partial_{W}^2\right) \calL r'&= \left(\ldots \partial_{W}^2 + \ldots \partial_{X} \cdot \partial_{W} + \ldots \partial_{X}^2\right) r'\,, \\
r_{\text{tr}}\big( (S\cdot \partial_{W})^2 \partial_{X}^2 - 2 ( S \cdot \partial_{X})( S \cdot \partial_{W})(\partial_{X} \cdot \partial_{W})& \\
+  (S\cdot \partial_X)^2 \partial_{W}^2\big) \calL r'&= \left(\ldots \partial_{W}^2 + \ldots \partial_{X} \cdot \partial_{W} + \ldots \partial_{X}^2\right) r' \,,
\eea{eq:Dg_map_examples_inverted}
or when acting on a $r'' \in R' \cap R''$
\bea
r_{\text{tr}}\partial_{W}^2 \calL r''&= 0 \,,\\
r_{\text{tr}}\left(( S\cdot \partial_{W})(\partial_{X} \cdot \partial_{W}) -  (S \cdot \partial_{X}) \partial_{W}^2\right) \calL r''&= 0\,, \\
r_{\text{tr}}\big( (S\cdot \partial_{W})^2 \partial_{X}^2 - 2 ( S \cdot \partial_{X})( S \cdot \partial_{W})(\partial_{X} \cdot \partial_{W})+  (S\cdot \partial_X)^2 \partial_{W}^2\big) \calL r''&= 0 \,,
\eea{eq:Dg_map_examples_inverted_Rpp}
where $r_{\text{tr}}$ is the transformed $r$ from \eqref{eq:Dg_map_examples}.
Since this needs to hold for any $r \in R$, we conclude that it implies \eqref{eq:WS_in_RpRpp}.
Thus, the $\calL$ are the weight shifting operators for AdS harmonic functions!

Next we write down the operators $\calL$ which are related to the
CFT weight shifting operators in the vector representation \cite{Karateev:2017jgd}\footnote{
We chose a different normalization for the last operator: 
$\big( \calD_{+0}^a(P,Z)\big)_{\text{here}} = 2 \big( \calD_{+0}^a(P,Z)\big)_{\text{[46]}}$.
}
\bea
\calD_{-0}^a(P,Z) ={}& P^a \,,\\
\calD_{0+}^a(P,Z) ={}& \Big((J+\D) \delta_b^a + P^a \partial_{P b}\Big) Z^b \,,\\
\calD_{0-}^a(P,Z) ={}&\Big( (\D -d +2 - J) \delta_b^a + P^a \partial_{P b} \Big)
\Big( (d-4+2 J) \delta_c^b -Z^b \partial_{Z c} \Big) \partial_{Z}^c \,,\\
\calD_{+0}^a(P,Z) ={}& \Big( c_1 \delta^a_b + P^a \partial_{P b} \Big)
\Big( c_2 \delta^b_c + Z^b \partial_{Z c} \Big)
\Big( c_3 \delta^c_d - \partial^c_{Z} Z_{d}  \Big) \partial^d_{P} \,,
\eea{eq:ws_operators}
where
\beq
c_1 = 2 - d + 2 \De\,, \qquad
c_2 = 2 - d + \De - J\,, \qquad
c_3 = \De + J\,.
\label{eq:c_i}
\eeq
Performing the three steps listed above we can directly read off the operators
\bea
\calL_{0-}^a(X,W) ={}& \partial_{W}^a\,,\\
\calL_{+0}^a(X,W) ={}& \partial_{X}^b\Big((-\D - J + 1) \de_b^a + W_b \partial_{W}^a \Big) \,,\\
\calL_{-0}^a(X,W) ={}& X^c\Big( (d-2\D) \delta_c^b -X_c \partial_{X}^b \Big)
\Big( (-J -d +1 +\D) \delta_b^a + W_{b} \partial_{W}^a \Big)\,,\\
\calL_{0+}^a(X,W) ={}& W^d
\Big( c'_3 \delta^c_d - \partial_{X d} X^{c}  \Big)
\Big( c'_2 \delta^b_c + X_c \partial_{X}^b \Big)
\Big( c'_1 \delta^a_b + W_{b} \partial_{W}^a \Big) \,,
\eea{eq:ws_operators_inverted}
where
\beq
c_i' = c_i |_{\D \to -J, \, J \to -\D+1}\,.
\eeq
A slightly modified version of these operators (which is spelled out in appendix \ref{sec:projectors}) can be used to relate different projectors to traceless mixed-symmetry tensors for two-row Young diagrams. These operators are known in the higher spin literature as sigma or cell operators.

\subsection{Mixed-symmetry tensors}
\label{sec:mixed-symmetry}

In this section we  generalize the previous construction from traceless symmetric tensors to traceless mixed-symmetry tensors. We derive previously unknown weight shifting operators for conformal structures and their transformations into AdS weight shifting operators. The operators derived here will not be used explicitly in the remainder of the paper, they may however appear implicitly in  general considerations. The reader may therefore wish to skip this section on a first reading.

Conformal correlators of mixed-symmetry tensors can be implemented in embedding space by using additional polarization vectors, one for each row of the Young diagram labelling the representation.
For simplicity we will focus on Young diagrams with two rows of lengths $J_1$ and $J_2$. We  encode these tensors with the two polarizations 
$Z_1$ and $Z_2$.
The construction of conformal correlators involving mixed-symmetry tensors was treated in  \cite{Costa:2014rya}. In short, they can be constructed in embedding space by requiring that products of the three vectors associated to the same operator vanish, that is
\beq
P^2 = Z_1^2 = Z_2^2 = P \cdot Z_1 = P \cdot Z_2 = Z_1 \cdot Z_2 = 0\,,
\label{eq:ms_product_constraints}
\eeq
together with transverseness
\beq
P \cdot \partial_{Z_1} \< \cO_{\D,J_1,J_2} (P,Z_1,Z_2) \ldots \> =
P \cdot \partial_{Z_2} \< \cO_{\D,J_1,J_2} (P,Z_1,Z_2) \ldots \>=0\,,
\label{eq:ms_transverseness}
\eeq
and the following constraint which enforces mixed symmetry (see also \cite{Costa:2016hju})
\beq
Z_1 \cdot \partial_{Z_2} \< \cO_{\D,J_1,J_2} (P,Z_1,Z_2) \ldots \>=0\,.
\label{eq:mixed-symmetry}
\eeq
We start by constructing the weight shifting operators which relate such conformal structures.
This is significantly easier than constructing the weight shifting operators for the bulk directly, since conformal structures are simpler functions due to the constraints \eqref{eq:ms_product_constraints}.
The construction can be done by repeating the steps that were used in \cite{Karateev:2017jgd} to find the operators \eqref{eq:ws_operators}. One starts by constructing an ansatz. There are only a finite number of possible terms, since the homogeneity of the operators in each vector is fixed and the only independent contractions of vectors and derivatives are
\beq
Z_1 \cdot \partial_P,\  
Z_2 \cdot \partial_P,\ 
Z_2 \cdot \partial_{Z_1},\ 
\partial_P^2, \partial_{Z_1}^2,\ 
\partial_{Z_2}^2,\ 
\partial_P \cdot \partial_{Z_1},\  
\partial_P \cdot \partial_{Z_2},\  
\partial_{Z_1} \cdot \partial_{Z_2}\,.
\eeq
Our ansatz is (already choosing an overall normalization)
\begin{align}
\calD_{-00}^a(P,Z_1,Z_2) ={}& P^a \,,
\nonumber\\
\calD_{0+0}^a(P,Z_1,Z_2) ={}& \Big( (J_1+\D) \delta_b^a + P^a \partial_{P b} \Big) Z_1^b  \,,
\nonumber\\
\calD_{00+}^a(P,Z_1,Z_2) ={}&\Big( \a_1 \delta_b^a + P^a \partial_{P b} \Big) 
\Big( \a_2 \delta^b_c + Z_1^b \partial_{Z_1 c} \Big) Z_2^c\,,
\label{eq:ms_ws_operators}
\\
\calD_{00-}^a(P,Z_1,Z_2) ={}& \Big( \b_1 \delta^a_b + P^a \partial_{P b} \Big)
\Big( \b_2 \delta^b_c + Z_1^b \partial_{Z_1 c} \Big)
\Big( \b_3 \delta^c_d + Z_2^c \partial_{Z_2 d} \Big) \partial^d_{Z_2} \,,
\nonumber\\
\calD_{0-0}^a(P,Z_1,Z_2) ={}& \Big( \g_1 \delta^a_b + P^a \partial_{P b} \Big)
\Big( \g_2 \delta^b_c + Z_1^b \partial_{Z_1 c} \Big)
\Big( \g_3 \delta^c_d + Z_2^c \partial_{Z_2 d} \Big)
\Big( \g_4 \delta^d_e + \partial^d_{Z_2} Z_{2 e}  \Big) \partial^e_{Z_1} \,,
\nonumber\\
\calD_{+00}^a(P,Z_1,Z_2) ={}& \Big( \e_1 \delta^a_b + P^a \partial_{P b} \Big)
\Big( \e_2 \delta^b_c + Z_1^b \partial_{Z_1 c} \Big)
\Big( \e_3 \delta^c_d + Z_2^c \partial_{Z_2 d} \Big)
\Big( \e_4 \delta^d_e + \partial^d_{Z_2} Z_{2 e}  \Big)
\nonumber\\
&\Big( \e_5 \delta^e_f + \partial^e_{Z_1} Z_{1 f}  \Big) \partial^f_{P} \,.
\nonumber
\end{align}
The coefficients can be fixed by checking that the operators satisfy $\calD R \subseteq R$ and $\calD (R \cap I) \subseteq (R \cap I)$ 
for a sufficient number of test functions,
where now $R$ is the ring of functions of $P$, $Z_1$ and $Z_2$ that are killed by $\{P \cdot \partial_{Z_1}, P \cdot \partial_{Z_2}, Z_1 \cdot \partial_{Z_2}\}$, and $I$ is the ideal generated by $\{P^2, Z_1^2, Z_2^2, P \cdot Z_1, P \cdot Z_2, Z_1 \cdot Z_2\}$. One finds the coefficients
\be
\a_1 &= \De + J_2 -1 \,,
&\a_2 &= J_2 - J_1 \,, & & & &\nn\\
\b_1 &= 3 - d + \De - J_2 \,,
&\b_2 &= 4 - d - J_1 - J_2 \,,
&\b_3 &= 6 - d - 2 J_2\,, & &\nn\\
\g_1 &= 2 - d + \De - J_1\,,
&\g_2 &= 4 - d - 2 J_1\,,
&\g_3 &= 4 - d- J_1 - J_2\,,
&\g_4 &= J_1 - J_2 \,,\nn\\
\e_1 &= 2 - d +2 \De\,,
&\e_2 &= 2 - d + \De - J_1 \,,
&\e_3 &= 3 - d + \De - J_2 \,, & &\nn\\
\e_4 &= 1 - \De - J_2 \,,
&\e_5 &= - \De - J_1\,. & & & &
\label{eq:ms_coefficients}
\ee

For the harmonic functions on AdS we rely again on the observation that they are the analytic continuations of projectors to $SO(d+2)$ traceless mixed-symmetry tensors, this time for Young diagrams with three rows. That means \cite{Costa:2016hju} they should satisfy
\beq
D \, \O_{\De,J_1,J_2} (X, W_1, W_2, X', W'_{1},W'_{2}) = 0\,,
 \qquad \forall D \in S_1 \cup S_2\,,
\label{eq:ms_bulk_constraints}
\eeq
where $S_1$ and $S_2$ are the sets of operators
\bea
S_1&=\big\{X \cdot \partial_{W_1},\,  X \cdot \partial_{W_2},\,  W_1 \cdot \partial_{W_2}\big\}\,,\\
S_2&=\big\{\partial_X^2,\, \partial_{W_1}^2,\, \partial_{W_2}^2,\, \partial_X \cdot \partial_{W_1},\, \partial_X \cdot \partial_{W_2},\, \partial_{W_1} \cdot \partial_{W_2} \big\}\,.
\eea{eq:ms_S}
Next we repeat the three steps described in the previous section to turn weight shifting operators for conformal structures into operators for harmonic functions. In the first step we now do the replacement
  \beq
   P \to \partial_{W_2}\,, \quad
   Z_1 \to \partial_{W_1}\,, \quad
   Z_2 \to \partial_X\,, \quad
   \partial_P \to W_2\,, \quad
   \partial_{Z_1} \to W_1\,, \quad
   \partial_{Z_2} \to X\,.
   \eeq
This is the unique choice that relates the constraints for conformal structures (\ref{eq:ms_product_constraints}-\ref{eq:mixed-symmetry}) to the ones for AdS harmonic functions \eqref{eq:ms_bulk_constraints}.
After replacing the vectors and inverting the equation any operator $ \calD_{\delta_{\De},\de_{J_1},\delta_{J_2}} (P,Z_1,Z_2)$ becomes
\beq
 \calL_{\delta_{J_2},-\de_{J_1},\delta_\D} (X,W_1,W_2): g_{\D, J_1,J_2}(X,W_1,W_2) \to g'_{\D+\delta_{J_2}, J_1-\de_{J_1},J_2+\delta_\D}(X,W_1,W_2)\,.
\eeq
By acting on explicitly known projectors to mixed-symmetry tensors from \cite{Costa:2016hju} (as explained in appendix \ref{sec:projectors}), 
one finds the following  required shifts of the parameters
   \beq
   \D \to -J_2 - \delta_\D +2\,, \qquad
   J_1 \to J_1- \delta_{J_1}\,, \qquad
   J_2 \to -\D- \delta_{J_2} + 2\,.
   \eeq
These shifts happen to be the right shifts to transform the object (which has homogeneity 0 in $P$, $Z_1$ and $Z_2$)
\bea
&g_{-J_2 - \delta_{J_2} +2, J_1 + \de_{J_1}, -\D- \delta_\D+2}(\partial_{Z_2},\partial_{Z_1},\partial_{P})\\
&\times\big[ Z_1 \cdot \partial_{Z_2} P \cdot \partial_{Z_2} P \cdot \partial_{Z_1}, \calD_{\delta_\D, \delta_{J_1}, \delta_{J_2}} \big]
f_{\D, J_1, J_2}(P,Z_1,Z_2) \,,
\eea{eq:mixed-sym-commutators-g-f}
into
\bea
&f_{-J_2 - \delta_{\De} +2, J_1 - \de_{J_1}, -\D- \delta_{J_2}+2}(\partial_{W_2},\partial_{W_1},\partial_{X})\\
&\times\big[ \calL_{\delta_{J_2},-\delta_{J_1}, \delta_{\De}},  W_1 \cdot \partial_{W_2} X \cdot \partial_{W_2} X \cdot \partial_{W_1} \big]
g_{\D, J_1, J_2}(X,W_1,W_2) \,,
\eea{eq:mixed-sym-commutators-f-g}
where $g_{\D, J_1, J_2}$ and $f_{\D, J_1, J_2}$ are homogeneous of degree $(-\De,J_1,J_2)$ in their arguments.
The weight shifting operators for AdS harmonic functions obtained in this way are
\bea
\calL_{00-}^a(X,W_1,W_2) ={}& \partial_{W_2}^a \,,\\
\calL_{0-0}^a(X,W_1,W_2) ={}& \partial_{W_1}^b
\Big( (J_1-J_2+1) \delta_b^a +  W_{2 b} \partial_{W_2}^a \Big) \,,\\
\calL_{+00}^a(X,W_1,W_2) ={}& \partial_X^c
\Big( \a'_2 \delta^b_c + W_{1 c} \partial_{W_1}^b \Big)
\Big( \a'_1 \delta^a_b + W_{2 b} \partial_{W_2}^a \Big) \,,\\
\calL_{-00}^a(X,W_1,W_2) ={}& X^d
\Big( \b'_3 \delta^c_d + X_d \partial_{X}^c \Big)
\Big( \b'_2 \delta^b_c + W_{1 c} \partial_{W_1}^b \Big)
\Big( \b'_1 \delta^a_b + W_{2 b} \partial_{W_2}^a \Big)  \,,\\
\calL_{0+0}^a(X,W_1,W_2) ={}& W_1^e
\Big( \g'_4 \delta^d_e + \partial_{X e} X^{d}  \Big)
\Big( \g'_3 \delta^c_d + X_d \partial_{X}^c \Big)
\Big( \g'_2 \delta^b_c + W_{1 c} \partial_{W_1}^b \Big)
\Big( \g'_1 \delta^a_b + W_{2 b} \partial_{W_2}^a \Big)  \,,\\
\calL_{00+}^a(X,W_1,W_2) ={}& 
W_2^f
\Big( \e'_5 \delta^e_f + \partial_{W_1 f} W_{1}^e  \Big)
\Big( \e'_4 \delta^d_e + \partial_{X e} X^{d}  \Big)
\Big( \e'_3 \delta^c_d + X_d \partial_{X}^c \Big)
\Big( \e'_2 \delta^b_c + W_{1 c} \partial_{W_1}^b \Big)\\
&\Big( \e'_1 \delta^a_b + W_{2 b} \partial_{W_2}^a \Big)  \,,
\eea{eq:ms_ws_L_operators}
where
\bea
\a_i' &= \a_i |_{   \D \to -J_2 +2, J_2 \to -\D + 1}\,,\\
\b_i' &= \b_i |_{   \D \to -J_2 +2, J_2 \to -\D + 3}\,,\\
\g_i' &= \g_i |_{   \D \to -J_2 +2, J_1 \to J_1+1, J_2 \to -\D + 2}\,,\\
\e_i' &= \e_i |_{   \D \to -J_2 +1, J_2 \to -\D + 2}\,.\\
\eea{eq:ms_coeffs_shifted}
Given the clear structure in the ansatz \eqref{eq:ms_ws_operators}, repeating this exercise for even more general mixed-symmetry tensors should not be too hard.
For the sake of clarity of exposition from now on we will focus on symmetric traceless fields.

\subsection{Diagrammatic language}

For the remainder of this paper it will be convenient to extend the diagrammatic notation of \cite{Karateev:2017jgd} to include bulk points.
In addition to  using  a single line to denote a representation of the conformal group labeled by the corresponding operator ${\cal O}$, 
we will denote representations in the bulk of AdS by double lines
and label them by $\f$. To denote functions of two points, more concretely,
conformal two-point functions, bulk-to-boundary propagators and bulk harmonic functions we use the following diagrams
\bea
\text{boundary-boundary:}& \quad
\< \cO (P_1,Z_1) \cO(P_2,Z_2) \> 
&=&
\quad
\diagramEnvelope{\begin{tikzpicture}[anchor=base,baseline]
	\node (opO) at (-1,0) [left] {$\calO$};
	\node (opOprime) at (1,0) [right] {$\calO^\dag$};
	\node (vert) at (0,0) [twopt] {};
	\draw [spinning] (vert) -- (opO);
	\draw [spinning] (vert) -- (opOprime);
\end{tikzpicture}},\\
\text{bulk-boundary:}& \quad
\Pi_{\phi}(X_1,P_2;W_1,Z_2) &=&
\quad
\diagramEnvelope{\begin{tikzpicture}[anchor=base,baseline]
	\node (opO) at (-1,0) [left] {$\phi$};
	\node (opOprime) at (1,0) [right] {$\cO^\dag$};
	\node (vert) at (0,0) [twopt] {};
	\draw [spinning bulk] (vert) -- (opO);
	\draw [spinning] (vert) -- (opOprime);
\end{tikzpicture}},\\
\text{bulk-bulk:}& \quad
\Omega_{\phi}(X_1,X_2;W_1,W_2) &=&
\quad
\diagramEnvelope{\begin{tikzpicture}[anchor=base,baseline]
	\node (opO) at (-1,0) [left] {$\phi$};
	\node (opOprime) at (1,0) [right] {$\phi^\dag$};
	\node (vert) at (0,0) [twopt] {};
	\draw [spinning bulk] (vert) -- (opO);
	\draw [spinning bulk] (vert) -- (opOprime);
\end{tikzpicture}},
\eea{eq:2pt-diagrams}
where at each point we wrote explicitly the dependence on a single polarization vector, as appropriate for symmetric traceless fields,
and for simplicity leave mixed symmetry tensors out of the discussion.
We can think of the dot in these diagrams as separating the representation at one point from the representation at the other point. For example, everything to the left of the dot in the first diagram, that follows the direction of its arrow, is at point $P_1$. This is important for the more elaborated diagrams that follow.
Notice that the reason the harmonic function is used, rather than the bulk-to-bulk propagator, is
that the weight shifting operators act naturally on these and not on the bulk-to-bulk propagator. We will express the bulk-to-bulk propagator in the diagrammatic language in section \ref{sec:bulk-to-bulk-propagator-diagram} below.

In a given diagram the weight shifting operators are determined based on which kind of line they act on, according to
\begin{equation}
\cD^{a}_{(m)}(P,Z) \quad=\quad
\diagramEnvelope{\begin{tikzpicture}[anchor=base,baseline]
	\node (vert) at (0,0) [threept] {$m$};
	\node (opO) at (-0.5,-1) [below] {$\cO$};
	\node (opOprime) at (-0.5,1) [above] {$\cO'$};
	\node (opFinite) at (1,0) [right] {$\cW$};	
	\draw [spinning] (vert)-- (opOprime);
	\draw [spinning] (opO) -- (vert);
	\draw [finite with arrow] (vert) -- (opFinite);
\end{tikzpicture}}\,, \qquad \qquad 
\cL^{a}_{(m)}(X,W) \quad=\quad
\diagramEnvelope{\begin{tikzpicture}[anchor=base,baseline]
	\node (vert) at (0,0) [threept] {$m$};
	\node (opO) at (-0.5,-1) [below] {$\phi$};
	\node (opOprime) at (-0.5,1) [above] {$\phi'$};
	\node (opFinite) at (1,0) [right] {$\cW$};	
	\draw [spinning bulk] (vert)-- (opOprime);
	\draw [spinning bulk] (opO) -- (vert);
	\draw [finite with arrow] (vert) -- (opFinite);
\end{tikzpicture}},
\label{eq:diffoppicture}
\end{equation}
where $\cW$ is a finite-dimensional irreducible representation of $SO(d+1,1)$ as discussed in \cite{Karateev:2017jgd}.
Here the index $a$  runs over a basis of $\cW$. In this paper we will be considering compositions of the weight shifting operators (\ref{eq:ws_operators}, \ref{eq:ws_operators_inverted}) which are themselves in the vector representation $\calV$. The
index $m$ labels the possible operators in the $\cW$ representation, here the operators transform $\cO$ into $\cO'$ and $\f$ into $\f'$ as indicated by the direction of the arrows.

Finally, scalar representations are denoted by dashed lines. 

\section{Crossing relations}
\label{sec:crossing}

In this section we extend the crossing relations for two- and three-point functions derived in \cite{Karateev:2017jgd} to the case of 
bulk harmonic functions and bulk-to-boundary propagators. We shall use the diagrammatic language introduced in the previous section. When translating 
such diagrams to equations we specify, for simplicity, to the case of symmetric traceless operators. 

\subsection{Crossing for two-point functions}
\label{sec:2pt-6j-symbols}

It is easy to guess (and to check) that the crossing equation relating the two-point functions of the operators $\calO$ and $\calO'$ takes the form \cite{Karateev:2017jgd} 
\be
\diagramEnvelope{\begin{tikzpicture}[anchor=base,baseline]
	\node (vertU) at (0,0) [twopt] {};
	\node (vertD) at (1,-0.08) [threept] {$m$};
	\node (opO1) at (-1,0) [left] {$\cO^\dag$};
	\node (opO3) at (2,0) [right] {$\cO'$};
	\node (opW) at (1,-1) [below] {$\cW$};	
	\node at (0.6,0.1) [above] {$\cO$};
	\draw [spinning] (vertU)-- (opO1);
	\draw [spinning] (vertU)-- (vertD);
	\draw [spinning] (vertD)-- (opO3);
	\draw [finite with arrow] (vertD)-- (opW);
\end{tikzpicture}}
	\quad=\quad
\left\{ \begin{matrix}
\cO^\dag\\ \cO'
\end{matrix} \right\}^{m}_{\bar{m}} 
\diagramEnvelope{\begin{tikzpicture}[anchor=base,baseline]
	\node (vertL) at (1,0) [twopt] {};
	\node (vertR) at (0,-0.12) [threept] {$\bar{m}$};
	\node (opO1) at (-1,0) [left] {$\cO^\dag$};
	\node (opO3) at (2,0) [right] {$\cO'$};
	\node (opW) at (0,-1) [below] {$\cW$};
	\node at (0.6,0.1) [above] {$\cO'^\dag$};
	\draw [spinning] (vertR)-- (opO1);
	\draw [spinning] (vertL)-- (vertR);
	\draw [spinning] (vertL)-- (opO3);
	\draw [finite with arrow] (vertR)-- (opW);
\end{tikzpicture}},
\label{eq:twoptcrossing}
\ee
with $\bar{m}$ denoting the weight shifting operator for the shift opposite to $m$.
Here $\cO^\dag$ indicates the dual-reflected representation with respect to $\cO$.
Since all our examples consider real bosons, we often omit the daggers in the remainder of the paper.
For symmetric traceless operators this diagram encodes the equation
\beq
\calD_{(m)}^a(P_2,Z_2)
\< \cO (P_1,Z_1) \cO(P_2,Z_2) \> 
=
\left\{ \begin{matrix}
\cO^\dag\\ \cO'
\end{matrix} \right\}^{m}_{\bar{m}} 
\calD_{(\bar{m})}^a(P_1,Z_1)
\< \cO' (P_1,Z_1) \cO'(P_2,Z_2) \>  \,.
\label{eq:twoptcrossing_formula}
\eeq
An analogous equation holds in the case of bulk harmonic functions
\be
\diagramEnvelope{\begin{tikzpicture}[anchor=base,baseline]
	\node (vertU) at (0,0) [twopt] {};
	\node (vertD) at (1,-0.08) [threept] {$m$};
	\node (opO1) at (-1,0) [left] {$\f^\dag$};
	\node (opO3) at (2,0) [right] {$\f'$};
	\node (opW) at (1,-1) [below] {$\cW$};	
	\node at (0.6,0.1) [above] {$\f$};
	\draw [spinning bulk] (vertU)-- (opO1);
	\draw [spinning bulk] (vertU)-- (vertD);
	\draw [spinning bulk] (vertD)-- (opO3);
	\draw [finite with arrow] (vertD)-- (opW);
\end{tikzpicture}}
	\quad=\quad
\left\{ \begin{matrix}
\f^\dag \\ \f'
\end{matrix} \right\}^{m}_{\bar{m}}
\diagramEnvelope{\begin{tikzpicture}[anchor=base,baseline]
	\node (vertL) at (1,0) [twopt] {};
	\node (vertR) at (0,-0.12) [threept] {$\bar{m}$};
	\node (opO1) at (-1,0) [left] {$\f^\dag$};
	\node (opO3) at (2,0) [right] {$\f'$};
	\node (opW) at (0,-1) [below] {$\cW$};
	\node at (0.4,0.1) [above] {$\f'^\dag$};
	\draw [spinning bulk] (vertR)-- (opO1);
	\draw [spinning bulk] (vertL)-- (vertR);
	\draw [spinning bulk] (vertL)-- (opO3);
	\draw [finite with arrow] (vertR)-- (opW);
\end{tikzpicture}},
\label{eq:twoptcrossing_bulk}
\ee
which, for symmetric, traceless and transverse bulk fields, reads
\beq
\cL_{(m)}^a(X_2,W_2)\,
\Omega_{\f}(X_1,X_2;W_1,W_2) =
\left\{ \begin{matrix}
\f^\dag\\ \f'
\end{matrix} \right\}^{m}_{\bar{m}} 
\cL_{(\bar{m})}^a(X_1,W_1)\,
\Omega_{\f'}(X_1,X_2;W_1,W_2)\,.
\label{eq:twoptcrossing_bulk_formula}
\eeq
One can also mix $\cD$ and $\cL$ operators and obtain a crossing equation for bulk-to-boundary propagators
\be
\diagramEnvelope{\begin{tikzpicture}[anchor=base,baseline]
	\node (vertU) at (0,0) [twopt] {};
	\node (vertD) at (1,-0.08) [threept] {$m$};
	\node (opO1) at (-1,0) [left] {$\cO^\dag$};
	\node (opO3) at (2,0) [right] {$\f'$};
	\node (opW) at (1,-1) [below] {$\cW$};	
	\node at (0.6,0.1) [above] {$\f$};
	\draw [spinning] (vertU)-- (opO1);
	\draw [spinning bulk] (vertU)-- (vertD);
	\draw [spinning bulk] (vertD)-- (opO3);
	\draw [finite with arrow] (vertD)-- (opW);
\end{tikzpicture}}
	\quad=\quad
\left\{ \begin{matrix}
\cO^\dag\\ \f'
\end{matrix} \right\}^{m}_{\bar{m}} 
\diagramEnvelope{\begin{tikzpicture}[anchor=base,baseline]
	\node (vertL) at (1,0) [twopt] {};
	\node (vertR) at (0,-0.12) [threept] {$\bar{m}$};
	\node (opO1) at (-1,0) [left] {$\cO^\dag$};
	\node (opO3) at (2,0) [right] {$\f'$};
	\node (opW) at (0,-1) [below] {$\cW$};
	\node at (0.4,0.1) [above] {$\cO'^\dag$};
	\draw [spinning] (vertR)-- (opO1);
	\draw [spinning] (vertL)-- (vertR);
	\draw [spinning bulk] (vertL)-- (opO3);
	\draw [finite with arrow] (vertR)-- (opW);
\end{tikzpicture}},
\label{eq:twoptcrossing_bulk_boundary}
\ee
that is, 
\beq
\cL_{(m)}^a(X_2,W_2)\,
\Pi_{\f}(P_1,X_2;Z_1,W_2) =
\left\{ \begin{matrix}
\cO^\dag\\ \f'
\end{matrix} \right\}^{m}_{\bar{m}} 
\calD_{(\bar{m})}^a(P_1,Z_1)\,
\Pi_{\f'}(P_1,X_2;Z_1,W_2)\,.
\label{eq:twoptcrossing_bulk_boundary_formula}
\eeq
The coefficients $\{ \ \}$ defined in the equations above are given in appendix \ref{app:2pt-6j-symbols} for the weight shifting operators in the vector representation (\ref{eq:ws_operators}, \ref{eq:ws_operators_inverted}).

\subsection{Bubbles}
\label{sec:bubbles}

When two weight shifting operators acting on the same leg are contracted,
this corresponds to a bubble diagram which simplifies   \cite{Karateev:2017jgd}
\be
\cD^{(m)}_{a}(P_1,Z_1) \cD^{(n)a} (P_1,Z_1)\ =\ 
\diagramEnvelope{\begin{tikzpicture}[anchor=base,baseline]
\node (top) at (0,1.8) [above] {$\cO_1'$};
\node (topmid) at (0,0.9) [threept] {$n$};
\node (botmid) at (0,-0.9) [threept] {$m$};
\node (bot) at (0,-1.8) [below] {$\cO_1''$};
\node () at (0.7,0) [right] {$\cW$};
\node () at (-0.7,0) [left] {$\cO_1$};
\draw [spinning] (top) -- (topmid);
\draw [spinning] (botmid) -- (bot);
\draw [finite with arrow] (topmid) to[out=-20,in=20] (botmid);
\draw [spinning] (topmid) to[out=-160,in=160] (botmid);
\end{tikzpicture}}
&=
\begin{pmatrix}
\cO_1'\\
\cO_1\ \cW
\end{pmatrix}^{mn} \de_{\cO_1'\cO_1''}
\diagramEnvelope{\begin{tikzpicture}[anchor=base,baseline]
\node (top) at (0,1) [above] {$\cO_1'$};
\node (bot) at (0,-1) [below] {$\cO_1'$};
\draw [spinning] (top) -- (bot);
\end{tikzpicture}}\,.
\label{eq:bubble}
\ee
The same  happens in the bulk
\be
\cL^{(m)}_{a}(X_1,W_1) \cL^{(n)a} (X_1,W_1)\ =\ 
\diagramEnvelope{\begin{tikzpicture}[anchor=base,baseline]
\node (top) at (0,1.8) [above] {$\f_1'$};
\node (topmid) at (0,0.9) [threept] {$n$};
\node (botmid) at (0,-0.9) [threept] {$m$};
\node (bot) at (0,-1.8) [below] {$\f_1''$};
\node () at (0.7,0) [right] {$\cW$};
\node () at (-0.7,0) [left] {$\f_1$};
\draw [spinning bulk] (top) -- (topmid);
\draw [spinning bulk] (botmid) -- (bot);
\draw [finite with arrow] (topmid) to[out=-20,in=20] (botmid);
\draw [spinning bulk] (topmid) to[out=-160,in=160] (botmid);
\end{tikzpicture}}
&=
\begin{pmatrix}
\f_1'\\
\f_1\ \cW
\end{pmatrix}^{mn} \de_{\f_1'\f_1''}
\diagramEnvelope{\begin{tikzpicture}[anchor=base,baseline]
\node (top) at (0,1) [above] {$\f_1'$};
\node (bot) at (0,-1) [below] {$\f_1'$};
\draw [spinning bulk] (top) -- (bot);
\end{tikzpicture}}\,.
\label{eq:bubble_bulk}
\ee
The coefficients defined by these two equations are computed in appendix \ref{app:bubbles}.
Combining this with the crossing equation 
\eqref{eq:twoptcrossing_bulk}, different harmonic functions are related by
\be
\diagramEnvelope{\begin{tikzpicture}[anchor=base,baseline]
	\node (opLeft) at (-2,0) [left] {$\f$};
	\node (vertLeft) at (-1,-0.08) [threept,inner sep=1pt] {$m$};
	\node (vertCenter) at (0,0) [twopt] {};
	\node (vertRight) at (1,-0.08) [threept,inner sep=1pt] {$m$};
	\node (opRight) at (2,0) [right] {$\f^\dag$};
	\node at (0,-0.9) [below] {$\cW$};
	\node at (0,0.2) [above] {$\f'$};
	\draw [spinning bulk] (vertLeft) -- (opLeft);
	\draw [spinning bulk] (vertCenter) -- (vertLeft);
	\draw [spinning bulk] (vertRight) -- (opRight);
	\draw [spinning bulk] (vertCenter) -- (vertRight);
	\draw [finite with arrow] (vertLeft) to[out=-90,in=-90] (vertRight);
\end{tikzpicture}}
&=
	\left\{
		\begin{matrix}
		\f' \\
		\f^\dag
		\end{matrix}
	\right\}^{m}_{\bar{m}}
\diagramEnvelope{\begin{tikzpicture}[anchor=base,baseline]
	\node (opLeft) at (-2,0) [left] {$\f$};
	\node (shad) at (1.5,0) [twopt] {};
	\node (vert1) at (-1,-0.08) [threept,inner sep=1pt] {$m$};
	\node (vert2) at (0.5,-0.12) [threept,inner sep=1pt] {$\bar{m}$};
	\node (opRight) at (2.5,0) [right] {$\f^\dag$};
	\node at (-0.2,-0.8) [below] {$\cW$};
	\node at (-0.2,0.2) [above] {$\f'$};
	\draw [spinning bulk] (vert1) -- (opLeft);
	\draw [spinning bulk] (shad) -- (vert2);
	\draw [spinning bulk] (vert2) -- (vert1);
	\draw [spinning bulk] (shad) -- (opRight);
	\draw [finite with arrow] (vert1) to[out=-90,in=-90] (vert2);
\end{tikzpicture}}
\nn\\
&=
	\left\{
		\begin{matrix}
		\f'  \\
		\f^\dag
		\end{matrix}
	\right\}^{m}_{\bar{m}}
\begin{pmatrix}
\f\\
\f' \cW
\end{pmatrix}^{m \bar{m}}
\diagramEnvelope{\begin{tikzpicture}[anchor=base,baseline]
	\node (opLeft) at (-1,0) [left] {$\f$};
	\node (shad) at (0,0) [twopt] {};
	\node (opRight) at (1,0) [right] {$\f^\dag$};
	\draw [spinning bulk] (shad) -- (opLeft);
	\draw [spinning bulk] (shad) -- (opRight);
\end{tikzpicture}}.
\label{eq:harmonic_relation}
\ee
Analogous relations hold for conformal two-point structures and bulk-to-boundary propagators.

\subsection{Shadow integrals}

Let us here review the definition of the shadow integral (which allows us to "glue" two CFT representations) and how it enters operations 
with weight shifting operators \cite{Karateev:2017jgd}.
The diagrammatic notation for the shadow integral for spin $J$ operators is given by
\be
\diagramEnvelope{\begin{tikzpicture}[anchor=base,baseline]
	\node (opO) at (-1,0) [left] {$\cO$};
	\node (opOprime) at (1,0) [right] {$\cO$};
	\node (vert) at (0,0) [cross] {};
	\draw [spinning] (opO) -- (vert);
	\draw [spinning] (opOprime) -- (vert);
\end{tikzpicture}}
\,
= \ 
&\frac{1}{(J! (h-1)_J)^2\cN_{\De,J}} \int D^d P_1\, D^dP_2 \big|\calO(P_1, D_{Z_1})\big\>
\nonumber
\\
& \big\< \cO(P_1, Z_1) \cO(P_2,D_{Z_2})\big\>\Big|_{\D \to d-\D} \big\<\calO(P_2,Z_2)\big|
\,,
\label{eq:gluing_spin}
\ee
where $D_Z$ is the differential operator implementing a traceless symmetric contraction of indices
\be
D_Z^a = \left( h-1+ Z\cdot \partial_Z \right) \partial_Z^a - \frac{1}{2} Z^a \partial_Z \cdot \partial_Z\,,
\ee
and the normalization constant is fixed to
\be
\cN_{\De,J} &= \frac{\pi^d (\D-1) (d-\D-1) \G(\De- d/2)\G(d/2- \De)}{(\D-1+J) (d-\D-1+J)\G(\De)\G(d-\De)}\,,
\label{eq:N}
\ee
by requiring that the shadow integral acting on a two-point function gives the identity transformation
\be
\diagramEnvelope{\begin{tikzpicture}[anchor=base,baseline]
	\node (opO) at (-1,0) [left] {$\cO$};
	\node (vert) at (0,0) [twopt] {};
	\node (shad) at (1,0) [cross] {};
	\node (opOprime) at (2,0) [right] {$\cO$};
	\draw [spinning] (vert) -- (opO);
	\draw [spinning] (vert) -- (shad);
	\draw [spinning] (opOprime) -- (shad);
\end{tikzpicture}}
&\quad=\quad
\diagramEnvelope{\begin{tikzpicture}[anchor=base,baseline]
	\node (opO) at (-1,0) [left] {$\cO$};
	\node (opOprime) at (0,0) [right] {$\cO$};
	\draw [spinning] (opOprime) -- (opO);
\end{tikzpicture}}.
\label{eq:shadownormalization}
\ee
This combination of shadow integral and two-point function is equivalent to performing two consecutive shadow transformations, which should give the identity.
The shadow transformation is defined by
\be
\tilde{\calO}(P,Z)
= \frac{1}{J! (h-1)_J}
\ 
\int D^d P_0\,  \big\< \cO(P, Z) \cO(P_0,D_{Z_0})\big\>\Big|_{\D \to d-\D} \calO(P_0,Z_0)\,.
\label{eq:shadow_transform}
\ee
It was shown in \cite{Karateev:2017jgd}
that the condition \eqref{eq:shadownormalization} is enough to show that weight shifting operators can be commuted past shadow integrals,
trivially performing an integration by parts,
\be
\diagramEnvelope{\begin{tikzpicture}[anchor=base,baseline]
	\node (vertU) at (1,0) [cross] {};
	\node (vertD) at (0,-0.08) [threept] {$m$};
	\node (opO1) at (-1,0) [left] {$\cO$};
	\node (opO3) at (2,0) [right] {$\cO'^\dag$};
	\node (opW) at (0,-1) [below] {$\cW$};	
	\node at (0.6,0.1) [above] {$\cO'$};
	\draw [spinning] (opO1)-- (vertD);
	\draw [spinning] (vertD)-- (vertU);
	\draw [spinning] (opO3)-- (vertU);
	\draw [finite with arrow] (vertD)-- (opW);
\end{tikzpicture}}
	\quad=\quad
	\left\{
		\begin{matrix}
		\cO^\dag \\
		\cO'
		\end{matrix}
	\right\}^{m}_{\bar{m}}
\diagramEnvelope{\begin{tikzpicture}[anchor=base,baseline]
	\node (vertL) at (0,0) [cross] {};
	\node (vertR) at (1,-0.12) [threept, inner sep=1pt] {$\bar{m}$};
	\node (opO1) at (-1,0) [left] {$\cO$};
	\node (opO3) at (2,0) [right] {$\cO'^\dag$};
	\node (opW) at (1,-1) [below] {$\cW$};
	\node at (0.4,0.1) [above] {$\cO^\dag$};
	\draw [spinning] (opO1)-- (vertL);
	\draw [spinning] (vertR)-- (vertL);
	\draw [spinning] (opO3)-- (vertR);
	\draw [finite with arrow] (vertR)-- (opW);
\end{tikzpicture}}
\,.
\label{eq:integrationbyparts}
\ee
In particular, shadow integrals for different representations are related by
\be
\diagramEnvelope{\begin{tikzpicture}[anchor=base,baseline]
	\node (opLeft) at (-2,0) [left] {$\cO$};
	\node (vertLeft) at (-1,-0.08) [threept,inner sep=1pt] {$m$};
	\node (vertCenter) at (0,0) [cross] {};
	\node (vertRight) at (1,-0.08) [threept,inner sep=1pt] {$m$};
	\node (opRight) at (2,0) [right] {$\cO^\dag$};
	\node at (0,-0.9) [below] {$\cW$};
	\node at (0,0.1) [above] {$\cO'$};
	\draw [spinning] (opLeft) -- (vertLeft);
	\draw [spinning] (vertLeft) -- (vertCenter);
	\draw [spinning] (opRight) -- (vertRight);
	\draw [spinning] (vertRight) -- (vertCenter);
	\draw [finite with arrow] (vertLeft) to[out=-90,in=-90] (vertRight);
\end{tikzpicture}}
&=
	\left\{
		\begin{matrix}
		\cO^\dag\\
		\cO'
		\end{matrix}
	\right\}^{m}_{\bar{m}}
\diagramEnvelope{\begin{tikzpicture}[anchor=base,baseline]
	\node (opLeft) at (-2,0) [left] {$\cO$};
	\node (shad) at (-1,0) [cross] {};
	\node (vert1) at (0,-0.12) [threept,inner sep=1pt] {$\bar{m}$};
	\node (vert2) at (1.5,-0.08) [threept,inner sep=1pt] {$m$};
	\node (opRight) at (2.5,0) [right] {$\cO^\dag$};
	\node at (0.8,-0.8) [below] {$\cW$};
	\node at (0.8,0.2) [above] {$\cO'^\dag$};
	\draw [spinning] (opLeft) -- (shad);
	\draw [spinning] (vert1) -- (shad);
	\draw [spinning] (vert2) -- (vert1);
	\draw [spinning] (opRight) -- (vert2);
	\draw [finite with arrow] (vert1) to[out=-90,in=-90] (vert2);
\end{tikzpicture}}
\nn\\
&=
	\left\{
		\begin{matrix}
		\cO^\dag \\
		\cO'
		\end{matrix}
	\right\}^{m}_{\bar{m}}
\begin{pmatrix}
\cO^\dag\\
\cO'^\dag \cW
\end{pmatrix}^{\bar{m} m}
\diagramEnvelope{\begin{tikzpicture}[anchor=base,baseline]
	\node (opLeft) at (-1,0) [left] {$\cO$};
	\node (shad) at (0,0) [cross] {};
	\node (opRight) at (1,0) [right] {$\cO^\dag$};
	\draw [spinning] (opLeft) -- (shad);
	\draw [spinning] (opRight) -- (shad);
\end{tikzpicture}}.
\label{eq:shadow_relation}
\ee

\subsection{Split representation}

The harmonic functions have a representation as an integral over the AdS boundary
of a product of two bulk-to-boundary propagators. It is termed the split representation, which for scalar fields 
takes the form
\beq
\Omega_{\D} (X_1,X_2) = -\frac{(\D-\frac{d}{2})^2}{\pi}\int_{\partial} dP \ \Pi_{\D} (X_1,P) \,{\tilde\Pi}_{\D} (X_2,P)\,.
\label{eq:split_omega_scalar}
\eeq
We define the ``shadow'' bulk-to-boundary propagator with the tilde to be a propagator for a dimension $d-\D$ field,
 with the correct homogeneity in $X_2$ expected for $\Omega_{\D} (X_1,X_2)$,
\beq
{\tilde\Pi}_{\D} (X,P) \equiv (-X^2)^{\frac{d}{2}-\D} {\Pi}_{d-\D} (X,P)\,.
\eeq
This shadow propagator is related to ${\Pi}_{\D}$ by a shadow transformation
\beq
{\tilde\Pi}_{\D} (X,P) =  \frac{\Gamma(\D)}{\pi^{\frac{d}{2}}\Gamma(\D-\tfrac{d}{2})}
\frac{\cC_{d-\D}}{\cC_\D} \int_\partial dP_0 \, \frac{1}{(-2 P\cdot P_0)^{(d-\D)}} \,{\Pi}_{\D} (X,P_0)\,.
\label{eq:shadow_transform_propagator}
\eeq

Combining \eqref{eq:split_omega_scalar}, \eqref{eq:shadow_transform_propagator} 
we can now relate the harmonic function to the shadow  integral \eqref{eq:gluing_spin}  dressed with two bulk-to-boundary propagators
\be
\diagramEnvelope{\begin{tikzpicture}[anchor=base,baseline]
	\node (opO) at (-1,0) [left] {$\phi_\D$};
	\node (opOprime) at (1,0) [right] {$\phi_\D$};
	\node (vert) at (0,0) [twopt] {};
	\draw [scalar bulk] (vert) -- (opO);
	\draw [scalar bulk] (vert) -- (opOprime);
\end{tikzpicture}}
\quad = \ \cT_\D \ 
\diagramEnvelope{\begin{tikzpicture}[anchor=base,baseline]
	\node (opO) at (-2,0) [left] {$\phi_{\D}$};
	\node (opOprime) at (2,0) [right] {$\phi_{\D}$};
	\node (propO) at (-1,0) [twopt] {};
	\node (propOprime) at (1,0) [twopt] {};
	\node (vert) at (0,0) [cross] {};
	\node at (0,0.2) [above] {$\cO_{\D}$};
	\draw [scalar] (propO) -- (vert);
	\draw [scalar] (propOprime) -- (vert);
	\draw [scalar bulk] (propO) -- (opO);
	\draw [scalar bulk] (propOprime) -- (opOprime);
\end{tikzpicture}}
\,,
\label{eq:propagator_to_shadow_integral}
\ee
where dashed lines indicate scalar representations and we defined
\beq
\cT_\D = \pi^{\frac{d}{2}-1} \frac{\left(\D-\frac{d}{2}\right)^2}{\left(\D-\frac{d}{2}\right)_\frac{d}{2} }\,.
\eeq
An analogous relation holds for the spin $J$ harmonic function
\be
\diagramEnvelope{\begin{tikzpicture}[anchor=base,baseline]
	\node (opLeft) at (-1,0) [left] {$\f_{\D,J}$};
	\node (shad) at (0,0) [twopt] {};
	\node (opRight) at (1,0) [right] {$\f_{\D,J}$};
	\draw [spinning bulk] (shad) -- (opLeft);
	\draw [spinning bulk] (shad) -- (opRight);
\end{tikzpicture}} 
=
\ \cT_{\D,J} \ 
\diagramEnvelope{\begin{tikzpicture}[anchor=base,baseline]
	\node (opLeft) at (-2,0) [left] {$\f_{\D,J}$};
	\node (propLeft) at (-1,0) [twopt] {};
	\node (vertCenter) at (0,0) [cross] {};
	\node (propRight) at (1,0) [twopt] {};
	\node (opRight) at (2,0) [right] {$\f_{\D,J}$};
	\node at (0,0.2) [above] {$\cO_{\D,J}$};
	\draw [spinning bulk] (propLeft) -- (opLeft);
	\draw [spinning] (propLeft) -- (vertCenter);
	\draw [spinning bulk] (propRight) -- (opRight);
	\draw [spinning] (propRight) -- (vertCenter);
\end{tikzpicture}}
\,.
\label{eq:spinning_omega_to_shadow}
\ee
The coefficient can be computed easily with the help of weight shifting operators.
Using \eqref{eq:harmonic_relation} and then \eqref{eq:propagator_to_shadow_integral},
the harmonic function for spin $J$ can also be written in terms of a scalar shadow integral
\be
&\diagramEnvelope{\begin{tikzpicture}[anchor=base,baseline]
	\node (opLeft) at (-1,0) [left] {$\f_{\D,J}$};
	\node (shad) at (0,0) [twopt] {};
	\node (opRight) at (1,0) [right] {$\f_{\D,J}$};
	\draw [spinning bulk] (shad) -- (opLeft);
	\draw [spinning bulk] (shad) -- (opRight);
\end{tikzpicture}}=
\frac{1}{\{ \ldots \} (\ldots)} \ 
\diagramEnvelope{\begin{tikzpicture}[anchor=base,baseline]
	\node (opLeft) at (-2,0) [left] {$\f_{\D,J}$};
	\node (vertLeft) at (-1,-0.1) [threept] {$m$};
	\node (vertCenter) at (0,0) [twopt] {};
	\node (vertRight) at (1,-0.1) [threept] {$m$};
	\node (opRight) at (2,0) [right] {$\f_{\D,J}$};
	\node at (0,-0.9) [below] {$\cW$};
	\node at (0,0.2) [above] {$\f_\D$};
	\draw [spinning bulk] (vertLeft) -- (opLeft);
	\draw [scalar bulk] (vertCenter) -- (vertLeft);
	\draw [spinning bulk] (vertRight) -- (opRight);
	\draw [scalar bulk] (vertCenter) -- (vertRight);
	\draw [finite with arrow] (vertLeft) to[out=-90,in=-90] (vertRight);
\end{tikzpicture}}
\nn\\
&=\frac{\cT_\D}{\{ \ldots \} (\ldots)} \quad
\diagramEnvelope{\begin{tikzpicture}[anchor=base,baseline]
	\node (opLeft) at (-3,0) [left] {$\f_{\D,J}$};
	\node (vertLeft) at (-2,-0.1) [threept] {$m$};
	\node (propLeft) at (-1,0) [twopt] {};
	\node (vertCenter) at (0,0) [cross] {};
	\node (propRight) at (1,0) [twopt] {};
	\node (vertRight) at (2,-0.1) [threept] {$m$};
	\node (opRight) at (3,0) [right] {$\f_{\D,J}$};
	\node at (0,-0.9) [below] {$\cW$};
	\node at (0,0.2) [above] {$\cO_{\D}$};
	\node at (1.5,0.2) [above] {$\f_{\D}$};
	\node at (-1.5,0.2) [above] {$\f_{\D}$};
	\draw [spinning bulk] (vertLeft) -- (opLeft);
	\draw [scalar bulk] (propLeft) -- (vertLeft);
	\draw [scalar] (propLeft) -- (vertCenter);
	\draw [spinning bulk] (vertRight) -- (opRight);
	\draw [scalar bulk] (propRight) -- (vertRight);
	\draw [scalar] (propRight) -- (vertCenter);
	\draw [finite with arrow] (vertLeft) to[out=-90,in=-90] (vertRight);
\end{tikzpicture}}
\label{eq:spinning_omega_to_scalar_shadow}
\ee
where $m = (0,J)$ and
\bea
\{ \ldots \} (\ldots)
&\equiv
\left\{\begin{matrix}
\f_{\D}\\ \f_{\D,J} 
\end{matrix} \right\}^{m}_{\bar{m}}
\begin{pmatrix}
\f_{\D,J}\\
\f_{\D} \cW
\end{pmatrix}^{m \bar{m}}\,.
\eea{eq:spinning_omega_factor}
We can relate this to the spin $J$ shadow integral defined in \eqref{eq:gluing_spin}
by commuting the weight shifting operators past the bulk-to-boundary propagators and the shadow integrals
using \eqref{eq:twoptcrossing_bulk_boundary} and \eqref{eq:shadow_relation}
\be
\eqref{eq:spinning_omega_to_scalar_shadow}
&=
\frac{\cT_\D}{\{ \ldots \} (\ldots)}
 \{ \ldots \}^{2}
\diagramEnvelope{\begin{tikzpicture}[anchor=base,baseline]
	\node (opLeft) at (-3,0) [left] {$\f_{\D,J}$};
	\node (vertLeft) at (-1,-0.11) [threept] {$\bar{m}$};
	\node (propLeft) at (-2,0) [twopt] {};
	\node (vertCenter) at (0,0) [cross] {};
	\node (propRight) at (2,0) [twopt] {};
	\node (vertRight) at (1,-0.11) [threept] {$\bar{m}$};
	\node (opRight) at (3,0) [right] {$\f_{\D,J}$};
	\node at (0,-0.9) [above] {$\cW$};
	\node at (0,0.2) [above] {$\cO_{\D}$};
	\node at (1.5,0.2) [above] {$\f_{\D,J}$};
	\node at (-1.5,0.2) [above] {$\f_{\D,J}$};
	\draw [spinning bulk] (propLeft) -- (opLeft);
	\draw [spinning] (propLeft) -- (vertLeft);
	\draw [scalar] (vertLeft) -- (vertCenter);
	\draw [spinning bulk] (propRight) -- (opRight);
	\draw [spinning] (propRight) -- (vertRight);
	\draw [scalar] (vertRight) -- (vertCenter);
	\draw [finite with arrow] (vertLeft) to[out=-90,in=-90] (vertRight);
\end{tikzpicture}}
\label{eq:harmonic_to_spinning_shadow_integral}\\
&=
\frac{\cT_\D}{\{ \ldots \} (\ldots)}
 \{ \ldots \}^{3} (\ldots)
\diagramEnvelope{\begin{tikzpicture}[anchor=base,baseline]
	\node (opLeft) at (-2,0) [left] {$\f_{\D,J}$};
	\node (propLeft) at (-1,0) [twopt] {};
	\node (vertCenter) at (0,0) [cross] {};
	\node (propRight) at (1,0) [twopt] {};
	\node (opRight) at (2,0) [right] {$\f_{\D,J}$};
	\node at (0,0.2) [above] {$\cO_{\D,J}$};
	\draw [spinning bulk] (propLeft) -- (opLeft);
	\draw [spinning] (propLeft) -- (vertCenter);
	\draw [spinning bulk] (propRight) -- (opRight);
	\draw [spinning] (propRight) -- (vertCenter);
\end{tikzpicture}} \nn
\ee
where
\bea
\{ \ldots \}^{3} (\ldots) ={}&
\left(
\left\{\begin{matrix}
 \calO_{\D}\\
 \f_{\D,J} 
\end{matrix} \right\}^{m}_{\bar{m}}
\right)^2
\left\{\begin{matrix}
\calO_{\D,J}\\	\calO_{\D}
\end{matrix} \right\}^{\bar{m}}_{m}
\begin{pmatrix}
\calO_{\D,J}\\
\calO_{\D} \cW
\end{pmatrix}^{m \bar{m}}\,.
\eea{eq:6jcoeffs_1}
Combining \eqref{eq:spinning_omega_to_scalar_shadow} and \eqref{eq:harmonic_to_spinning_shadow_integral} we 
see that the coefficient in \eqref{eq:spinning_omega_to_shadow} is
\be
\cT_{\D,J} \equiv  \cT_\D  \, \frac{\{ \ldots \}^{3} (\ldots)}{\{ \ldots \}(\ldots)}
= \cT_\D  \, \frac{\D-1}{\D-1+J}\,.
\ee

\subsection{Crossing for three-point functions}

Following \cite{Karateev:2017jgd}  a conformally invariant three-point structure is denoted by
\begin{equation}
\big\<\cO_1 (P_1,Z_1)\cO_2 (P_2,Z_2) \cO_3 (P_3,Z_3)\big\>^{(a)} \quad=\quad
\diagramEnvelope{\begin{tikzpicture}[anchor=base,baseline]
	\node (vert) at (0,-0.08) [threept] {$a$};
	\node (opO1) at (-0.5,-1) [below] {$\cO_1$};
	\node (opO2) at (-0.5,1) [above] {$\cO_2$};
	\node (opO3) at (1,0) [right] {$\cO_3$};	
	
	\draw [spinning] (vert)-- (opO1);
	\draw [spinning] (vert)-- (opO2);
	\draw [spinning] (vert)-- (opO3);
\end{tikzpicture}}.
\label{eq:3pt_function}
\end{equation}
Here the arrows point away from the center of the diagram, so each operator is inserted at a different point,
just as for two-point functions in (\ref{eq:2pt-diagrams}).
The number of independent structures in a three-point function depends on the representations of the operators, as  determined in \cite{Costa:2011mg,Costa:2014rya,Kravchuk:2016qvl}.
The index $a$ in \eqref{eq:3pt_function} labels these structures.
These three-point structures satisfy the following crossing equation, allowing to move weight shifting operators
$\cD$ from one leg to another  \cite{Karateev:2017jgd} 
\be
\diagramEnvelope{\begin{tikzpicture}[anchor=base,baseline]
	\node (vertL) at (0,0) [threept] {$a$};
	\node (vertR) at (2,0) [threept] {$m$};
	\node (opO1) at (-0.5,-1) [below] {$\cO_1$};
	\node (opO2) at (-0.5,1) [above] {$\cO_2$};
	\node (opO3) at (2.5,1) [above] {$\cO_3$};
	\node (opW) at (2.5,-1) [below] {$\cW$};	
	\node at (1,0.1) [above] {$\cO_3'$};	
	\draw [spinning] (vertL)-- (opO1);
	\draw [spinning] (vertL)-- (opO2);
	\draw [spinning] (vertL)-- (vertR);
	\draw [spinning] (vertR)-- (opO3);
	\draw [finite with arrow] (vertR)-- (opW);
\end{tikzpicture}}
	\quad=\quad
	\sum_{\cO_1',b,n}
	\left\{
		\begin{matrix}
		\cO_1 & \cO_2 & \cO_1' \\
		\cO_3 & \cW & \cO_3'
		\end{matrix}
	\right\}^{am}_{bn}
\diagramEnvelope{\begin{tikzpicture}[anchor=base,baseline]
	\node (vertU) at (0,0.7) [threept] {$b$};
	\node (vertD) at (0,-0.7) [threept] {$n$};
	\node (opO1) at (-1,-1.5) [below] {$\cO_1$};
	\node (opO2) at (-1,1.5) [above] {$\cO_2$};
	\node (opO3) at (1,1.5) [above] {$\cO_3$};
	\node (opW) at (1,-1.5) [below] {$\cW$};	
	\node at (0.1,0) [right] {$\cO_1'$};	
	\draw [spinning] (vertD)-- (opO1);
	\draw [spinning] (vertU)-- (opO2);
	\draw [spinning] (vertU)-- (vertD);
	\draw [spinning] (vertU)-- (opO3);
	\draw [finite with arrow] (vertD)-- (opW);
\end{tikzpicture}}\,.
\label{eq:6jdefinition}
\ee
The corresponding equation has the form
\bea
{}&
\cD^{c}_{(m)}(P_3,Z_3)\big\<\cO_1 (P_1,Z_1)\cO_2 (P_2,Z_2) \cO'_3 (P_3,Z_3)\big\>^{(a)}=
\\
{}&
\sum_{\cO_1',p,q}
	\left\{
		\begin{matrix}
		\cO_1 & \cO_2 & \cO_1' \\
		\cO_3 & \cW & \cO_3'
		\end{matrix}
	\right\}^{am}_{bn}
\cD^{c}_{(n)}(P_1,Z_1)\big\<\cO'_1 (P_1,Z_1)\cO_2 (P_2,Z_2) \cO_3 (P_3,Z_3)\big\>^{(b)}\,.
\eea{eq:3ptCFTcrossing}
Note that \eqref{eq:6jdefinition} reduces to the two-point crossing \eqref{eq:twoptcrossing} when taking $\cO_2$ to be the identity. 
As a consequence the coefficients, which are  called $6j$ symbols, are related by
\be
	\left\{
		\begin{matrix}
		\cO_1 & \mathbf{1} & \cO_3 \\
		\cO_3 & \cW & \cO_1
		\end{matrix}
	\right\}^{\uniq m}_{\uniq \bar{m}}
=
\left\{ \begin{matrix}
\cO_1 \\ \cO_3
\end{matrix} \right\}^{m}_{\bar{m}}\,.
\label{eq:shorthand_6j_2pt}
\ee

Whenever two of the operators in a three-point function are scalars, the third operator has to be a symmetric tensor\footnote{We are considering only parity even structures for simplicity.} and there exists only the  tensor structure \eqref{eq:3pt_correlator}. Indicating scalars by dashed lines, this structure has the diagram\footnote{The ordering of the scalar operators on the left hand side is not obvious from the diagram. We will always order the scalars by their labels when translating from diagrams to the formula \eqref{eq:3pt_correlator}.}
\begin{equation}
\big\<\cO_1 (P_1)\cO_2 (P_2) \cO_3 (P_3,Z_3)\big\> \quad=\quad
\diagramEnvelope{\begin{tikzpicture}[anchor=base,baseline]
	\node (vert) at (0,0) [twopt] {};
	\node (opO1) at (-0.5,-1) [below] {$\cO_1$};
	\node (opO2) at (-0.5,1) [above] {$\cO_2$};
	\node (opO3) at (1,0) [right] {$\cO_3$};	
	
	\draw [scalar] (vert)-- (opO1);
	\draw [scalar] (vert)-- (opO2);
	\draw [spinning] (vert)-- (opO3);
\end{tikzpicture}}.
\end{equation}
For explicit examples we will mostly use this case for simplicity,
in particular, the $6j$ symbols in \eqref{eq:6jdefinition} are computed in appendix \ref{app:3pt-6j-symbols} for the cases where the three-point structures are unique and the weight shifting operators are in the vector representation.
This unique structure satisfies another useful crossing relation, where
all weight shifting operators on the right hand side change only the dimension $\De$ but not the spin $J$. 
For this to work one needs to allow the weight shifting operators on the right hand side to act on both legs 1 and 2,
\be
\diagramEnvelope{\begin{tikzpicture}[anchor=base,baseline]
	\node (vertL) at (0,0) [twopt] {};
	\node (vertR) at (1.5,-0.12) [threept] {$m$};
	\node (opO1) at (-0.5,-1) [below] {$\cO_1$};
	\node (opO2) at (-0.5,1) [above] {$\cO_2$};
	\node (opO3) at (2,1) [above] {$\cO_3$};
	\node (opW) at (2,-1) [below] {$\cW$};	
	\node at (0.75,0.1) [above] {$\cO_3'$};	
	\draw [scalar] (vertL)-- (opO1);
	\draw [scalar] (vertL)-- (opO2);
	\draw [spinning] (vertL)-- (vertR);
	\draw [spinning] (vertR)-- (opO3);
	\draw [finite with arrow] (vertR)-- (opW);
\end{tikzpicture}}
	\hspace{-0.3cm}=
	\sum_{\cO_1',n}
	\left[
		\begin{matrix}
		\cO_1 & \cO_2 & \cO_1' \\
		\cO_3 & \cW & \cO_3'
		\end{matrix}
	\right]^{m}_{n}
\hspace{-0.8cm}
\diagramEnvelope{\begin{tikzpicture}[anchor=base,baseline]
	\node (vertU) at (0,0.7) [twopt] {};
	\node (vertD) at (0,-0.7) [threept] {$n$};
	\node (opO1) at (-1,-1.5) [below] {$\cO_1$};
	\node (opO2) at (-1,1.5) [above] {$\cO_2$};
	\node (opO3) at (1,1.5) [above] {$\cO_3$};
	\node (opW) at (1,-1.5) [below] {$\cW$};	
	\node at (0.1,0) [right] {$\cO_1'$};	
	\draw [scalar] (vertD)-- (opO1);
	\draw [scalar] (vertU)-- (opO2);
	\draw [scalar] (vertU)-- (vertD);
	\draw [spinning] (vertU)-- (opO3);
	\draw [finite with arrow] (vertD)-- (opW);
\end{tikzpicture}}
\hspace{-0.3cm}+ (-1)^{J_3-J_3'}
	\sum_{\cO_2',n}
	\left[
		\begin{matrix}
		\cO_2 & \cO_1 & \cO_2' \\
		\cO_3 & \cW & \cO_3'
		\end{matrix}
	\right]^{m}_{n}
\hspace{-0.8cm}
\diagramEnvelope{\begin{tikzpicture}[anchor=base,baseline]
	\node (vertU) at (0,-0.7) [twopt] {};
	\node (vertD) at (0,0.7) [threept] {$n$};
	\node (opO1) at (-1,1.5) [above] {$\cO_2$};
	\node (opO2) at (-1,-1.5) [below] {$\cO_1$};
	\node (opO3) at (1,-1.5) [below] {$\cO_3$};
	\node (opW) at (1,1.5) [above] {$\cW$};	
	\node at (0.1,0) [right] {$\cO_2'$};	
	\draw [scalar] (vertD)-- (opO1);
	\draw [scalar] (vertU)-- (opO2);
	\draw [scalar] (vertU)-- (vertD);
	\draw [spinning] (vertU)-- (opO3);
	\draw [finite with arrow] (vertD)-- (opW);
\end{tikzpicture}}
\hspace{-0.3cm}.
\label{eq:6jdefinition_single_ts}
\ee
Here we defined new symbols for the coefficients and it is indeed enough to take $\cO_1'$ and $\cO_2'$ to be scalar operators, as indicated by the dashed lines. The factor $(-1)^{J_3-J_3'}$ depending on the difference of the spins of $\cO_3$ and $\cO_3'$ is required to ensure compatibility of both sides of the equation under exchange of the labels $1$ and $2$. The three-point function on the left hand side
changes by a factor $(-1)^{J_3'}$ when exchanging $\cO_1$ and $\cO_2$ while the three-point functions
on the right hand side change by $(-1)^{J_3}$. 
The coefficients are computed in appendix \ref{app:3pt-6j-symbols}.

\subsection{Local AdS couplings}

The last missing ingredient to compute amplitudes in AdS are of course the local couplings.
We define the diagram with three incoming bulk fields to denote a local cubic coupling.
If all three lines are scalars this is easy. The common point is simply integrated over AdS
\begin{equation}
\diagramEnvelope{\begin{tikzpicture}[anchor=base,baseline]
	\node (vert) at (0,0) [twopt] {};
	\node (opO1) at (-0.5,-1) [below] {$\f_1$};
	\node (opO2) at (-0.5,1) [above] {$\f_2$};
	\node (opO3) at (1,0) [right] {$\f_3$};	
	
	\draw [scalar bulk] (opO1)-- (vert);
	\draw [scalar bulk] (opO2)-- (vert);
	\draw [scalar bulk] (opO3)-- (vert);
\end{tikzpicture}}
\quad=\quad
\int_{\rm AdS} dX \ \< \f_1 (X)| \ \< \f_2 (X)| \ \< \f_3 (X)| 
\,.
\end{equation} 
We wrote the integrand in terms of bras because we did not specify 
yet what are the objects we are integrating over, bulk-to-boundary propagators or harmonic functions.  
We can always use the split representation \eqref{eq:spinning_omega_to_shadow} to write harmonic functions in terms of bulk-to-boundary propagators, so
it is enough to consider AdS integrals where the incoming lines belong to bulk-to-boundary propagators. Such integrals have the form of conformal three point structures. For three scalars
\bea
 \int_{\rm AdS} dX \,\Pi_{\De_1} (X, P_1)\, \Pi_{\De_2} (X, P_2) \,\Pi_{\De_3} (X, P_3)
= b(\cO_{\De_1},\cO_{\De_2},\cO_{\De_3} ) \< \cO_{1} (P_1) \cO_{2} (P_2) \cO_{3} (P_3) \>
\,,
\eea{eq:AdS_integral}
with
\beq
\label{eq:b_def}
 \frac{b(\cO_{\De_1},\cO_{\De_2},\cO_{\De_3} )}{\cC_{\D_1,0}\, \cC_{\D_2,0} \,\cC_{\D_3,0}} = \frac{\pi^{\frac{d}{2}}
 \Gamma\Big(\frac{\Delta_1+\Delta_2+\Delta_3-d}{2}\Big)
\Gamma\Big(\frac{\Delta_1+\Delta_2-\Delta_3}{2}\Big)
\Gamma\Big( \frac{\Delta_1+\Delta_3-\Delta_2}{2}\Big)
 \Gamma\Big(\frac{\Delta_3+\Delta_2-\Delta_1}{2}\Big)}{
 2\, \Gamma(\Delta_1) \,\Gamma(\Delta_2)\, \Gamma(\Delta_3)} \,,
\eeq
where $\cC_{\D,0}$ is defined in (\ref{eq:normalizationboundary}).
The same equation written in diagrammatic language reads 
\begin{equation}
\diagramEnvelope{\begin{tikzpicture}[anchor=base,baseline]
	\node (vert) at (-0.1,0) [twopt] {};
	\node (prop1) at (-0.4,-0.8) [twopt] {};
	\node (prop2) at (-0.4,0.8) [twopt] {};
	\node (prop3) at (0.8,0) [twopt] {};	
	\node (opO1) at (-0.8,-1.6) [below] {$\cO_{\De_1}$};
	\node (opO2) at (-0.8,1.6) [above] {$\cO_{\De_2}$};
	\node (opO3) at (1.6,0) [right] {$\cO_{\De_3}$};		
	\draw [scalar bulk] (prop1)-- (vert);
	\draw [scalar bulk] (prop2)-- (vert);
	\draw [scalar bulk] (prop3)-- (vert);
	\draw [scalar] (prop1)-- (opO1);
	\draw [scalar] (prop2)-- (opO2);
	\draw [scalar] (prop3)-- (opO3);
\end{tikzpicture}}
\quad=\quad b(\cO_{\De_1},\cO_{\De_2},\cO_{\De_3})
\quad
\diagramEnvelope{\begin{tikzpicture}[anchor=base,baseline]
	\node (vert) at (0,0) [twopt] {};
	\node (opO1) at (-0.5,-1) [below] {$\cO_{\De_1}$};
	\node (opO2) at (-0.5,1) [above] {$\cO_{\De_2}$};
	\node (opO3) at (1,0) [right] {$\cO_{\De_3}$};	
	
	\draw [scalar] (vert)-- (opO1);
	\draw [scalar] (vert)-- (opO2);
	\draw [scalar] (vert)-- (opO3);
\end{tikzpicture}}.
\label{eq:ads_integral}
\end{equation}
When one or more of the incoming lines carry spin, all the indices of the spinning fields need to be contracted. 
Take for example the single coupling of one spin $J$ and two scalar fields\footnote{This formula is unchanged when replacing all the $\partial_X$ (which act on a scalar propagator or harmonic function) with $\nabla_X$, for the reason explained in footnote \ref{fn:covariant_derivative}.}\footnote{We will  let the derivatives act on the scalar with the smaller label when translating  diagrams to a formula.}
\begin{equation}
\diagramEnvelope{\begin{tikzpicture}[anchor=base,baseline]
	\node (vert) at (0,0) [twopt] {};
	\node (opO1) at (-0.5,-1) [below] {$\f_1$};
	\node (opO2) at (-0.5,1) [above] {$\f_2$};
	\node (opO3) at (1,0) [right] {$\f_3$};	
	
	\draw [scalar bulk] (opO1)-- (vert);
	\draw [scalar bulk] (opO2)-- (vert);
	\draw [spinning bulk] (opO3)-- (vert);
\end{tikzpicture}}
=
\int_{\rm AdS} dX \ \< \f_2 (X)| \, \partial_{X \left\{a_1\right. }  \ldots \partial_{ X \left.a_J\right\}}  \< \f_1 (X)| \, \partial_{W}^{\{a_1} \ldots \partial_{W}^{a_J\}} \< \f_3 (X,W)| 
\,.
\label{eq:unique_coupling}
\end{equation}
Here all  indices of the field $\f_3$ are contracted to derivatives of $\f_1$. The curly brackets denote a traceless and transverse contraction. Up to normalization this is the only way to couple a spinning field to two scalars \cite{Berends:1985xx,Metsaev:2005ar}. A vertex with derivatives on $\f_2$ or with more derivatives can be related to this one by integration by parts and using the 
(linear) equations of motion \cite{Costa:2014kfa}.

The AdS integral for any local cubic coupling dressed with bulk-to-boundary propagators can be reduced to the simple case \eqref{eq:ads_integral} by using weight shifting operators. Let us demonstrate this for the coupling \eqref{eq:unique_coupling}.
As a first step we have to replace the derivatives by weight shifting operators. This is discussed for a more general expression in appendix \ref{app:bulk-to-bulk-diagram}. We can use that result by setting $l=J$ in (\ref{eq:3pt_with_Ls}),  obtaining 
\be
\label{eq:unique_coupling2}
\diagramEnvelope{\begin{tikzpicture}[anchor=base,baseline]
	\node (vert) at (-0.1,0) [twopt] {};
	\node (prop1) at (-0.4,-0.8) [twopt] {};
	\node (prop2) at (-0.4,0.8) [twopt] {};
	\node (prop3) at (0.8,0) [twopt] {};	
	\node (opO1) at (-0.8,-1.6) [below] {$\cO_{\De_1}$};
	\node (opO2) at (-0.8,1.6) [above] {$\cO_{\De_2}$};
	\node (opO3) at (1.6,0) [right] {$\cO_{\De,J}$};	
	\draw [scalar bulk] (prop1)-- (vert);
	\draw [scalar bulk] (prop2)-- (vert);
	\draw [spinning bulk] (prop3)-- (vert);
	\draw [scalar] (prop1)-- (opO1);
	\draw [scalar] (prop2)-- (opO2);
	\draw [spinning] (prop3)-- (opO3);
\end{tikzpicture}}
\quad=\quad
\beta_{J,J,0}^{\D_1 \D}
\diagramEnvelope{\begin{tikzpicture}[anchor=base,baseline]
	\node (vertL) at (-2,0) [twopt] {};
	\node (opO1) at (-3,1.8) [above] {$\cO_{\De_2}$};
	\node (opO2) at (-3.2,-2.2) [below] {$\cO_{\De_1}$};
	\node (prop1) at (-2.4,0.8) [twopt] {};
	\node (prop2) at (-2.8,-1.6) [twopt] {};
	\node (prop3) at (0,0) [twopt] {};
	\node (shad) at (1,0) [right] {$\cO_{\D,J}$};
	\node (ws2) at (-2.4,-0.9) [threept] {$n$};
	\node (ws22) at (-1,-0.08) [threept,inner sep=1.5pt] {$m$};
	\draw [scalar] (prop1)-- (opO1);
	\draw [scalar] (prop2)-- (opO2);
	\draw [scalar bulk] (prop1)-- (vertL);
	\draw [scalar bulk] (prop2)-- (ws2);
	\draw [scalar bulk] (ws2)-- (vertL);
	\draw [scalar bulk] (ws22)-- (vertL);
	\draw [spinning bulk] (prop3)-- (ws22);
	\draw [spinning] (prop3)-- (shad);
	\draw [finite with arrow] (ws2) to[out=0,in=240] (ws22);
\end{tikzpicture}},
\ee
where $\beta_{J,J,0}^{\D_1 \D}$ is defined in (\ref{eq:b_coeff}).
The weight shifting operators appearing here are $n=(J,0)$ and $m=(0,-J)$, i.e.\ $J$ copies of the operators $(+0)$ and $(0-)$ respectively.
Next use two-point crossing \eqref{eq:twoptcrossing_bulk_boundary} to ensure that no weight shifting operators act on integrated AdS points,
\be
\label{eq:unique_coupling3}
\eqref{eq:unique_coupling2} =
\beta_{J,J,0}^{\D_1 \D}
\left\{ \begin{matrix}
\cO_{\De_1}\\ \f_{\De_1+J}
\end{matrix} \right\}^{n}_{\bar{n}} 
\left\{ \begin{matrix}
\cO_{\De,J}\\ \f_{\De}
\end{matrix} \right\}^{m}_{\bar{m}} 
\diagramEnvelope{\begin{tikzpicture}[anchor=base,baseline]
	\node (vertL) at (-2,0) [twopt] {};
	\node (opO1) at (-3,1.8) [above] {$\cO_{\De_2}$};
	\node (opO2) at (-3.2,-2.2) [below] {$\cO_{\De_1}$};
	\node (prop1) at (-2.4,0.8) [twopt] {};
	\node (prop2) at (-2.4,-0.8) [twopt] {};
	\node (prop3) at (-1,0) [twopt] {};
	\node (shad) at (1,0) [right] {$\cO_{\D,J}$};
	\node (ws2) at (-2.8,-1.76) [threept] {$\bar{n}$};
	\node (ws22) at (0,-0.12) [threept,inner sep=1.5pt] {$\bar{m}$};
	\draw [scalar] (prop1)-- (opO1);
	\draw [scalar] (ws2)-- (opO2);
	\draw [scalar bulk] (prop1)-- (vertL);
	\draw [scalar] (prop2)-- (ws2);
	\draw [scalar bulk] (prop2)-- (vertL);
	\draw [scalar] (prop3)-- (ws22);
	\draw [spinning] (ws22)-- (shad);
	\draw [scalar bulk] (prop3)-- (vertL);
	\draw [finite with arrow] (ws2) to[out=0,in=240] (ws22);
\end{tikzpicture}},
\ee
allowing us to use \eqref{eq:ads_integral} to perform the AdS integral,
\be
\label{eq:unique_coupling4}
\eqref{eq:unique_coupling3} =
\beta_{J,J,0}^{\D_1 \D} \{ \ldots \}^2 b(\cO_{\De_1+J}, \cO_{\De_2} ,\cO_{\De})
\diagramEnvelope{\begin{tikzpicture}[anchor=base,baseline]
	\node (vertL) at (-2,0) [twopt] {};
	\node (opO1) at (-2.6,1) [above] {$\cO_{\De_2}$};
	\node (opO2) at (-3.0,-1.8) [below] {$\cO_{\De_1}$};
	\node (shad) at (0,0) [right] {$\cO_{\D,J}$};
	\node (ws2) at (-2.4,-0.96) [threept] {$\bar{n}$};
	\node (ws22) at (-1,-0.12) [threept,inner sep=1.5pt] {$\bar{m}$};
	\draw [scalar] (vertL)-- (opO1);
	\draw [scalar] (ws2)-- (opO2);
	\draw [scalar] (vertL)-- (ws2);
	\draw [scalar] (vertL)-- (ws22);
	\draw [spinning] (ws22)-- (shad);
	\draw [finite with arrow] (ws2) to[out=0,in=240] (ws22);
\end{tikzpicture}}.
\ee
Finally, move all weight shifting operators to the same leg using  the crossing equation for CFT three-point functions \eqref{eq:6jdefinition},
\be
\label{eq:unique_coupling5}
\eqref{eq:unique_coupling4} =
\beta_{J,J,0}^{\D_1 \D} \{ \ldots \}^2 b(\ldots)
	\left\{
		\begin{matrix}
		\cO_{\De_1+J} & \cO_{\De_2} & \cO_{\De_1} \\
		\cO_{\De,J} & \cW & \cO_{\De}
		\end{matrix}
	\right\}^{\uniq \bar{m}}_{\uniq n}
\diagramEnvelope{\begin{tikzpicture}[anchor=base,baseline]
	\node (vertL) at (-2,0) [twopt] {};
	\node (opO1) at (-2.6,1) [above] {$\cO_{\De_2}$};
	\node (opO2) at (-3.3,-2.4) [below] {$\cO_{\De_1}$};
	\node (shad) at (-0.8,0) [right] {$\cO_{\D,J}$};
	\node (ws2) at (-2.8,-1.76) [threept] {$\bar{n}$};
	\node (ws22) at (-2.4,-0.96) [threept] {$n$};
	\draw [scalar] (vertL)-- (opO1);
	\draw [scalar] (ws2)-- (opO2);
	\draw [scalar] (ws22)-- (ws2);
	\draw [scalar] (vertL)-- (ws22);
	\draw [spinning] (vertL)-- (shad);
	\draw [finite with arrow] (ws2) to[out=340,in=340] (ws22);
\end{tikzpicture}},
\ee
 and remove bubbles using \eqref{eq:bubble},
\be
\label{eq:unique_coupling6}
\eqref{eq:unique_coupling5} =
\beta_{J,J,0}^{\D_1 \D} \{ \ldots \}^3 b(\ldots)
\begin{pmatrix}
\cO_{\De_1}\\
\cO_{\De_1+J}\ \cW
\end{pmatrix}^{\bar{n}n} 
\diagramEnvelope{\begin{tikzpicture}[anchor=base,baseline]
	\node (vertL) at (-2,0) [twopt] {};
	\node (opO1) at (-2.6,-1) [below] {$\cO_{\De_1}$};
	\node (opO2) at (-2.6,1) [above] {$\cO_{\De_2}$};
	\node (shad) at (-0.8,0) [right] {$\cO_{\D,J}$};
	\draw [scalar] (vertL)-- (opO1);
	\draw [scalar] (vertL)-- (opO2);
	\draw [spinning] (vertL)-- (shad);
\end{tikzpicture}}.
\ee
We have just computed the coefficient\footnote{
This $b(\cO_{\De_1}, \cO_{\De_2} ,\cO_{\De,J})$ equals the expression $b(\D_2,\D_1,\D,J)$ from \cite{Costa:2014kfa}.}
\beq
\frac{
b(\cO_{\De_1}, \cO_{\De_2} ,\cO_{\De,J})
}{
b(\cO_{\De_1+J}, \cO_{\De_2} ,\cO_{\De}) }
=
\beta_{J,J,0}^{\D_1 \D}
\left\{ \begin{matrix}
\cO_{\De_1}\\ \f_{\De_1+J}
\end{matrix} \right\}^{n}_{\bar{n}} 
\left\{ \begin{matrix}
\cO_{\De,J}\\ \f_{\De}
\end{matrix} \right\}^{m}_{\bar{m}} 
	\left\{
		\begin{matrix}
		\cO_{\De_1+J} & \cO_{\De_2} & \cO_{\De_1} \\
		\cO_{\De,J} & \cW & \cO_{\De}
		\end{matrix}
	\right\}^{\uniq \bar{m}}_{\uniq n}
\begin{pmatrix}
\cO_{\De_1}\\
\cO_{\De_1+J}\ \cW
\end{pmatrix}^{\bar{n}n} 
\,,
\eeq
relating a local AdS coupling, dressed with bulk-to-boundary propagators,
with the corresponding conformal three-point structure,
\begin{equation}
\diagramEnvelope{\begin{tikzpicture}[anchor=base,baseline]
	\node (vert) at (-0.1,0) [twopt] {};
	\node (prop1) at (-0.4,-0.8) [twopt] {};
	\node (prop2) at (-0.4,0.8) [twopt] {};
	\node (prop3) at (0.8,0) [twopt] {};	
	\node (opO1) at (-0.8,-1.6) [below] {$\cO_{\De_1}$};
	\node (opO2) at (-0.8,1.6) [above] {$\cO_{\De_2}$};
	\node (opO3) at (1.6,0) [right] {$\cO_{\De,J}$};	
	
	\draw [scalar bulk] (prop1)-- (vert);
	\draw [scalar bulk] (prop2)-- (vert);
	\draw [spinning bulk] (prop3)-- (vert);
	\draw [scalar] (prop1)-- (opO1);
	\draw [scalar] (prop2)-- (opO2);
	\draw [spinning] (prop3)-- (opO3);
\end{tikzpicture}}
\quad=\quad b(\cO_{\De_1},\cO_{\De_2},\cO_{\De,J})
\quad
\diagramEnvelope{\begin{tikzpicture}[anchor=base,baseline]
	\node (vert) at (0,0) [twopt] {};
	\node (opO1) at (-0.5,-1) [below] {$\cO_{\De_1}$};
	\node (opO2) at (-0.5,1) [above] {$\cO_{\De_2}$};
	\node (opO3) at (1,0) [right] {$\cO_{\De,J}$};	
	
	\draw [scalar] (vert)-- (opO1);
	\draw [scalar] (vert)-- (opO2);
	\draw [spinning] (vert)-- (opO3);
\end{tikzpicture}}.
\label{eq:ads_integral_unique}
\end{equation}

In general there can be multiple independent ways to arrange derivatives. In such cases we introduce a label 
to enumerate the different local couplings
\be
\diagramEnvelope{\begin{tikzpicture}[anchor=base,baseline]
	\node (vert) at (0,-0.08) [threept] {$a$};
	\node (opO1) at (-0.5,-1) [below] {$\f_1$};
	\node (opO2) at (-0.5,1) [above] {$\f_2$};
	\node (opO3) at (1,0) [right] {$\f_3$};	
	\draw [spinning bulk] (opO1)-- (vert);
	\draw [spinning bulk] (opO2)-- (vert);
	\draw [spinning bulk] (opO3)-- (vert);
\end{tikzpicture}},
\ee
however we do not construct more general couplings explicitly here.
The number of such structures is known to be the same as the number of conformal three point structures involving the same representations.
It is also clear that a version of \eqref{eq:ads_integral} holds in the general case
\begin{equation}
\diagramEnvelope{\begin{tikzpicture}[anchor=base,baseline]
	\node (vert) at (-0.1,-0.08) [threept] {$a$};
	\node (prop1) at (-0.4,-0.8) [twopt] {};
	\node (prop2) at (-0.4,0.8) [twopt] {};
	\node (prop3) at (0.8,0) [twopt] {};	
	\node (opO1) at (-0.8,-1.6) [below] {$\cO_1$};
	\node (opO2) at (-0.8,1.6) [above] {$\cO_2$};
	\node (opO3) at (1.6,0) [right] {$\cO_3$};	
	
	\draw [spinning bulk] (prop1)-- (vert);
	\draw [spinning bulk] (prop2)-- (vert);
	\draw [spinning bulk] (prop3)-- (vert);
	\draw [spinning] (prop1)-- (opO1);
	\draw [spinning] (prop2)-- (opO2);
	\draw [spinning] (prop3)-- (opO3);
\end{tikzpicture}}
\quad=\quad \sum_c b\big(\cO_1, \cO_2 ,\cO_3\big)^a_c
\quad
\diagramEnvelope{\begin{tikzpicture}[anchor=base,baseline]
	\node (vert) at (0,-0.1) [threept] {$c$};
	\node (opO1) at (-0.5,-1) [below] {$\cO_1$};
	\node (opO2) at (-0.5,1) [above] {$\cO_2$};
	\node (opO3) at (1,0) [right] {$\cO_3$};	
	
	\draw [spinning] (vert)-- (opO1);
	\draw [spinning] (vert)-- (opO2);
	\draw [spinning] (vert)-- (opO3);
\end{tikzpicture}},
\label{eq:ads_integral_general}
\end{equation}
where $b\big(\cO_1, \cO_2, \cO_3\big)^a_b$ is a square matrix.
An example illustrating this equation is worked out in appendix \ref{app:non-unique_couplings}.
One can derive a crossing equation for local couplings by inverting this basis change and inserting it into the crossing equation for conformal three-point structures \eqref{eq:6jdefinition}. The resulting linear system of equations can be solved for any of the  local couplings. Next one commutes the weight shifting operators past the bulk-to-boundary propagators and removes the propagators from all legs on both sides of the equation to find\footnote{For equations with incoming double lines like this one we generally assume that these lines belong to either bulk-to-boundary propagators or harmonic functions. Equation \eqref{eq:Bulk6jdefinition} is derived with bulk-to-boundary propagators but also holds with harmonic functions, as can be seen by writing them in the split representation \eqref{eq:spinning_omega_to_shadow}.}
\be
\diagramEnvelope{\begin{tikzpicture}[anchor=base,baseline]
	\node (vertL) at (0,0) [threept] {$a$};
	\node (vertR) at (2,0) [threept,inner sep=1pt] {$m$};
	\node (opO1) at (-0.5,-1) [below] {$\f_1$};
	\node (opO2) at (-0.5,1) [above] {$\f_2$};
	\node (opO3) at (2.5,1) [above] {$\f_3$};
	\node (opW) at (2.5,-1) [below] {$\cW$};	
	\node at (1,0.1) [above] {$\f_3'$};	
	\draw [spinning bulk] (opO1)-- (vertL);
	\draw [spinning bulk] (opO2)-- (vertL);
	\draw [spinning bulk] (vertR)-- (vertL);
	\draw [spinning bulk] (opO3)-- (vertR);
	\draw [finite with arrow] (vertR)-- (opW);
\end{tikzpicture}}
	\quad=\quad
	\sum_{\f_1',b,n}
	\left\{
		\begin{matrix}
		\f_1' & \f_2 & \f_1 \\
		\f_3' & \cW & \f_3
		\end{matrix}
	\right\}^{am}_{bn}
\diagramEnvelope{\begin{tikzpicture}[anchor=base,baseline]
	\node (vertU) at (0,0.7) [threept] {$b$};
	\node (vertD) at (0,-0.7) [threept] {$n$};
	\node (opO1) at (-1,-1.5) [below] {$\f_1$};
	\node (opO2) at (-1,1.5) [above] {$\f_2$};
	\node (opO3) at (1,1.5) [above] {$\f_3$};
	\node (opW) at (1,-1.5) [below] {$\cW$};	
	\node at (0.1,0) [right] {$\f_1'$};	
	\draw [spinning bulk] (opO1)-- (vertD);
	\draw [spinning bulk] (opO2)-- (vertU);
	\draw [spinning bulk] (vertD)-- (vertU);
	\draw [spinning bulk] (opO3)-- (vertU);
	\draw [finite with arrow] (vertD)-- (opW);
\end{tikzpicture}}\,,
\label{eq:Bulk6jdefinition}
\ee
defining a new set of $6j$ symbols.
This procedure of computing the bulk $6j$ symbols in terms of the $6j$
symbols for conformal structures is demonstrated in appendix \ref{app:computing_bulk_6j}.
In the same way, the special crossing relation \eqref{eq:6jdefinition_single_ts}
can also be translated to an equation for local AdS couplings
\be
\diagramEnvelope{\begin{tikzpicture}[anchor=base,baseline]
	\node (vertL) at (0,0) [twopt] {};
	\node (vertR) at (1.5,-0.12) [threept] {$m$};
	\node (opO1) at (-0.5,-1) [below] {$\f_1$};
	\node (opO2) at (-0.5,1) [above] {$\f_2$};
	\node (opO3) at (2,1) [above] {$\f_3$};
	\node (opW) at (2,-1) [below] {$\cW$};	
	\node at (0.75,0.1) [above] {$\f_3'$};	
	\draw [scalar bulk] (opO1)-- (vertL);
	\draw [scalar bulk] (opO2)-- (vertL);
	\draw [spinning bulk] (vertR)-- (vertL);
	\draw [spinning bulk] (opO3)-- (vertR);
	\draw [finite with arrow] (vertR)-- (opW);
\end{tikzpicture}}
	\hspace{-0.3cm}=
	\sum_{\f_1',n}
	\left[
		\begin{matrix}
		\f_1' & \f_2 & \f_1 \\
		\f_3' & \cW & \f_3
		\end{matrix}
	\right]^{m}_{n}
\hspace{-0.8cm}
\diagramEnvelope{\begin{tikzpicture}[anchor=base,baseline]
	\node (vertU) at (0,0.7) [twopt] {};
	\node (vertD) at (0,-0.7) [threept] {$n$};
	\node (opO1) at (-1,-1.5) [below] {$\f_1$};
	\node (opO2) at (-1,1.5) [above] {$\f_2$};
	\node (opO3) at (1,1.5) [above] {$\f_3$};
	\node (opW) at (1,-1.5) [below] {$\cW$};	
	\node at (0.1,0) [right] {$\f_1'$};	
	\draw [scalar bulk] (opO1)-- (vertD);
	\draw [scalar bulk] (opO2)-- (vertU);
	\draw [scalar bulk] (vertD)-- (vertU);
	\draw [spinning bulk] (opO3)-- (vertU);
	\draw [finite with arrow] (vertD)-- (opW);
\end{tikzpicture}}
\hspace{-0.3cm}+  (-1)^{J_3-J_3'}
	\sum_{\f_2',n}
	\left[
		\begin{matrix}
		\f_2' & \f_1 & \f_2 \\
		\f_3' & \cW & \f_3
		\end{matrix}
	\right]^{m}_{n}
\hspace{-0.8cm}
\diagramEnvelope{\begin{tikzpicture}[anchor=base,baseline]
	\node (vertU) at (0,-0.7) [twopt] {};
	\node (vertD) at (0,0.7) [threept] {$n$};
	\node (opO1) at (-1,1.5) [above] {$\f_2$};
	\node (opO2) at (-1,-1.5) [below] {$\f_1$};
	\node (opO3) at (1,-1.5) [below] {$\f_3$};
	\node (opW) at (1,1.5) [above] {$\cW$};	
	\node at (0.1,0) [right] {$\f_2'$};	
	\draw [scalar bulk] (opO1)-- (vertD);
	\draw [scalar bulk] (opO2)-- (vertU);
	\draw [scalar bulk] (vertD)-- (vertU);
	\draw [spinning bulk] (opO3)-- (vertU);
	\draw [finite with arrow] (vertD)-- (opW);
\end{tikzpicture}}
\hspace{-0.3cm}.
\label{eq:6jdefinition_bulk_single_ts}
\ee

To conclude this section we comment on the compatibility of the definitions of bulk $6j$ symbols in \eqref{eq:twoptcrossing_bulk} and \eqref{eq:Bulk6jdefinition}. For CFT correlators the $6j$ symbols for two-point functions are simply the $6j$ symbols for three-point functions where one of the operators is the identity \eqref{eq:shorthand_6j_2pt}. For the bulk we decided to define three-point crossing in terms of local couplings, different from the two-point crossing which is defined for harmonic functions. In the appendix \ref{app:compatibility} we show that the analogous relation to  \eqref{eq:shorthand_6j_2pt}, namely
\beq
	\left\{
		\begin{matrix}
		\f_3 & \mathbf{1} & \f_1 \\
		\f_1 & \cW & \f_3
		\end{matrix}
	\right\}^{\uniq \bar{m}}_{\uniq m}
=
\left\{ \begin{matrix}
\f_3\\ \f_1
\end{matrix} \right\}^{\bar{m}}_{m} \,,
\label{eq:compatibility}
\eeq
holds nevertheless.

\subsection{Weight shifting operator basis for local couplings}
\label{sec:differential-basis}

Another useful concept in order to simplify conformal blocks is to relate different three-point functions by weight shifting operators. Let us first recall how this works for conformal structures and then apply it to local AdS couplings. Different conformal three-point structures can be related by acting with weight shifting operators on two legs and contracting them
\be
\diagramEnvelope{\begin{tikzpicture}[anchor=base,baseline]
	\node (vertL) at (0.3,0) [threept] {$a$};
	\node (vertR) at (1.7,-0.05) [threept] {$m$};
	\node (opO1) at (0.2,-0.3) [below] {$\cO_1$};
	\node (opO2) at (-0.7,0.8) [above] {$\cO_2$};
	\node (opO3) at (2.7,0.8) [above] {$\cO_3$};
	\node (opW) at (1.8,-0.3) [below] {$\cW$};	
	\node (vertB) at (1,-1.2) [threept] {$n$};
	\node (bot) at (1,-2) [below] {$\cO_1'$};
	\node at (1,0.2) [above] {$\cO_3'$};	
	\draw [spinning] (vertL)-- (vertB);
	\draw [spinning] (vertL)-- (opO2);
	\draw [spinning] (vertL)-- (vertR);
	\draw [spinning] (vertR)-- (opO3);
	\draw [finite with arrow] (vertR)-- (vertB);
	\draw [spinning] (vertB) -- (bot);
\end{tikzpicture}}
	\quad&=\quad
	\sum_{b,p}
	\left\{
		\begin{matrix}
		\cO_1 & \cO_2 & \cO_1' \\
		\cO_3 & \cW & \cO_3'
		\end{matrix}
	\right\}^{am}_{bp}
	\begin{pmatrix}
\cO_1'\\
\cO_1\ \cW
\end{pmatrix}^{np}
\diagramEnvelope{\begin{tikzpicture}[anchor=base,baseline]
	\node (vertU) at (0,0) [threept] {$b$};
	\node (vertD) at (0,-1) [below] {$\cO_1'$};
	\node (opO2) at (-1,0.8) [above] {$\cO_2$};
	\node (opO3) at (1,0.8) [above] {$\cO_3$};
	\draw [spinning] (vertU)-- (opO2);
	\draw [spinning] (vertU)-- (vertD);
	\draw [spinning] (vertU)-- (opO3);
\end{tikzpicture}},
\label{eq:spinningcbtrick}
\ee
where \eqref{eq:6jdefinition} and \eqref{eq:bubble} were used.
This relation can often be inverted to express three-point structures in terms of simpler structures with weight shifting operators acting on them. For example, one can express any three-point structure of three symmetric traceless tensors $\cO_1$, $\cO_2$, $\cO_3$ in terms of the simple structure with two scalars \eqref{eq:3pt_correlator}
\be
\label{eq:differentialbasistrick}
\diagramEnvelope{\begin{tikzpicture}[anchor=base,baseline]
	\node (vert) at (0,-0.1) [threept] {$a$};
	\node (opO1) at (-0.5,-1) [below] {$\cO_1$};
	\node (opO2) at (-0.5,1) [above] {$\cO_2$};
	\node (opO3) at (1,0) [right] {$\cO_3$};	
	\draw [spinning] (vert)-- (opO1);
	\draw [spinning] (vert)-- (opO2);
	\draw [spinning] (vert)-- (opO3);
\end{tikzpicture}}
&=
\sum_{\cW,m,n}
K_{a,\cW,m,n}^{\cO_1 \cO_2 \cO_3}
\diagramEnvelope{\begin{tikzpicture}[anchor=base,baseline]
\node (tl) at (-1.6,1.6) [above] {$\cO_2$};
\node (vertr) at (0.6,0) [twopt] {};
\node (verttl) at (-0.7,0.7) [threept] {$n$};
\node (vertbl) at (-0.7,-0.7) [threept] {$m$};
\node (bl) at (-1.6,-1.6) [below] {$\cO_1$};
\node (r) at (1.6,0) [right] {$\cO_3$};
\node (w) at (-0.7,0) [left] {$\cW$};
\node (f1) at (0.3,0.35) [above] {$\cO_2'$};
\node (f2) at (0.3,-0.3) [below] {$\cO_1'$};
\draw [spinning] (verttl) -- (tl);
\draw [spinning] (vertbl) -- (bl);
\draw [spinning] (vertr) -- (r);
\draw [scalar] (vertr) -- (verttl);
\draw [scalar] (vertr) -- (vertbl);
\draw [finite with arrow] (verttl) -- (vertbl);
\end{tikzpicture}},
\ee
for some coefficients $K_{a,\cW,m,n}^{\cO_1 \cO_2 \cO_3}$. This relation was first found in \cite{Costa:2011dw} and rephrased in terms of weight shifting operators in \cite{Karateev:2017jgd}.
One can expect that \eqref{eq:spinningcbtrick} can generally be inverted as long as one includes sufficiently general three-point structures and finite dimensional representations $\cW$ in the sum in \eqref{eq:differentialbasistrick}. 

A similar relation holds for local AdS couplings.
For simplicity let us choose the basis for conformal three-point structures and local AdS couplings such that the matrix of coefficients in \eqref{eq:ads_integral_general} is proportional to unity 
$b\big(\cO_1 , \cO_2 ,\cO_3\big)^a_b = b\big(\cO_1, \cO_2 ,\cO_3\big) \de^a_b$, i.e.\ 
\begin{equation}
\diagramEnvelope{\begin{tikzpicture}[anchor=base,baseline]
	\node (vert) at (-0.1,-0.08) [threept] {$a$};
	\node (prop1) at (-0.4,-0.8) [twopt] {};
	\node (prop2) at (-0.4,0.8) [twopt] {};
	\node (prop3) at (0.8,0) [twopt] {};	
	\node (opO1) at (-0.8,-1.6) [below] {$\cO_1$};
	\node (opO2) at (-0.8,1.6) [above] {$\cO_2$};
	\node (opO3) at (1.6,0) [right] {$\cO_3$};	
	
	\draw [spinning bulk] (prop1)-- (vert);
	\draw [spinning bulk] (prop2)-- (vert);
	\draw [spinning bulk] (prop3)-- (vert);
	\draw [spinning] (prop1)-- (opO1);
	\draw [spinning] (prop2)-- (opO2);
	\draw [spinning] (prop3)-- (opO3);
\end{tikzpicture}}
\quad=\quad b\big(\cO_1, \cO_2 ,\cO_3\big)
\quad
\diagramEnvelope{\begin{tikzpicture}[anchor=base,baseline]
	\node (vert) at (0,-0.1) [threept] {$a$};
	\node (opO1) at (-0.5,-1) [below] {$\cO_1$};
	\node (opO2) at (-0.5,1) [above] {$\cO_2$};
	\node (opO3) at (1,0) [right] {$\cO_3$};	
	
	\draw [spinning] (vert)-- (opO1);
	\draw [spinning] (vert)-- (opO2);
	\draw [spinning] (vert)-- (opO3);
\end{tikzpicture}}.
\label{eq:ads_integral_general_diagonal}
\end{equation}
This relation can be used to rewrite \eqref{eq:differentialbasistrick} in terms of AdS couplings,
\be
\label{eq:differentialbasistrick-bulk-1}
\diagramEnvelope{\begin{tikzpicture}[anchor=base,baseline]
	\node (vert) at (-0.1,-0.08) [threept] {$a$};
	\node (prop1) at (-0.4,-0.8) [twopt] {};
	\node (prop2) at (-0.4,0.8) [twopt] {};
	\node (prop3) at (0.8,0) [twopt] {};	
	\node (opO1) at (-0.8,-1.6) [below] {$\cO_1$};
	\node (opO2) at (-0.8,1.6) [above] {$\cO_2$};
	\node (opO3) at (1.6,0) [right] {$\cO_3$};	
	\draw [spinning bulk] (prop1)-- (vert);
	\draw [spinning bulk] (prop2)-- (vert);
	\draw [spinning bulk] (prop3)-- (vert);
	\draw [spinning] (prop1)-- (opO1);
	\draw [spinning] (prop2)-- (opO2);
	\draw [spinning] (prop3)-- (opO3);
\end{tikzpicture}}
&=
\sum_{\cW,m,n}
K_{a,\cW,m,n}^{\cO_1 \cO_2 \cO_3}
\frac{b\big(\cO_1, \cO_2 ,\cO_3\big)}{b\big(\cO_1', \cO_2', \cO_3\big)}
\diagramEnvelope{\begin{tikzpicture}[anchor=base,baseline]
\node (tl) at (-2.3,2.3) [above] {$\cO_2$};
\node (vertr) at (0.95,0) [twopt] {};
\node (verttl) at (-1,0.92) [threept] {$n$};
\node (vertbl) at (-1,-1.08) [threept] {$m$};
\node (proptl) at (0,0.5) [twopt] {};
\node (propbl) at (0,-0.5) [twopt] {};
\node (propr) at (1.8,0) [twopt] {};
\node (bl) at (-2.3,-2.3) [below] {$\cO_1$};
\node (r) at (2.6,0) [right] {$\cO_3$};
\node (w) at (-1,0) [left] {$\cW$};
\node (f1) at (0.55,0.5) [above] {$\f_2'$};
\node (f2) at (0.55,-0.45) [below] {$\f_1'$};
\draw [spinning] (verttl) -- (tl);
\draw [spinning] (vertbl) -- (bl);
\draw [spinning] (propr) -- (r);
\draw [spinning bulk] (propr) -- (vertr);
\draw [scalar] (proptl) -- (verttl);
\draw [scalar] (propbl) -- (vertbl);
\draw [scalar bulk] (proptl) -- (vertr);
\draw [scalar bulk] (propbl) -- (vertr);
\draw [finite with arrow] (verttl) -- (vertbl);
\end{tikzpicture}}.
\ee
The weight shifting operators are now commuted past the bulk-to-boundary propagators which can then be removed to obtain the weight shifting operator basis for the cubic couplings of three traceless symmetric tensors \eqref{eq:3pt_correlator},
\be
\label{eq:differentialbasistrick-bulk}
\diagramEnvelope{\begin{tikzpicture}[anchor=base,baseline]
	\node (vert) at (0,-0.1) [threept] {$a$};
	\node (opO1) at (-0.5,-1) [below] {$\f_1$};
	\node (opO2) at (-0.5,1) [above] {$\f_2$};
	\node (opO3) at (1,0) [right] {$\f_3$};	
	\draw [spinning bulk] (opO1)-- (vert);
	\draw [spinning bulk] (opO2)-- (vert);
	\draw [spinning bulk] (opO3)-- (vert);
\end{tikzpicture}}
&=
\sum_{\cW,m,n}
\frac{K_{a,\cW,m,n}^{\cO_1 \cO_2 \cO_3} b\big(\cO_1, \cO_2 ,\cO_3\big)}{b\big(\cO_1' \cO_2' \cO_3\big)
\left\{ \begin{matrix}
\cO_1\\ \f_1'
\end{matrix} \right\}^{\bar{m}}_{m}
\left\{ \begin{matrix}
\cO_2\\ \f_2'
\end{matrix} \right\}^{\bar{n}}_{n}}
\diagramEnvelope{\begin{tikzpicture}[anchor=base,baseline]
\node (tl) at (-1.6,1.6) [above] {$\f_2$};
\node (vertr) at (0.6,0) [twopt] {};
\node (verttl) at (-0.7,0.7) [threept] {$\bar{n}$};
\node (vertbl) at (-0.7,-0.7) [threept] {$\bar{m}$};
\node (bl) at (-1.6,-1.6) [below] {$\f_1$};
\node (r) at (1.6,0) [right] {$\f_3$};
\node (w) at (-0.7,0) [left] {$\cW$};
\node (f1) at (0.3,0.35) [above] {$\f_2'$};
\node (f2) at (0.3,-0.3) [below] {$\f_1'$};
\draw [spinning bulk] (tl)-- (verttl);
\draw [spinning bulk] (bl)-- (vertbl);
\draw [spinning bulk] (r)-- (vertr);
\draw [scalar bulk] (verttl)-- (vertr);
\draw [scalar bulk] (vertbl)-- (vertr);
\draw [finite with arrow] (verttl) -- (vertbl);
\end{tikzpicture}}.
\ee
Defining new coefficients more suitable for bulk computations we rewrite this equation as
\be
\label{eq:differentialbasistrickshorthand}
\diagramEnvelope{\begin{tikzpicture}[anchor=base,baseline]
	\node (vert) at (0,-0.1) [threept] {$a$};
	\node (opO1) at (-0.5,-1) [below] {$\f_1$};
	\node (opO2) at (-0.5,1) [above] {$\f_2$};
	\node (opO3) at (1,0) [right] {$\f_3$};	
	\draw [spinning bulk] (opO1)-- (vert);
	\draw [spinning bulk] (opO2)-- (vert);
	\draw [spinning bulk] (opO3)-- (vert);
\end{tikzpicture}}
&=
\sum_{\cW,m,n}
K_{a,\cW,m,n}^{\f_1 \f_2 \f_3} 
\diagramEnvelope{\begin{tikzpicture}[anchor=base,baseline]
\node (tl) at (-1.6,1.6) [above] {$\f_2$};
\node (vertr) at (0.6,0) [twopt] {};
\node (verttl) at (-0.7,0.7) [threept] {$n$};
\node (vertbl) at (-0.7,-0.7) [threept] {$m$};
\node (bl) at (-1.6,-1.6) [below] {$\f_1$};
\node (r) at (1.6,0) [right] {$\f_3$};
\node (w) at (-0.7,0) [left] {$\cW$};
\node (f1) at (0.3,0.35) [above] {$\f_2'$};
\node (f2) at (0.3,-0.3) [below] {$\f_1'$};
\draw [spinning bulk] (tl)-- (verttl);
\draw [spinning bulk] (bl)-- (vertbl);
\draw [spinning bulk] (r)-- (vertr);
\draw [scalar bulk] (verttl)-- (vertr);
\draw [scalar bulk] (vertbl)-- (vertr);
\draw [finite with arrow] (verttl) -- (vertbl);
\end{tikzpicture}}.
\ee

\section{Witten diagrams}
\label{sec:WittenDiagrams}

In this section we will illustrate how to use the ingredients collected above to simplify the computation of Witten diagrams. We will see that the bulk-to-bulk propagators and most of the bulk-to-boundary propagators in any tree-level or one-loop Witten diagram with cubic couplings can easily be reduced to scalar fields, leading to a big simplification in the computations.
For the sake of brevity we will sometimes use symbols $\{\ldots\}$ to denote $6j$ symbols that appear in crossing equations and $(\ldots)$ for bubble coefficients defined by (\ref{eq:bubble},\ref{eq:bubble_bulk}). This way we avoid writing overcrowded equations. The interested reader can easily follow the steps by taking note of which $6j$ symbols appear in the equations we are simplifying. Similarly we will sometimes abbreviate the coefficients appearing in \eqref{eq:differentialbasistrickshorthand} by $K$. Powers of such abbreviations indicate a number of different factors of the same type.

\subsection{Bulk-to-bulk propagator}
\label{sec:bulk-to-bulk-propagator-diagram}

To compute Witten diagrams we would also like to translate the spectral representation of the bulk-to-bulk propagator \eqref{eq:SlipRepStart} into the diagrammatic notation.  Let us start by defining a diagram for the bulk-to-bulk propagator,
\be
\Pi_{\D,J}(X_1,X_2;W_1,W_2) \quad = \quad 
\diagramEnvelope{\begin{tikzpicture}[anchor=base,baseline]
	\node (opO) at (-1,0) [left] {$\phi$};
	\node (opOprime) at (1,0) [right] {$\phi^\dag$};
	\node (vert) at (0,0) [bulkprop] {};
	\draw [spinning bulk] (vert) -- (opO);
	\draw [spinning bulk] (vert) -- (opOprime);
\end{tikzpicture}}\,.
\ee
The right hand side of \eqref{eq:SlipRepStart} by itself cannot be written in terms of harmonic functions and weight shifting operators. It is however possible to write it in terms of diagrams with harmonic functions if we sandwich the equation between two local couplings
\be
\diagramEnvelope{\begin{tikzpicture}[anchor=base,baseline]
	\node (vertL) at (-1,-0.1) [threept] {$a$};
	\node (vertR) at (1,-0.12) [threept,inner sep=1pt] {$b$};
	\node (opO1) at (-1.5,-1) [below] {$\f_1$};
	\node (opO2) at (-1.5,1) [above] {$\f_2$};
	\node (opO3) at (1.5,-1) [below] {$\f_3$};
	\node (opO4) at (1.5,1) [above] {$\f_4$};
	\node (shad) at (0,0) [bulkprop] {};
	\node at (0,-0.1) [below] {$\f_{\De,J}$};	
	\draw [spinning bulk] (opO1)-- (vertL);
	\draw [spinning bulk] (opO2)-- (vertL);
	\draw [spinning bulk] (shad)-- (vertL);
	\draw [spinning bulk] (shad)-- (vertR);
	\draw [spinning bulk] (opO3)-- (vertR);
	\draw [spinning bulk] (opO4)-- (vertR);
\end{tikzpicture}}
= \quad \sum_{l=0}^J  \int d\nu \sum_{c,d} \a_{J,l}(\nu)^{ab}_{cd} 
\diagramEnvelope{\begin{tikzpicture}[anchor=base,baseline]
	\node (vertL) at (-1,-0.1) [threept] {$c$};
	\node (vertR) at (1,-0.12) [threept,inner sep=1pt] {$d$};
	\node (opO1) at (-1.5,-1) [below] {$\f_1$};
	\node (opO2) at (-1.5,1) [above] {$\f_2$};
	\node (opO3) at (1.5,-1) [below] {$\f_3$};
	\node (opO4) at (1.5,1) [above] {$\f_4$};
	\node (shad) at (0,0) [twopt] {};
	\node at (0,-0.1) [below] {$\f_{\De(\nu),l}$};	
	\draw [spinning bulk] (opO1)-- (vertL);
	\draw [spinning bulk] (opO2)-- (vertL);
	\draw [spinning bulk] (shad)-- (vertL);
	\draw [spinning bulk] (shad)-- (vertR);
	\draw [spinning bulk] (opO3)-- (vertR);
	\draw [spinning bulk] (opO4)-- (vertR);
\end{tikzpicture}}\,.
\label{eq:bulk-to-bulk-diagram}
\ee
In this equation we defined new coefficients which depend on the local couplings on both sides of the equation.
It will be shown in section \ref{sec:cb_decomposition} below that they are closely related to the coefficients in the conformal block decomposition of single exchange Witten diagrams.
We will show how they can be computed by considering the case where all the local couplings are unique,
\be
\diagramEnvelope{\begin{tikzpicture}[anchor=base,baseline]
	\node (vertL) at (-1,0) [twopt] {};
	\node (vertR) at (1,0) [twopt] {};
	\node (opO1) at (-1.5,-1) [below] {$\f_1$};
	\node (opO2) at (-1.5,1) [above] {$\f_2$};
	\node (opO3) at (1.5,-1) [below] {$\f_3$};
	\node (opO4) at (1.5,1) [above] {$\f_4$};
	\node (shad) at (0,0) [bulkprop] {};
	\node at (0,-0.1) [below] {$\f_{\De,J}$};	
	\draw [scalar bulk] (opO1)-- (vertL);
	\draw [scalar bulk] (opO2)-- (vertL);
	\draw [spinning bulk] (shad)-- (vertL);
	\draw [spinning bulk] (shad)-- (vertR);
	\draw [scalar bulk] (opO3)-- (vertR);
	\draw [scalar bulk] (opO4)-- (vertR);
\end{tikzpicture}}\,.
\label{eq:bulk-to-bulk-diagram-unique}
\ee
It is shown in appendix \ref{app:bulk-to-bulk-diagram} that the combination of derivatives acting on the harmonic functions when combining \eqref{eq:SlipRepStart} with the local coupling to scalars \eqref{eq:unique_coupling} can be written in terms of weight shifting operators as
\be
\eqref{eq:bulk-to-bulk-diagram-unique}
= \int d\nu \sum_{l=0}^J a_{J,l}(\nu) \sum_{m_{1},m_2=0}^{J-l} \beta_{J,l,m_1}^{\De_1 \De(\nu)} \beta_{J,l,m_2}^{\De_3 \De(\nu)}
\hspace{-0.8cm}
\diagramEnvelope{\begin{tikzpicture}[anchor=base,baseline]
	\node (vertL) at (-2,0) [twopt] {};
	\node (vertR) at (2,0) [twopt] {};
	\node (opO1) at (-2.5,1) [above] {$\f_{\De_2}$};
	\node (opO2) at (-3,-2) [below] {$\f_{\De_1}$};
	\node (opO3) at (2.5,1) [above] {$\f_{\De_4}$};
	\node (opO4) at (3,-2) [below] {$\f_{\De_3}$};
	\node (shad) at (0,0) [twopt] {};
	\node (ws2) at (-2.5,-1.16) [threept] {$n$};
	\node (ws4) at (2.5,-1.16) [threept,inner sep=2pt] {$q$};
	\node (ws22) at (-1,-0.08) [threept,inner sep=2pt] {$m$};
	\node (ws42) at (1,-0.05) [threept,inner sep=2pt] {$p$};
	\node at (0,-0.1) [below] {$\f_{\D(\nu),l}$};
	\draw [scalar bulk] (opO1)-- (vertL);
	\draw [scalar bulk] (opO2)-- (ws2);
	\draw [scalar bulk] (ws2)-- (vertL);
	\draw [scalar bulk] (ws22)-- (vertL);
	\draw [spinning bulk] (shad)-- (ws22);
	\draw [spinning bulk] (shad)-- (ws42);
	\draw [scalar bulk] (ws42)-- (vertR);
	\draw [scalar bulk] (ws4)-- (vertR);
	\draw [scalar bulk] (opO3)-- (vertR);
	\draw [scalar bulk] (opO4)-- (ws4);
	\draw [finite with arrow] (ws2) to[out=0,in=240] (ws22);
	\draw [finite with arrow] (ws4) to[out=180,in=300] (ws42);
\end{tikzpicture}}
\hspace{-0.3cm},
\label{eq:bulk-to-bulk-diagram-unique-2}
\ee
where $a_{J,l}(\nu)$ is the spectral function for the bulk-to-bulk propagator introduced in (\ref{eq:SlipRepStart}). 
The weight shifting operators here are shifting the weights by
\bea
n = (J-m_1,0), \quad m=(J-l-m_1,-l), \quad
p = (J-l-m_2,-l), \quad q=(J-m_2,0),
\eea{eq:wso_propagator_diagram}
with the precise ordering of them being defined by \eqref{eq:3pt_with_Ls}.
We can use the crossing equation for local couplings \eqref{eq:Bulk6jdefinition} to move all of them to the internal leg, and then remove the bubbles 
using \eqref{eq:bubble_bulk} to arrive at 
\be
\eqref{eq:bulk-to-bulk-diagram-unique-2}
= \quad \sum_{l=0}^J  \int d\nu \ \a_{J,l}(\nu)
\diagramEnvelope{\begin{tikzpicture}[anchor=base,baseline]
	\node (vertL) at (-1,0) [twopt] {};
	\node (vertR) at (1,0) [twopt] {};
	\node (opO1) at (-1.5,-1) [below] {$\f_1$};
	\node (opO2) at (-1.5,1) [above] {$\f_2$};
	\node (opO3) at (1.5,-1) [below] {$\f_3$};
	\node (opO4) at (1.5,1) [above] {$\f_4$};
	\node (shad) at (0,0) [twopt] {};
	\node at (0,-0.1) [below] {$\f_{\De(\nu),l}$};	
	\draw [scalar bulk] (opO1)-- (vertL);
	\draw [scalar bulk] (opO2)-- (vertL);
	\draw [spinning bulk] (shad)-- (vertL);
	\draw [spinning bulk] (shad)-- (vertR);
	\draw [scalar bulk] (opO3)-- (vertR);
	\draw [scalar bulk] (opO4)-- (vertR);
\end{tikzpicture}}\,,
\ee
 with
\bea
{}&\a_{J,l}(\nu) = a_{J,l}(\nu) \sum_{m_{1},m_2=0}^{J-l} \beta_{J,l,m_1}^{\De_1 \De(\nu)} \beta_{J,l,m_2}^{\De_3 \De(\nu)}
\prod_{i=1,2}
\left\{\begin{matrix}
\f_{\D(\nu),l} & \f_{\D_{2i}} & \f_{\De(\nu)}\nn\\
\f_{\D_{2i-1}+l} & \cW & \f_{\D_{2i-1}}
\end{matrix} \right\}^{\uniq (l,0)}_{\uniq (0,l)}
\begin{pmatrix}
\f_{\D(\nu),l}\\
\f_{\D(\nu)}\ \cW
\end{pmatrix}^{(0,l)(0,-l)}\\
& \times \left\{\begin{matrix}
\f_{\D(\nu)} & \f_{\D_{2i}} & \f_{\De(\nu)+J-l-m_i}\nn\\
\f_{\D_{2i-1}+J-m_i} & \cW & \f_{\D_{2i-1}+l}
\end{matrix} \right\}^{\uniq (J-l-m_i,0)}_{\uniq (-J+l+m_i,0)}
\begin{pmatrix}
\f_{\D(\nu)}\\
\f_{\D(\nu)+J-l-m_i}\ \cW
\end{pmatrix}^{(-J+l+m_i,0)(J-l-m_i,0)}.
\eea{eq:cb_decomp_coeff}
Note that the crossing equation does not lead to additional sums because all the weight shifting operators end up in bubbles after the three-point crossing relations are used.

\subsection{Tree level single exchange diagrams}

Having expressed the bulk-to-bulk propagator in terms of harmonic functions and weight shifting operators, we can start simplifying Witten diagrams.
Consider an exchange Witten diagram with arbitrary spinning operators and  local couplings $a$ and $b$,
\be
\diagramEnvelope{\begin{tikzpicture}[anchor=base,baseline]
	\node (vertL) at (-0.8,0) [twopt] {};
	\node (vertR) at ( 0.8,0) [twopt] {};
	\node at (-1.1,-0.1) [] {$a$};
	\node at ( 1.1,-0.1) [] {$b$};
	\node (opO1) at (-1.6,-1.6) [] {};
	\node (opO2) at (-1.6, 1.6) [] {};
	\node (opO3) at ( 1.6,-1.6) [] {};
	\node (opO4) at ( 1.6, 1.6) [] {};
	\node at (-1.9,-1.9) {$\cO_1$};
	\node at (-1.9, 1.9) [] {$\cO_2$};
	\node at ( 1.9,-1.9) [] {$\cO_3$};
	\node at ( 1.9, 1.9) [] {$\cO_4$};
	\node at (0,-0.1) [below] {$\f_{\D,J}$};	
	\draw [spinning no arrow] (vertL)-- (opO1);
	\draw [spinning no arrow] (vertL)-- (opO2);
	\draw [spinning no arrow] (vertL)-- (vertR);
	\draw [spinning no arrow] (vertR)-- (opO3);
	\draw [spinning no arrow] (vertR)-- (opO4);
    \draw (0,0) circle (2.12);
\end{tikzpicture}}
\quad = \quad
\diagramEnvelope{\begin{tikzpicture}[anchor=base,baseline]
	\node (vertL) at (-1.1,-0.08) [threept] {$a$};
	\node (vertR) at (1.1,-0.11) [threept] {$b$};
	\node (opO1) at (-1.8,-1.6) [below] {$\cO_1$};
	\node (opO2) at (-1.8,1.6) [above] {$\cO_2$};
	\node (opO3) at (1.8,-1.6) [below] {$\cO_3$};
	\node (opO4) at (1.8,1.6) [above] {$\cO_4$};
	\node (prop1) at (-1.4,-0.8) [twopt] {};
	\node (prop2) at (-1.4,0.8) [twopt] {};
	\node (prop3) at (1.4,-0.8) [twopt] {};
	\node (prop4) at (1.4,0.8) [twopt] {};
	\node (prop) at (0,0) [bulkprop] {};
	\node at (0,-0.1) [below] {$\f_{\D,J}$};	
	\draw [spinning] (prop1)-- (opO1);
	\draw [spinning] (prop2)-- (opO2);
	\draw [spinning bulk] (prop1)-- (vertL);
	\draw [spinning bulk] (prop2)-- (vertL);
	\draw [spinning bulk] (prop)-- (vertL);
	\draw [spinning bulk] (prop)-- (vertR);
	\draw [spinning bulk] (prop3)-- (vertR);
	\draw [spinning bulk] (prop4)-- (vertR);
	\draw [spinning] (prop3)-- (opO3);
	\draw [spinning] (prop4)-- (opO4);
\end{tikzpicture}}\,.
\label{eq:W_exchange_1}
\ee
We will now insert the spectral representation of the bulk-to-bulk propagator \eqref{eq:bulk-to-bulk-diagram},
\be
\eqref{eq:W_exchange_1}
= \sum_{l=0}^J  \int d\nu \sum_{c,d} \a_{J,l}(\nu)^{ab}_{cd} 
\diagramEnvelope{\begin{tikzpicture}[anchor=base,baseline]
	\node (vertL) at (-1.1,-0.08) [threept] {$c$};
	\node (vertR) at (1.1,-0.11) [threept,inner sep=1pt] {$d$};
	\node (opO1) at (-1.8,-1.6) [below] {$\cO_1$};
	\node (opO2) at (-1.8,1.6) [above] {$\cO_2$};
	\node (opO3) at (1.8,-1.6) [below] {$\cO_3$};
	\node (opO4) at (1.8,1.6) [above] {$\cO_4$};
	\node (prop1) at (-1.4,-0.8) [twopt] {};
	\node (prop2) at (-1.4,0.8) [twopt] {};
	\node (prop3) at (1.4,-0.8) [twopt] {};
	\node (prop4) at (1.4,0.8) [twopt] {};
	\node (prop) at (0,0) [twopt] {};
	\node at (0,-0.1) [below] {$\f_{\D(\nu),l}$};	
	\draw [spinning] (prop1)-- (opO1);
	\draw [spinning] (prop2)-- (opO2);
	\draw [spinning bulk] (prop1)-- (vertL);
	\draw [spinning bulk] (prop2)-- (vertL);
	\draw [spinning bulk] (prop)-- (vertL);
	\draw [spinning bulk] (prop)-- (vertR);
	\draw [spinning bulk] (prop3)-- (vertR);
	\draw [spinning bulk] (prop4)-- (vertR);
	\draw [spinning] (prop3)-- (opO3);
	\draw [spinning] (prop4)-- (opO4);
\end{tikzpicture}}\,,
\label{eq:W_exchange_2}
\ee
and write each harmonic function in terms of the scalar one using \eqref{eq:harmonic_relation},
\be
\eqref{eq:W_exchange_2}
= \sum_{l=0}^J  \int d\nu \sum_{c,d} \frac{\a_{J,l}(\nu)^{ab}_{cd}}{\{\ldots\} (\ldots)} 
\diagramEnvelope{\begin{tikzpicture}[anchor=base,baseline]
	\node (vertL) at (-2.1,-0.08) [threept] {$c$};
	\node (vertR) at (2.1,-0.11) [threept,inner sep=1pt] {$d$};
	\node (opO1) at (-2.8,-1.6) [below] {$\cO_1$};
	\node (opO2) at (-2.8,1.6) [above] {$\cO_2$};
	\node (opO3) at (2.8,-1.6) [below] {$\cO_3$};
	\node (opO4) at (2.8,1.6) [above] {$\cO_4$};
	\node (prop1) at (-2.4,-0.8) [twopt] {};
	\node (prop2) at (-2.4,0.8) [twopt] {};
	\node (prop3) at (2.4,-0.8) [twopt] {};
	\node (prop4) at (2.4,0.8) [twopt] {};
	\node (prop) at (0,0) [twopt] {};
	\node (wsL) at (-1,-0.08) [threept] {$n$};
	\node (wsR) at (1,-0.08) [threept] {$n$};
	\node at (0,-0.1) [below] {$\f_{\D(\nu)}$};	
	\node at (-1.5,-0.1) [below] {$\f_{\D(\nu),l}$};	
	\node at (1.5,-0.1) [below] {$\f_{\D(\nu),l}$};	
	\node at (0,1) [above] {$\cW$};	
	\draw [spinning] (prop1)-- (opO1);
	\draw [spinning] (prop2)-- (opO2);
	\draw [spinning bulk] (prop1)-- (vertL);
	\draw [spinning bulk] (prop2)-- (vertL);
	\draw [scalar bulk] (prop)-- (wsL);
	\draw [scalar bulk] (prop)-- (wsR);
	\draw [spinning bulk] (wsL)--  (vertL);
	\draw [spinning bulk] (wsR)--  (vertR);
	\draw [spinning bulk] (prop3)-- (vertR);
	\draw [spinning bulk] (prop4)-- (vertR);
	\draw [spinning] (prop3)-- (opO3);
	\draw [spinning] (prop4)-- (opO4);
	\draw [finite with arrow] (wsL) to[out=90,in=90] (wsR);
\end{tikzpicture}}\,,
\label{eq:W_exchange_3}
\ee
where $n=(0,l)$.
Next we move the weight shifting operators to the external legs using the crossing equation for local couplings \eqref{eq:Bulk6jdefinition},
\be
\eqref{eq:W_exchange_3}
= \sum_{l=0}^J  \int d\nu \sum_{c,d} \frac{\a_{J,l}(\nu)^{ab}_{cd}}{\{\ldots\} (\ldots)} 
\sum_{\f_2',\f_4',e,f,p,q} \{\ldots\}^2
\diagramEnvelope{\begin{tikzpicture}[anchor=base,baseline]
	\node (vertL) at (-1,-0.08) [threept] {$e$};
	\node (vertR) at (1,-0.11) [threept,inner sep=1pt] {$f$};
	\node (opO1) at (-1.9,-1.6) [below] {$\cO_1$};
	\node (opO2) at (-2.3,2.3) [above] {$\cO_2$};
	\node (opO3) at (1.9,-1.6) [below] {$\cO_3$};
	\node (opO4) at (2.3,2.3) [above] {$\cO_4$};
	\node (prop1) at (-1.4,-0.8) [twopt] {};
	\node (prop2) at (-1.8,1.6) [twopt] {};
	\node (prop3) at (1.4,-0.8) [twopt] {};
	\node (prop4) at (1.8,1.6) [twopt] {};
	\node (prop) at (0,0) [twopt] {};
	\node (wsL) at (-1.4,0.8) [threept,inner sep=1pt] {$p$};
	\node (wsR) at (1.4,0.8) [threept,inner sep=1pt] {$q$};
	\node at (0,-0.1) [below] {$\f_{\D(\nu)}$};	
	\node at (-1.8,0.3) [] {$\f_2'$};	
	\node at (1.8,0.3) [] {$\f_4'$};	
	\node at (0,1) [above] {$\cW$};	
	\draw [spinning] (prop1)-- (opO1);
	\draw [spinning] (prop2)-- (opO2);
	\draw [spinning bulk] (prop1)-- (vertL);
	\draw [spinning bulk] (prop2)-- (wsL);
	\draw [spinning bulk] (wsL)-- (vertL);
	\draw [scalar bulk] (prop)-- (vertL);
	\draw [scalar bulk] (prop)-- (vertR);
	\draw [spinning bulk] (prop3)-- (vertR);
	\draw [spinning bulk] (prop4)-- (wsR);
	\draw [spinning bulk] (wsR)-- (vertR);
	\draw [spinning] (prop3)-- (opO3);
	\draw [spinning] (prop4)-- (opO4);
	\draw [finite with arrow] (wsL) to[out=0,in=180] (wsR);
\end{tikzpicture}}\,.
\label{eq:W_exchange_4}
\ee
The local couplings can be written in the weight shifting operator basis \eqref{eq:differentialbasistrickshorthand}, thus
\be
\eqref{eq:W_exchange_4}
= \sum_{l=0}^J  \int d\nu \sum_{c,d} \frac{\a_{J,l}(\nu)^{ab}_{cd}}{\{\ldots\} (\ldots)} 
\sum_{\substack{\f_2',\f_4',\\ e,f,p,q}} \{\ldots\}^2
\sum_{\substack{\cW',r,s,\\\cW'',t,u}} K^2
\diagramEnvelope{\begin{tikzpicture}[anchor=base,baseline]
	\node (vertL) at (-1,0) [twopt] {};
	\node (vertR) at (1,0) [twopt] {};
	\node (opO1) at (-2.3,-2.3) [below] {$\cO_1$};
	\node (opO2) at (-2.7,3.1) [above] {$\cO_2$};
	\node (opO3) at (2.3,-2.3) [below] {$\cO_3$};
	\node (opO4) at (2.7,3.1) [above] {$\cO_4$};
	\node (prop1) at (-1.8,-1.6) [twopt] {};
	\node (prop2) at (-2.2,2.4) [twopt] {};
	\node (prop3) at (1.8,-1.6) [twopt] {};
	\node (prop4) at (2.2,2.4) [twopt] {};
	\node (db1) at (-1.4,-0.8-0.05) [threept] {$r$};
	\node (db2) at (-1.4,0.8-0.05) [threept] {$s$};
	\node (db3) at (1.4,-0.8-0.08) [threept] {$t$};
	\node (db4) at (1.4,0.8-0.05) [threept] {$u$};
	\node at (-2.4,0) [] {$\cW'$};
	\node at (2.4,0) [] {$\cW''$};
	\node (prop) at (0,0) [twopt] {};
	\node (wsL) at (-1.8,1.6) [threept] {$p$};
	\node (wsR) at (1.8,1.6) [threept] {$q$};
	\node at (0,-0.1) [below] {$\f_{\D(\nu)}$};	
	\node at (-2,0.9) [] {$\f_2'$};	
	\node at (2,0.9) [] {$\f_4'$};	
	\node at (0,1) [above] {$\cW$};	
	\draw [spinning] (prop1)-- (opO1);
	\draw [spinning] (prop2)-- (opO2);
	\draw [spinning bulk] (prop1)-- (db1);
	\draw [spinning bulk] (prop2)-- (wsL);
	\draw [spinning bulk] (wsL)-- (db2);
	\draw [scalar bulk] (db1)-- (vertL);
	\draw [scalar bulk] (db2)-- (vertL);
	\draw [scalar bulk] (prop)-- (vertL);
	\draw [scalar bulk] (prop)-- (vertR);
	\draw [scalar bulk] (db3)-- (vertR);
	\draw [scalar bulk] (db4)-- (vertR);
	\draw [spinning bulk] (prop3)-- (db3);
	\draw [spinning bulk] (prop4)-- (wsR);
	\draw [spinning bulk] (wsR)-- (db4);
	\draw [spinning] (prop3) -- (opO3);
	\draw [spinning] (prop4) -- (opO4);
	\draw [finite with arrow] (db1) to[out=160,in=210] (db2);
	\draw [finite with arrow] (db3) to[out=30,in=330] (db4);
	\draw [finite with arrow] (wsL) to[out=0,in=180] (wsR);
\end{tikzpicture}},
\label{eq:W_exchange_5}
\ee
where $K^2$ is a placeholder for
\beq
K^2 = K_{e,\cW',r,s}^{\f_1 \f_2' \f_{\De(\nu)}} K_{f,\cW'',t,u}^{\f_3 \f_4' \f_{\De(\nu)}} \,.
\eeq
Moving the weight shifting operators to the boundary operators  our final result is
\be
\eqref{eq:W_exchange_5}
= \sum_{l=0}^J  \int d\nu \sum_{c,d} \frac{\a_{J,l}(\nu)^{ab}_{cd}}{\{\ldots\} (\ldots)} 
\sum_{\substack{\f_2',\f_4',\\ g, h,p,q}}
\sum_{\substack{\cW',r,s,\\\cW'',t,u}} K^2 \{\ldots\}^8
\diagramEnvelope{\begin{tikzpicture}[anchor=base,baseline]
	\node (vertL) at (-1,0) [twopt] {};
	\node (vertR) at (1,0) [twopt] {};
	\node (opO1) at (-2.3,-2.3) [below] {$\cO_1$};
	\node (opO2) at (-2.7,3.1) [above] {$\cO_2$};
	\node (opO3) at (2.3,-2.3) [below] {$\cO_3$};
	\node (opO4) at (2.7,3.1) [above] {$\cO_4$};
	\node (prop1) at (-1.4,-0.8) [twopt] {};
	\node (prop2) at (-1.4,0.8) [twopt] {};
	\node (prop3) at (1.4,-0.8) [twopt] {};
	\node (prop4) at (1.4,0.8) [twopt] {};
	\node (db1) at (-1.8,-1.6-0.08) [threept] {$\bar{r}$};
	\node (db2) at (-1.8,1.6-0.08) [threept] {$\bar{s}$};
	\node (db3) at (1.8,-1.6-0.1) [threept] {$\bar{t}$};
	\node (db4) at (1.8,1.6-0.08) [threept] {$\bar{u}$};
	\node at (-2.4,0) [] {$\cW'$};
	\node at (2.4,0) [] {$\cW''$};
	\node (prop) at (0,0) [twopt] {};
	\node (wsL) at (-2.2,2.4) [threept,inner sep=1pt] {$\bar{p}$};
	\node (wsR) at (2.2,2.4) [threept,inner sep=1pt] {$\bar{q}$};
	\node at (0,-0.1) [below] {$\f_{\D(\nu)}$};	
	\node at (-2.4,1.8) [] {$\f_2'$};	
	\node at (2.4,1.8) [] {$\f_4'$};	
	\node at (0,1.8) [above] {$\cW$};	
	\draw [spinning] (db1)-- (opO1);
	\draw [spinning] (wsL)-- (opO2);
	\draw [scalar] (prop1)-- (db1);
	\draw [scalar] (prop2)-- (db2);
	\draw [spinning] (db2)-- (wsL);
	\draw [scalar bulk] (prop1)-- (vertL);
	\draw [scalar bulk] (prop2)-- (vertL);
	\draw [scalar bulk] (prop)-- (vertL);
	\draw [scalar bulk] (prop)-- (vertR);
	\draw [scalar bulk] (prop3)-- (vertR);
	\draw [scalar bulk] (prop4)-- (vertR);
	\draw [scalar] (prop3)-- (db3);
	\draw [scalar] (prop4)-- (db4);
	\draw [spinning] (db4)-- (wsR);
	\draw [spinning] (db3)-- (opO3);
	 \draw [spinning] (wsR)-- (opO4);
	\draw [finite with arrow] (db1) to[out=160,in=210] (db2);
	\draw [finite with arrow] (db3) to[out=30,in=330] (db4);
	\draw [finite with arrow] (wsL) to[out=0,in=180] (wsR);
\end{tikzpicture}}\,,
\label{eq:W_exchange_6}
\ee
expressing a generic four-point exchange diagram in terms of all-scalar
exchange diagrams with weight shifting operators acting on them.

\subsection{Conformal block decomposition of single exchange  diagrams}
\label{sec:cb_decomposition}

As a check of the computations of Witten diagrams with AdS weight shifting operators, in this section we will 
reproduce the decomposition of the scalar Witten diagram with  spin $J$ exchange,
in terms of scalar conformal blocks exchanging operators with spin $l\le J$, which includes double trace operators \cite{Costa:2014kfa}.
The Witten diagram we want to decompose is
\be
\diagramEnvelope{\begin{tikzpicture}[anchor=base,baseline]
	\node (vertL) at (-0.8,0) [twopt] {};
	\node (vertR) at ( 0.8,0) [twopt] {};
	\node (opO1) at (-1.6,-1.6) [] {};
	\node (opO2) at (-1.6, 1.6) [] {};
	\node (opO3) at ( 1.6,-1.6) [] {};
	\node (opO4) at ( 1.6, 1.6) [] {};
	\node at (-1.9,-1.9) {$\cO_{\De_1}$};
	\node at (-1.9, 1.9) [] {$\cO_{\De_2}$};
	\node at ( 1.9,-1.9) [] {$\cO_{\De_3}$};
	\node at ( 1.9, 1.9) [] {$\cO_{\De_4}$};
	\node at (0,-0.1) [below] {$\f_{\D,J}$};	
	\draw [scalar no arrow] (vertL)-- (opO1);
	\draw [scalar no arrow] (vertL)-- (opO2);
	\draw [spinning no arrow] (vertL)-- (vertR);
	\draw [scalar no arrow] (vertR)-- (opO3);
	\draw [scalar no arrow] (vertR)-- (opO4);
    \draw (0,0) circle (2.12);
\end{tikzpicture}}
\quad = \quad
\diagramEnvelope{\begin{tikzpicture}[anchor=base,baseline]
	\node (vertL) at (-1.1,0) [twopt] {};
	\node (vertR) at (1.1,0) [twopt] {};
	\node (opO1) at (-1.8,-1.6) [below] {$\cO_1$};
	\node (opO2) at (-1.8,1.6) [above] {$\cO_2$};
	\node (opO3) at (1.8,-1.6) [below] {$\cO_3$};
	\node (opO4) at (1.8,1.6) [above] {$\cO_4$};
	\node (prop1) at (-1.4,-0.8) [twopt] {};
	\node (prop2) at (-1.4,0.8) [twopt] {};
	\node (prop3) at (1.4,-0.8) [twopt] {};
	\node (prop4) at (1.4,0.8) [twopt] {};
	\node (prop) at (0,0) [bulkprop] {};
	\node at (0,-0.1) [below] {$\f_{\D,J}$};	
	\draw [scalar] (prop1)-- (opO1);
	\draw [scalar] (prop2)-- (opO2);
	\draw [scalar bulk] (prop1)-- (vertL);
	\draw [scalar bulk] (prop2)-- (vertL);
	\draw [spinning bulk] (prop)-- (vertL);
	\draw [spinning bulk] (prop)-- (vertR);
	\draw [scalar bulk] (prop3)-- (vertR);
	\draw [scalar bulk] (prop4)-- (vertR);
	\draw [scalar] (prop3)-- (opO3);
	\draw [scalar] (prop4)-- (opO4);
\end{tikzpicture}}\,.
\label{eq:W_cb_1}
\ee
We insert the spectral representation of the bulk-to-bulk propagator \eqref{eq:bulk-to-bulk-diagram}, writing the harmonic functions in the split representation \eqref{eq:spinning_omega_to_shadow},
\be
\eqref{eq:W_cb_1}
= \sum_{l=0}^J  \int d\nu \, \a_{J,l}(\nu) \cT_{\D(\nu),l}
\diagramEnvelope{\begin{tikzpicture}[anchor=base,baseline]
	\node (vertL) at (-2.1,0) [twopt] {};
	\node (vertR) at (2.1,0) [twopt] {};
	\node (opO1) at (-2.8,-1.6) [below] {$\cO_1$};
	\node (opO2) at (-2.8,1.6) [above] {$\cO_2$};
	\node (opO3) at (2.8,-1.6) [below] {$\cO_3$};
	\node (opO4) at (2.8,1.6) [above] {$\cO_4$};
	\node (prop1) at (-2.4,-0.8) [twopt] {};
	\node (prop2) at (-2.4,0.8) [twopt] {};
	\node (prop3) at (2.4,-0.8) [twopt] {};
	\node (prop4) at (2.4,0.8) [twopt] {};
	\node (shad) at (0,0) [cross] {};
	\node (propL) at (-1,0) [twopt] {};
	\node (propR) at (1,0) [twopt] {};
	\node at (0,-0.1) [below] {$\f_{\D(\nu),l}$};	
	\draw [scalar] (prop1)-- (opO1);
	\draw [scalar] (prop2)-- (opO2);
	\draw [scalar bulk] (prop1)-- (vertL);
	\draw [scalar bulk] (prop2)-- (vertL);
	\draw [spinning bulk] (propL)-- (vertL);
	\draw [spinning] (propL)-- (shad);
	\draw [spinning] (propR)-- (shad);
	\draw [spinning bulk] (propR)-- (vertR);
	\draw [scalar bulk] (prop3)-- (vertR);
	\draw [scalar bulk] (prop4)-- (vertR);
	\draw [scalar] (prop3)-- (opO3);
	\draw [scalar] (prop4)-- (opO4);
\end{tikzpicture}}\,.
\label{eq:W_cb_2}
\ee
Now we can  simply perform the AdS integrals using \eqref{eq:ads_integral_unique}
and arrive at the conformal block decomposition
\be
\eqref{eq:W_cb_2}
=
\sum_{l=0}^J  \int d\nu \, \a_{J,l}(\nu) \cT_{\D(\nu),l} \,b(\cO_1, \cO_2 ,\cO_{\De(\nu),l})
\,b(\cO_3 ,\cO_4 ,\cO_{\De(\nu),l})
\diagramEnvelope{\begin{tikzpicture}[anchor=base,baseline]
	\node (vertL) at (-1,0) [twopt] {};
	\node (vertR) at (1,-0) [twopt] {};
	\node (opO1) at (-1.5,-1) [below] {$\cO_1$};
	\node (opO2) at (-1.5,1) [above] {$\cO_2$};
	\node (opO3) at (1.5,-1) [below] {$\cO_3$};
	\node (opO4) at (1.5,1) [above] {$\cO_4$};
	\node (shad) at (0,0) [cross] {};
	\node at (0,-0.1) [below] {$\cO_{\D(\nu),l}$};	
	\draw [scalar] (vertL)-- (opO1);
	\draw [scalar] (vertL)-- (opO2);
	\draw [spinning] (vertL)-- (shad);
	\draw [spinning] (vertR)-- (shad);
	\draw [scalar] (vertR)-- (opO3);
	\draw [scalar] (vertR)-- (opO4);
\end{tikzpicture}}.
\label{eq:W_cb_3}
\ee
The remaining diagram is the sum of the two scalar conformal blocks for exchange of the operators $\cO_{\D(\nu),l}$ and its shadow $\cO_{d-\D(\nu),l}$, and defined by
\be
G^{\cO_1 \cO_2 \cO_3 \cO_4 }_{\cO_{\D(\nu),l}} (P_i)
=\frac{1}{l! (h-1)_l\cN_{\De(\nu),l}}
\int D^d P\, \< \cO_1 \cO_2 \cO_{\D(\nu),l}(P, D_{Z})\> 
\<\cO_3 \cO_4 \tilde{\calO}_{\D(\nu),l}(P,Z) \>\,,
\ee
where the constant $\cN_{\De(\nu),l}$ was introduced in (\ref{eq:N}).

Let us compare \eqref{eq:W_cb_3} to the conformal block decomposition of the spin $J$ exchange diagram computed in \cite{Costa:2014kfa},
\be
W_{J}&=
\int d \nu \sum_{l=0}^J a_{J,l}(\nu)\,  \frac{\nu^2}{\pi} \,\cB_{\D_2,\D_1,\D(\nu),l,J} \cB_{\D_4,\D_3,d-\D(\nu),l,J} \cS_{d-\D(\nu),\D_{34}}^l G^{\cO_1 \cO_2 \cO_3 \cO_4 }_{\cO_{\D(\nu),l}} (P_i)\,.
\label{eq:cb_decomposition_spinning_ads_propagators}
\ee
The function $\mathcal{B}_{\D_2,\D_1,\D(\nu),l,J}$ was computed in \cite{Costa:2014kfa} for any $J$. The factor $\cS_{d-\D(\nu),\D_{34}}^l$ (where $\De_{34}=\De_3-\De_4$) arises because the conformal blocks were defined in \cite{Costa:2014kfa} as
\be
\frac{1}{l!(h-1)_l}
\int D^d P\, \< \cO_{1} \cO_{2} \cO_{\D(\nu),l}(P, D_{Z})\> 
\< \cO_{3} \cO_{4} \calO_{d-\D(\nu),l}(P,Z) \>
= \cS_{d-\D(\nu),\D_{34}}^l G^{\cO_1 \cO_2 \cO_3 \cO_4 }_{\cO_{\D(\nu),l}} (P_i)\,.
\label{eq:block_spinning_ads_propagators}
\ee
It is the factor that arises when performing a shadow transformation \eqref{eq:shadow_transform} of an operator in a three-point function,
\bea
\cS_{\D, \D_{34}}^J &\equiv \frac{\<\cO_{3}(P_3) \cO_{4}(P_4) \tilde{\calO}_{\D,J}(P_0,Z_0) \>}{\<\cO_{3}(P_3) \cO_{4}(P_4) \calO_{d-\D,J}(P_0,Z_0) \> }\\
&=\pi^{\frac{d}{2}} (\D-1)_J \frac{\Gamma(\D-\frac{d}{2}) \Gamma(\frac{1}{2}(d-\D+\D_{34}+J))\Gamma(\frac{1}{2}(d-\D-\D_{34}+J))}{\Gamma(d-\D+J) \Gamma(\frac{1}{2}(\D+\D_{34}+J))\Gamma(\frac{1}{2}(\D-\D_{34}+J))}\,,
\eea{eq:shadow_3pt_constant}
computed for instance in \cite{Costa:2016hju}.
To show \eqref{eq:block_spinning_ads_propagators} one further needs to use  $\cN_{\De,J} = \cS_{\D,\D_{34}}^J \cS_{d-\D,\D_{34}}^J$.
One can bring \eqref{eq:cb_decomposition_spinning_ads_propagators} into a more symmetric form by noticing that
\beq
 \frac{\nu^2}{\pi} \, \cB_{\D_4,\D_3,d-\D(\nu),l,J} \cS_{d-\D(\nu),\D_{34}}^l
= \cT_{\De,l} \cB_{\D_4,\D_3,\D(\nu),l,J}\,.
\eeq
So in order to compare \eqref{eq:W_cb_3} and \eqref{eq:cb_decomposition_spinning_ads_propagators} it is enough to check that
\bea
{}&\cB_{\D_2,\D_1,\D,l,J} = b(\cO_1, \cO_2, \cO_{\De,l})
\sum_{m=0}^{J-l} \beta_{J,l,m}^{\De_1 \De} 
\left\{\begin{matrix}
\f_{\D,l} & \f_{\D_{2}} & \f_{\De}\nn\\
\f_{\D_{1}+l} & \cW & \f_{\D_{1}}
\end{matrix} \right\}^{\uniq (l,0)}_{\uniq (0,l)}
\begin{pmatrix}
\f_{\D,l}\\
\f_{\D}\ \cW
\end{pmatrix}^{(0,l)(0,-l)}\\
& \times \left\{\begin{matrix}
\f_{\D} & \f_{\D_{2}} & \f_{\De+J-l-m}\nn\\
\f_{\D_{1}+J-m} & \cW & \f_{\D_{1}+l}
\end{matrix} \right\}^{\uniq (J-l-m,0)}_{\uniq (-J+l+m,0)}
\begin{pmatrix}
\f_{\D}\\
\f_{\D+J-l-m}\ \cW
\end{pmatrix}^{(-J+l+m,0)(J-l-m,0)},
\eea{eq:cB_comparison}
which is indeed satisfied.

\subsection{Loop diagrams}

Now we want to go one step further and show that the same techniques can be used to 
reduce the internal lines of loop level Witten diagrams to scalars. We will demonstrate this in terms of some simple examples, but it will become clear that the approach can be used at least for any one-loop diagram with only cubic couplings. Quartic couplings pose a challenge that we postpone for now, because the crossing equation for four-point correlators can involve functions of cross-ratios.
One should note that even the Witten diagrams involving only scalar particles that appear in our final results were in many cases never computed. One exception is the two-point one-loop diagram, which was recently computed in \cite{Giombi:2017hpr}.

\subsubsection{Reduction to mostly scalar diagrams}

For the two-point one-loop diagram we start with
\be
\diagramEnvelope{\begin{tikzpicture}[anchor=base,baseline]
	\node (vertL) at (-0.8,0) [twopt] {};
	\node (vertR) at ( 0.8,0) [twopt] {};
	\node at (-0.6,-0.1) [] {$a$};
	\node at ( 0.6,-0.1) [] {$b$};
	\node (opO1) at (-2.25,0) [] {};
	\node (opO2) at (2.25, 0) [] {};
	\node at (-1.5,-0.1) [below] {$\cO_1$};
	\node at ( 1.5,-0.1) [below] {$\cO_2$};
	\node at (0, 0.9) [above] {$\f_{\D,J}$};
	\node at (0,-0.9) [below] {$\f_{\bar{\D},\bar{J}}$};
	\draw [spinning no arrow] (vertL)-- (opO1);
	\draw [spinning no arrow] (vertR)-- (opO2);
    \draw (0,0) circle (2.12);
	\draw [spinning no arrow] (vertL) to[out=90,in=90,looseness=1.6] (vertR);
	\draw [spinning no arrow] (vertL) to[out=270,in=270,looseness=1.6] (vertR);
\end{tikzpicture}}
=
\diagramEnvelope{\begin{tikzpicture}[anchor=base,baseline]
	\node (vertL) at (-1.4,-0.08) [threept] {$a$};
	\node (vertR) at (1.4,-0.12) [threept] {$b$};
	\node (propL) at (-2.4,0) [twopt] {};
	\node (propR) at (2.4,0) [twopt] {};
	\node (opO1) at (-3.4,0) [left] {};
	\node (opO2) at (3.4,0) [right] {};
	\node at (-3.4,0) [below] {$\cO_1$};
	\node at (3.4,0) [below] {$\cO_2$};
	\node (shad) at (0,0.8) [bulkprop] {};
	\node (splitL) at (-1,0.8) [inner sep=0pt] {};
	\node (splitR) at (1,0.8) [inner sep=0pt] {};
	\node at (0,0.9) [above] {$\f_{\D,J}$};
	\node (shad2) at (0,-0.8) [bulkprop] {};
	\node (splitL2) at (-1,-0.8) [inner sep=0pt] {};
	\node (splitR2) at (1,-0.8) [inner sep=0pt] {};
	\node at (0,-0.9) [below] {$\f_{\bar{\D},\bar{J}}$};
	\draw [spinning] (propL)-- (opO1);
	\draw [spinning bulk] (propL)-- (vertL);
	\draw [spinning bulk] (splitL)-- (vertL);
	\draw [spinning bulk] (splitL2)-- (vertL);
	\draw [spinning] (propR)-- (opO2);
	\draw [spinning bulk] (propR)-- (vertR);
	\draw [spinning bulk] (splitR)-- (vertR);
	\draw [spinning bulk] (splitR2)-- (vertR);
	\draw [spinning bulk] (shad2)-- (splitL2);
	\draw [spinning bulk] (shad)-- (splitR);
	\draw [spinning bulk] (shad)-- (splitL);
	\draw [spinning bulk] (shad2)-- (splitR2);
\end{tikzpicture}}.
\label{eq:2pt-1loop-1}
\ee
In order to insert the spectral representation of the bulk-to-bulk propagator one has to
follow the procedure explained in section \ref{sec:bulk-to-bulk-propagator-diagram}. First one expresses the bulk-to-bulk propagators in terms of weight shifting operators acting on harmonic functions and then one can remove the weight shifting operators using crossing. In general it is not possible to just use \eqref{eq:bulk-to-bulk-diagram} twice  because that would assume one can use crossing when a bulk-to-bulk propagator is attached to two of the external legs of \eqref{eq:bulk-to-bulk-diagram}.
Instead one first needs to determine how the derivatives from the spectral representations of both
bulk-to-boundary propagators can be expressed in terms of weight shifting operators
before carrying on using crossing equations.
The Witten diagram can thus be written as
\be
\eqref{eq:2pt-1loop-1}
=
\sum_{l=0}^J \sum_{\bar{l}=0}^{\bar{J}} \int d\nu d\bar{\nu} \sum_{c,d} \a_{J,l,\bar{J},\bar{l}}(\nu,\bar{\nu})^{ab}_{cd} 
\diagramEnvelope{\begin{tikzpicture}[anchor=base,baseline]
	\node (vertL) at (-1.4,-0.08) [threept] {$c$};
	\node (vertR) at (1.4,-0.12) [threept] {$d$};
	\node (propL) at (-2.4,0) [twopt] {};
	\node (propR) at (2.4,0) [twopt] {};
	\node (opO1) at (-3.4,0) [left] {};
	\node (opO2) at (3.4,0) [right] {};
	\node at (-3.4,0) [below] {$\cO_1$};
	\node at (3.4,0) [below] {$\cO_2$};
	\node (shad) at (0,0.8) [twopt] {};
	\node (splitL) at (-1,0.8) [inner sep=0pt] {};
	\node (splitR) at (1,0.8) [inner sep=0pt] {};
	\node at (0,0.9) [above] {$\f_{\D(\nu),l}$};
	\node (shad2) at (0,-0.8) [twopt] {};
	\node (splitL2) at (-1,-0.8) [inner sep=0pt] {};
	\node (splitR2) at (1,-0.8) [inner sep=0pt] {};
	\node at (0,-0.9) [below] {$\f_{\D(\bar{\nu}),\bar{l}}$};
	\draw [spinning] (propL)-- (opO1);
	\draw [spinning bulk] (propL)-- (vertL);
	\draw [spinning bulk] (splitL)-- (vertL);
	\draw [spinning bulk] (splitL2)-- (vertL);
	\draw [spinning] (propR)-- (opO2);
	\draw [spinning bulk] (propR)-- (vertR);
	\draw [spinning bulk] (splitR)-- (vertR);
	\draw [spinning bulk] (splitR2)-- (vertR);
	\draw [spinning bulk] (shad2)-- (splitL2);
	\draw [spinning bulk] (shad)-- (splitR);
	\draw [spinning bulk] (shad)-- (splitL);
	\draw [spinning bulk] (shad2)-- (splitR2);
\end{tikzpicture}}.
\label{eq:2pt-1loop-3}
\ee
The functions $\a_{J,l,\bar{J},\bar{l}}(\nu,\bar{\nu})^{ab}_{cd}$ are defined by this equation.
To simplify this expression we would like to reduce as many lines as possible to scalars in the diagram
\be
\diagramEnvelope{\begin{tikzpicture}[anchor=base,baseline]
	\node (vertL) at (-1.4,-0.08) [threept] {$c$};
	\node (vertR) at (1.4,-0.12) [threept] {$d$};
	\node (propL) at (-2.4,0) [twopt] {};
	\node (propR) at (2.4,0) [twopt] {};
	\node (opO1) at (-3.4,0) [left] {};
	\node (opO2) at (3.4,0) [right] {};
	\node at (-3.4,0) [below] {$\cO_1$};
	\node at (3.4,0) [below] {$\cO_2$};
	\node (shad) at (0,0.8) [twopt] {};
	\node (splitL) at (-1,0.8) [inner sep=0pt] {};
	\node (splitR) at (1,0.8) [inner sep=0pt] {};
	\node at (0,0.9) [above] {$\f_{\D(\nu),l}$};
	\node (shad2) at (0,-0.8) [twopt] {};
	\node (splitL2) at (-1,-0.8) [inner sep=0pt] {};
	\node (splitR2) at (1,-0.8) [inner sep=0pt] {};
	\node at (0,-0.9) [below] {$\f_{\D(\bar{\nu}),\bar{l}}$};
	\draw [spinning] (propL)-- (opO1);
	\draw [spinning bulk] (propL)-- (vertL);
	\draw [spinning bulk] (splitL)-- (vertL);
	\draw [spinning bulk] (splitL2)-- (vertL);
	\draw [spinning] (propR)-- (opO2);
	\draw [spinning bulk] (propR)-- (vertR);
	\draw [spinning bulk] (splitR)-- (vertR);
	\draw [spinning bulk] (splitR2)-- (vertR);
	\draw [spinning bulk] (shad2)-- (splitL2);
	\draw [spinning bulk] (shad)-- (splitR);
	\draw [spinning bulk] (shad)-- (splitL);
	\draw [spinning bulk] (shad2)-- (splitR2);
\end{tikzpicture}}.
\label{eq:2pt-1loop-diagram-1}
\ee
To this end we start by writing one of the local couplings in the weight shifting operator basis
\eqref{eq:differentialbasistrickshorthand},
\be
\eqref{eq:2pt-1loop-diagram-1} =
\sum_{\cW,m,n}
K_{c,\cW,m,n}^{\f_1 \f_{\De(\nu),l} \f_{\D(\bar{\nu}),\bar{l}}} 
\diagramEnvelope{\begin{tikzpicture}[anchor=base,baseline]
	\node (vertL) at (-1.4,0) [twopt] {};
	\node (vertR) at (1.4,-0.12) [threept] {$d$};
	\node (propL) at (-3.4,0) [twopt] {};
	\node (propR) at (2.4,0) [twopt] {};
	\node (opO1) at (-4.4,0) [left] {};
	\node (opO2) at (3.4,0) [right] {};
	\node at (-4.4,0) [below] {$\cO_1$};
	\node at (3.4,0) [below] {$\cO_2$};
	\node (ws1L) at (-2.4,-0.1) [threept] {$m$};
	\node (ws1R) at (-1,0.8-0.08) [threept] {$n$};
	\node at (-1.8,0) [below] {$\f_1'$};
	\node at (-1.1,0.3) [right] {$\f'$};
	\node at (-2,0.8) [above] {$\cW$};
	\node (shad) at (0,0.8) [twopt] {};
	\node (splitR) at (1,0.8) [inner sep=0pt] {};
	\node at (0,0.9) [above] {$\f_{\D(\nu),l}$};
	\node (shad2) at (0,-0.8) [twopt] {};
	\node (splitL2) at (-1,-0.8) [inner sep=0pt] {};
	\node (splitR2) at (1,-0.8) [inner sep=0pt] {};
	\node at (0,-0.9) [below] {$\f_{\D(\bar{\nu}),\bar{l}}$};
	\draw [spinning] (propL)-- (opO1);
	\draw [spinning bulk] (propL)-- (ws1L);
	\draw [scalar bulk] (ws1L)-- (vertL);
	\draw [scalar bulk] (ws1R)-- (vertL);
	\draw [spinning bulk] (splitL2)-- (vertL);
	\draw [spinning] (propR)-- (opO2);
	\draw [spinning bulk] (propR)-- (vertR);
	\draw [spinning bulk] (splitR)-- (vertR);
	\draw [spinning bulk] (splitR2)-- (vertR);
	\draw [spinning bulk] (shad2)-- (splitL2);
	\draw [spinning bulk] (shad)-- (splitR);
	\draw [spinning bulk] (shad)-- (ws1R);
	\draw [spinning bulk] (shad2)-- (splitR2);
	\draw [finite with arrow] (ws1L) to[out=60,in=180] (ws1R);
\end{tikzpicture}},
\label{eq:2pt-1loop-diagram-2}
\ee
and move the weight shifting operator labelled $n$ past the harmonic function and the vertex to the other external leg
\be
\eqref{eq:2pt-1loop-diagram-2} =
\sum_{\cW,m,n} K \sum_{\f_2',e,o} \{ \ldots \}^2
\diagramEnvelope{\begin{tikzpicture}[anchor=base,baseline]
	\node (vertL) at (-1.4,0) [twopt] {};
	\node (vertR) at (1.4,-0.08) [threept] {$e$};
	\node (propL) at (-3.4,0) [twopt] {};
	\node (propR) at (3.4,0) [twopt] {};
	\node (opO1) at (-4.4,0) [left] {};
	\node (opO2) at (4.4,0) [right] {};
	\node at (-4.4,0) [below] {$\cO_1$};
	\node at (4.4,0) [below] {$\cO_2$};
	\node (ws1L) at (-2.4,-0.1) [threept] {$m$};
	\node (ws1R) at (2.4,-0.08) [threept] {$o$};
	\node at (-1.8,0) [below] {$\f_1'$};
	\node at (1.8,0) [below] {$\f_2'$};
	\node at (0,0.7) [below] {$\f'$};
	\node (shad) at (0,0.8) [twopt] {};
	\node (splitL) at (-1,0.8) [inner sep=0pt] {};
	\node (splitR) at (1,0.8) [inner sep=0pt] {};
	\node at (0,1.2) [above] {$\cW$};
	\node (shad2) at (0,-0.8) [twopt] {};
	\node (splitL2) at (-1,-0.8) [inner sep=0pt] {};
	\node (splitR2) at (1,-0.8) [inner sep=0pt] {};
	\node at (0,-0.9) [below] {$\f_{\D(\bar{\nu}),\bar{l}}$};
	\draw [spinning] (propL)-- (opO1);
	\draw [spinning bulk] (propL)-- (ws1L);
	\draw [scalar bulk] (ws1L)-- (vertL);
	\draw [scalar bulk] (splitL)-- (vertL);
	\draw [spinning bulk] (splitL2)-- (vertL);
	\draw [spinning] (propR)-- (opO2);
	\draw [spinning bulk] (propR)-- (ws1R);
	\draw [spinning bulk] (ws1R)-- (vertR);
	\draw [scalar bulk] (splitR)-- (vertR);
	\draw [spinning bulk] (splitR2)-- (vertR);
	\draw [spinning bulk] (shad2)-- (splitL2);
	\draw [scalar bulk] (shad)-- (splitR);
	\draw [scalar bulk] (shad)-- (splitL);
	\draw [spinning bulk] (shad2)-- (splitR2);
	\draw [finite with arrow] (ws1L) to[out=90,in=90] (ws1R);
\end{tikzpicture}}.
\label{eq:2pt-1loop-diagram-3}
\ee
The same steps repeated for the other vertex, yield
\be
{}&\eqref{eq:2pt-1loop-diagram-3} =
\sum_{\substack{\cW,m,n,\\\cW',p,q}} K^2 \sum_{\f_2',e,o} \{ \ldots \}^2
\diagramEnvelope{\begin{tikzpicture}[anchor=base,baseline]
	\node (vertL) at (-1.4,0) [twopt] {};
	\node (vertR) at (1.4,0) [twopt] {};
	\node (propL) at (-3.4,0) [twopt] {};
	\node (propR) at (4.4,0) [twopt] {};
	\node (opO1) at (-4.4,0) [left] {};
	\node (opO2) at (5.4,0) [right] {};
	\node at (-4.4,0) [below] {$\cO_1$};
	\node at (5.4,0) [below] {$\cO_2$};
	\node (ws1L) at (-2.4,-0.1) [threept] {$m$};
	\node (ws1R) at (3.4,-0.08) [threept] {$o$};
	\node at (-1.8,0) [below] {$\f_1'$};
	\node at (2.8,0) [below] {$\f_2'$};
	\node at (1.8,0) [above] {$\f_2''$};
	\node at (0,0.7) [below] {$\f'$};
	\node at (1.1,-0.3) [left] {$\f''$};
	\node (ws2L) at (1,-0.8-0.06) [threept] {$q$};
	\node (ws2R) at (2.4,-0.08) [threept] {$p$};
	\node at (2,-0.8) [below] {$\cW'$};
	\node (shad) at (0,0.8) [twopt] {};
	\node (splitL) at (-1,0.8) [inner sep=0pt] {};
	\node (splitR) at (1,0.8) [inner sep=0pt] {};
	\node at (0.5,1.5) [above] {$\cW$};
	\node (shad2) at (0,-0.8) [twopt] {};
	\node (splitL2) at (-1,-0.8) [inner sep=0pt] {};
	\node at (0,-0.9) [below] {$\f_{\D(\bar{\nu}),\bar{l}}$};
	\draw [spinning] (propL)-- (opO1);
	\draw [spinning bulk] (propL)-- (ws1L);
	\draw [scalar bulk] (ws1L)-- (vertL);
	\draw [scalar bulk] (splitL)-- (vertL);
	\draw [spinning bulk] (splitL2)-- (vertL);
	\draw [spinning] (propR)-- (opO2);
	\draw [spinning bulk] (propR)-- (ws1R);
	\draw [spinning bulk] (ws1R)-- (ws2R);
	\draw [scalar bulk] (ws2R)-- (vertR);
	\draw [scalar bulk] (splitR)-- (vertR);
	\draw [scalar bulk] (ws2L)-- (vertR);
	\draw [spinning bulk] (shad2)-- (splitL2);
	\draw [scalar bulk] (shad)-- (splitR);
	\draw [scalar bulk] (shad)-- (splitL);
	\draw [spinning bulk] (shad2)-- (ws2L);
	\draw [finite with arrow] (ws1L) to[out=90,in=90] (ws1R);
	\draw [finite with arrow] (ws2R) to[out=240,in=0] (ws2L);
\end{tikzpicture}}\nn\\
&=
\sum_{\substack{\cW,m,n,\\\cW',p,q}} K^2 \sum_{\substack{\f_2',e,o,\\\f_1'',r}} \{ \ldots \}^4
\diagramEnvelope{\begin{tikzpicture}[anchor=base,baseline]
	\node (vertL) at (-1.4,0) [twopt] {};
	\node (vertR) at (1.4,0) [twopt] {};
	\node (propL) at (-4.4,0) [twopt] {};
	\node (propR) at (4.4,0) [twopt] {};
	\node (opO1) at (-5.4,0) [left] {};
	\node (opO2) at (5.4,0) [right] {};
	\node at (-5.4,0) [below] {$\cO_1$};
	\node at (5.4,0) [below] {$\cO_2$};
	\node (ws1L) at (-3.4,-0.1) [threept] {$m$};
	\node (ws1R) at (3.4,-0.08) [threept] {$o$};
	\node at (-2.8,0) [below] {$\f_1'$};
	\node at (-1.8,0) [above] {$\f_1''$};
	\node at (2.8,0) [below] {$\f_2'$};
	\node at (1.8,0) [above] {$\f_2''$};
	\node at (0,0.7) [below] {$\f'$};
	\node (ws2L) at (-2.4,-0.08) [threept] {$r$};
	\node (ws2R) at (2.4,-0.06) [threept] {$p$};
	\node at (0,-1.2) [below] {$\cW'$};
	\node (shad) at (0,0.8) [twopt] {};
	\node (splitL) at (-1,0.8) [inner sep=0pt] {};
	\node (splitR) at (1,0.8) [inner sep=0pt] {};
	\node at (0,1.7) [above] {$\cW$};
	\node (shad2) at (0,-0.8) [twopt] {};
	\node (splitL2) at (-1,-0.8) [inner sep=0pt] {};
	\node (splitR2) at (1,-0.8) [inner sep=0pt] {};
	\draw [spinning] (propL)-- (opO1);
	\draw [spinning bulk] (propL)-- (ws1L);
	\draw [scalar bulk] (ws1L)-- (ws2L);
	\draw [spinning bulk] (ws2L)-- (vertL);
	\draw [scalar bulk] (splitL)-- (vertL);
	\draw [scalar bulk] (splitL2)-- (vertL);
	\draw [spinning] (propR)-- (opO2);
	\draw [spinning bulk] (propR)-- (ws1R);
	\draw [spinning bulk] (ws1R)-- (ws2R);
	\draw [scalar bulk] (ws2R)-- (vertR);
	\draw [scalar bulk] (splitR)-- (vertR);
	\draw [scalar bulk] (splitR2)-- (vertR);
	\draw [scalar bulk] (shad2)-- (splitL2);
	\draw [scalar bulk] (shad)-- (splitR);
	\draw [scalar bulk] (shad)-- (splitL);
	\draw [scalar bulk] (shad2)-- (splitR2);
	\draw [finite with arrow] (ws1L) to[out=90,in=90] (ws1R);
	\draw [finite with arrow] (ws2R) to[out=270,in=270] (ws2L);
\end{tikzpicture}}.
\label{eq:2pt-1loop-diagram-4}
\ee
Finally the weight shifting operators are moved past the bulk-to-boundary propagators on the external legs. A further simplification comes due to Schur's lemma, which implies $\f_1'' = \f_2''$ and $\f_2' = \f_1'$,
\be
\eqref{eq:2pt-1loop-diagram-4}=
\sum_{\substack{\cW,m,n,\\\cW',p,q}} K^2 \{ \ldots \}^8
\diagramEnvelope{\begin{tikzpicture}[anchor=base,baseline]
	\node (vertL) at (-1.4,0) [twopt] {};
	\node (vertR) at (1.4,0) [twopt] {};
	\node (propL) at (-2.4,0) [twopt] {};
	\node (propR) at (2.4,0) [twopt] {};
	\node (opO1) at (-5.4,0) [left] {};
	\node (opO2) at (5.4,0) [right] {};
	\node at (-5.4,0) [below] {$\cO_1$};
	\node at (5.4,0) [below] {$\cO_2$};
	\node (ws1L) at (-4.4,-0.1) [threept] {$\bar{m}$};
	\node (ws1R) at (4.4,-0.1) [threept] {$\bar{m}$};
	\node at (-3.8,-0.1) [below] {$\cO_1'$};
	\node at (-1.8,0) [above] {$\f_2''$};
	\node at (3.8,-0.1) [below] {$\cO_1'$};
	\node at (1.8,0) [above] {$\f_2''$};
	\node at (0,0.7) [below] {$\f'$};
	\node (ws2L) at (-3.4,-0.08) [threept] {$\bar{p}$};
	\node (ws2R) at (3.4,-0.08) [threept] {$\bar{p}$};
	\node at (0,-1.2) [below] {$\cW'$};
	\node (shad) at (0,0.8) [twopt] {};
	\node (splitL) at (-1,0.8) [inner sep=0pt] {};
	\node (splitR) at (1,0.8) [inner sep=0pt] {};
	\node at (0,1.7) [above] {$\cW$};
	\node (shad2) at (0,-0.8) [twopt] {};
	\node (splitL2) at (-1,-0.8) [inner sep=0pt] {};
	\node (splitR2) at (1,-0.8) [inner sep=0pt] {};
	\node at (0,-0.7) [above] {$\f''$};
	\draw [spinning] (ws1L)-- (opO1);
	\draw [scalar] (ws2L)-- (ws1L);
	\draw [scalar] (propL)-- (ws2L);
	\draw [scalar bulk] (propL)-- (vertL);
	\draw [scalar bulk] (splitL)-- (vertL);
	\draw [scalar bulk] (splitL2)-- (vertL);
	\draw [spinning] (ws1R)-- (opO2);
	\draw [scalar] (ws2R)-- (ws1R);
	\draw [scalar] (propR)-- (ws2R);
	\draw [scalar bulk] (propR)-- (vertR);
	\draw [scalar bulk] (splitR)-- (vertR);
	\draw [scalar bulk] (splitR2)-- (vertR);
	\draw [scalar bulk] (shad2)-- (splitL2);
	\draw [scalar bulk] (shad)-- (splitR);
	\draw [scalar bulk] (shad)-- (splitL);
	\draw [scalar bulk] (shad2)-- (splitR2);
	\draw [finite with arrow] (ws1L) to[out=60,in=120] (ws1R);
	\draw [finite with arrow] (ws2R) to[out=240,in=300] (ws2L);
\end{tikzpicture}}.
\label{eq:2pt-1loop-diagram-5}
\ee
Hence the computation is reduced to an all scalar loop diagram.

The same technique can be used for any ($n$-point) one-loop diagram with only cubic couplings.
Consider for example the generic  three-point one-loop diagram with cubic vertices,
\be
\diagramEnvelope{\begin{tikzpicture}[anchor=base,baseline]
	\node (vert1) at (-0.866,0.5) [twopt] {};
	\node (vert2) at ( 0.866,0.5) [twopt] {};
	\node (vert3) at ( 0, -1) [twopt] {};
	\node (opO1) at (-1.98,1.125) [] {};
	\node (opO2) at ( 1.98,1.125) [] {};
	\node (opO3) at ( 0,-2.25) [] {};
	\node at (-2.3,1.3) {$\cO_1$};
	\node at ( 2.3,1.3) {$\cO_2$};
	\node at ( 0,-2.6) {$\cO_3$};
	\node at (0,0.6) [above] {$\f_{\D_4,J_4}$};
	\node at (-0.5,-0.3) [left] {$\f_{\D_6,J_6}$};
	\node at (0.5,-0.3) [right] {$\f_{\D_5,J_5}$};
	\node at (-0.55,0.2) [] {$a$};
	\node at ( 0.55,0.2) [] {$b$};
	\node at ( 0, -0.7) [] {$c$};
	\draw [spinning no arrow] (vert1)-- (opO1);
	\draw [spinning no arrow] (vert2)-- (opO2);
	\draw [spinning no arrow] (vert3)-- (opO3);
	\draw [spinning no arrow] (vert1)-- (vert2);
	\draw [spinning no arrow] (vert2)-- (vert3);
	\draw [spinning no arrow] (vert3)-- (vert1);
    \draw (0,0) circle (2.12);
\end{tikzpicture}}
\quad = \quad
\diagramEnvelope{\begin{tikzpicture}[anchor=base,baseline]
	\node (vert1) at (-1,0.575-0.08) [threept] {$a$};
	\node (vert2) at ( 1,0.575-0.11) [threept] {$b$};
	\node (vert3) at ( 0, -1.15-0.08) [threept] {$c$};
	\node (prop1) at ( -1.74,1) [twopt] {};
	\node (opO1) at (-2.43,1.4) [] {};
	\node (prop2) at ( 1.74,1) [twopt] {};
	\node (opO2) at ( 2.43,1.4) [] {};
	\node (prop3) at ( 0,-2) [twopt] {};
	\node (opO3) at ( 0,-2.8) [] {};
	\node at (-2.6,1.5) {$\cO_1$};
	\node at ( 2.6,1.5) {$\cO_2$};
	\node at ( 0,-3.1) {$\cO_3$};
	\node (shad) at (0,0.575) [bulkprop] {};
	\node (shadL) at (-0.5,-0.3) [bulkprop] {};
	\node (shadR) at (0.5,-0.3) [bulkprop] {};
	\node at (0,0.7) [above] {$\f_{\D_4,J_4}$};
	\node at (-0.6,-0.4) [left] {$\f_{\D_6,J_6}$};
	\node at (0.6,-0.4) [right] {$\f_{\D_5,J_5}$};
	\draw [spinning]      (prop1)-- (opO1);
	\draw [spinning bulk] (prop1)-- (vert1);
	\draw [spinning]      (prop2)-- (opO2);
	\draw [spinning bulk] (prop2)-- (vert2);
	\draw [spinning]      (prop3)-- (opO3);
	\draw [spinning bulk] (prop3)-- (vert3);
	\draw [spinning bulk] (shad)--  (vert1);
	\draw [spinning bulk] (shadL)-- (vert1);
	\draw [spinning bulk] (shad)--  (vert2);
	\draw [spinning bulk] (shadR)-- (vert2);
	\draw [spinning bulk] (shadL)-- (vert3);
	\draw [spinning bulk] (shadR)-- (vert3);
\end{tikzpicture}}\,.
\label{eq:3pt-1loop-1}
\ee
We write the bulk-to-bulk propagators in the spectral representation (defining coefficients $\a$),
\be
\eqref{eq:3pt-1loop-1} =
\sum_{l_4,l_5,l_6=0}^{J_4,J_5,J_6} \int d\nu_4 d\nu_5 d\nu_6 \sum_{d,e,f} \a_{J_i,l_i}(\nu_4,\nu_5,\nu_6)^{abc}_{def} \!\!\!\!\!\!\!\!\!\!
\diagramEnvelope{\begin{tikzpicture}[anchor=base,baseline]
	\node (vert1) at (-1,0.575-0.1) [threept] {$d$};
	\node (vert2) at ( 1,0.575-0.08) [threept,inner sep=2.5pt] {$e$};
	\node (vert3) at ( 0, -1.15-0.1) [threept,inner sep=1.5pt] {$f$};
	\node (prop1) at ( -1.74,1) [twopt] {};
	\node (opO1) at (-2.43,1.4) [] {};
	\node (prop2) at ( 1.74,1) [twopt] {};
	\node (opO2) at ( 2.43,1.4) [] {};
	\node (prop3) at ( 0,-2) [twopt] {};
	\node (opO3) at ( 0,-2.8) [] {};
	\node at (-2.6,1.5) {$\cO_1$};
	\node at ( 2.6,1.5) {$\cO_2$};
	\node at ( 0,-3.1) {$\cO_3$};
	\node (shad) at (0,0.575) [twopt] {};
	\node (shadL) at (-0.5,-0.3) [twopt] {};
	\node (shadR) at (0.5,-0.3) [twopt] {};
	\node at (0,0.7) [above] {$\f_{\D(\nu_4),l_4}$};
	\node at (-0.6,-0.4) [left] {$\f_{\D(\nu_6),l_6}$};
	\node at (0.6,-0.4) [right] {$\f_{\D(\nu_5),l_5}$};
	\draw [spinning]      (prop1)-- (opO1);
	\draw [spinning bulk] (prop1)-- (vert1);
	\draw [spinning]      (prop2)-- (opO2);
	\draw [spinning bulk] (prop2)-- (vert2);
	\draw [spinning]      (prop3)-- (opO3);
	\draw [spinning bulk] (prop3)-- (vert3);
	\draw [spinning bulk] (shad)--  (vert1);
	\draw [spinning bulk] (shadL)-- (vert1);
	\draw [spinning bulk] (shad)--  (vert2);
	\draw [spinning bulk] (shadR)-- (vert2);
	\draw [spinning bulk] (shadL)-- (vert3);
	\draw [spinning bulk] (shadR)-- (vert3);
\end{tikzpicture}}.
\label{eq:3pt-1loop-2}
\ee
Following the same steps as for the two-point diagram and using \eqref{eq:differentialbasistrickshorthand} 
three times, the diagram in this equation can be brought into the form
\be
\sum_{\ldots} K^3 \{ \ldots \}^{12}
\diagramEnvelope{\begin{tikzpicture}[anchor=base,baseline]
	\node (vert1) at (-1,0.575) [twopt] {};
	\node (vert2) at ( 1,0.575) [twopt] {};
	\node (vert3) at ( 0, -1.15) [twopt] {};
	\node (prop1) at ( -1.74,1) [twopt] {};
	\node (prop2) at ( 1.74,1) [twopt] {};
	\node (prop3) at ( 0,-2) [twopt] {};
	\node (wsi1) at ( -2.48,1.425) [threept] {};
	\node (wsi2) at ( 2.48,1.425) [threept] {};
	\node (wsi3) at ( 0,-2.85) [threept] {};
	\node (wso1) at ( -3.22,1.85) [threept] {};
	\node (wso2) at ( 3.22,1.85) [threept] {};
	\node (wso3) at ( 0,-3.7) [threept] {};
	\node (opO1) at (-3.96,2.275) [] {};
	\node (opO2) at ( 3.96,2.275) [] {};
	\node (opO3) at ( 0,-4.55) [] {};
	\node at (-4.1,2.4) {$\cO_1$};
	\node at ( 4.1,2.4) {$\cO_2$};
	\node at ( 0,-4.7) {$\cO_3$};
	\node (shad) at (0,0.575) [twopt] {};
	\node (shadL) at (-0.5,-0.3) [twopt] {};
	\node (shadR) at (0.5,-0.3) [twopt] {};
	\draw [spinning]      (prop1)-- (wsi1);
	\draw [spinning bulk] (prop1)-- (vert1);
	\draw [scalar]      (prop2)-- (wsi2);
	\draw [scalar bulk] (prop2)-- (vert2);
	\draw [scalar]      (prop3)-- (wsi3);
	\draw [scalar bulk] (prop3)-- (vert3);
	\draw [scalar]      (wsi1)-- (wso1);
	\draw [spinning]      (wsi2)-- (wso2);
	\draw [spinning]      (wsi3)-- (wso3);
	\draw [spinning]      (wso1)-- (opO1);
	\draw [spinning]      (wso2)-- (opO2);
	\draw [spinning]      (wso3)-- (opO3);
	\draw [scalar bulk] (shad)--  (vert1);
	\draw [scalar bulk] (shadL)-- (vert1);
	\draw [scalar bulk] (shad)--  (vert2);
	\draw [scalar bulk] (shadR)-- (vert2);
	\draw [scalar bulk] (shadL)-- (vert3);
	\draw [scalar bulk] (shadR)-- (vert3);
	\draw [finite with arrow] (wso1) -- (wso2);
	\draw [finite with arrow] (wsi2) -- (wso3);
	\draw [finite with arrow] (wsi3) -- (wsi1);
\end{tikzpicture}}\,.
\label{eq:3pt-1loop-diagram-2}
\ee
This time we cannot invoke Schur's lemma to claim that the last remaining spinning external leg is scalar. For $n$-point one loop Witten diagrams this technique will always reduce all but one external leg to  scalars.

\subsubsection{Expressions without weight shifting operators}

For generic two- and three-point one-loop Witten diagrams it is also possible to reduce the internal lines to scalars and arrive at expressions without weight shifting operators. 
For the two-point function we would like to simplify again \eqref{eq:2pt-1loop-3}.
Assuming $\cO_1$ is a traceless symmetric tensor we can use \eqref{eq:differentialbasistrickshorthand}
such that weight shifting operators act on both internal lines
\be
\eqref{eq:2pt-1loop-diagram-1}
=
\sum_{\cW,m,n} K_{c,\cW,m,n}^{\f_{\De(\nu),l} \f_{\D(\bar{\nu}),\bar{l}} \f_1} 
\diagramEnvelope{\begin{tikzpicture}[anchor=base,baseline]
	\node (vertL) at (-1.4,0) [twopt] {};
	\node (vertR) at (1.4,-0.12) [threept] {$d$};
	\node (propL) at (-2.4,0) [twopt] {};
	\node (propR) at (2.4,0) [twopt] {};
	\node (opO1) at (-3.4,0) [left] {};
	\node (opO2) at (3.4,0) [right] {};
	\node at (-3.4,0) [below] {$\cO_1$};
	\node at (3.4,0) [below] {$\cO_2$};
	\node (shad) at (0,0.9) [twopt] {};
	\node (splitL) at (-1,0.82) [threept] {$m$};
	\node (splitR) at (1,0.9) [inner sep=0pt] {};
	\node at (0,1) [above] {$\f_{\D(\nu),l}$};
	\node (shad2) at (0,-0.92) [twopt] {};
	\node (splitL2) at (-1,-0.98) [threept] {$n$};
	\node (splitR2) at (1,-0.9) [inner sep=0pt] {};
	\node at (0,-1) [below] {$\f_{\D(\bar{\nu}),\bar{l}}$};
	\node at (0,0) [left] {$\cW$};
	\node at (-1.2,0.5) [left] {$\f'$};
	\node at (-1.2,-0.5) [left] {$\f''$};
	\draw [spinning] (propL)-- (opO1);
	\draw [spinning bulk] (propL)-- (vertL);
	\draw [scalar bulk] (splitL)-- (vertL);
	\draw [scalar bulk] (splitL2)-- (vertL);
	\draw [spinning] (propR)-- (opO2);
	\draw [spinning bulk] (propR)-- (vertR);
	\draw [spinning bulk] (splitR)-- (vertR);
	\draw [spinning bulk] (splitR2)-- (vertR);
	\draw [spinning bulk] (shad2)-- (splitL2);
	\draw [spinning bulk] (shad)-- (splitR);
	\draw [spinning bulk] (shad)-- (splitL);
	\draw [spinning bulk] (shad2)-- (splitR2);
	\draw [finite with arrow] (splitL) to[out=300,in=60] (splitL2);
\end{tikzpicture}}.
\label{eq:2pt-1loop-nowso-1}
\ee
After moving both weight shifting operators to the right past the harmonic functions, and then one of them past the vertex to remove the bubble, the internal lines are reduced to scalars and there are no weight shifting operators left in the diagram
\be
\eqref{eq:2pt-1loop-nowso-1}=
\sum_{\cW,m,n} K_{c,\cW,m,n}^{\f_{\De(\nu),l} \f_{\D(\bar{\nu}),\bar{l}} \f_1} 
\{\ldots\}^3 (\ldots)
\diagramEnvelope{\begin{tikzpicture}[anchor=base,baseline]
	\node (vertL) at (-1.4,0) [twopt] {};
	\node (vertR) at (1.4,0) [twopt] {};
	\node (propL) at (-2.4,0) [twopt] {};
	\node (propR) at (2.4,0) [twopt] {};
	\node (opO1) at (-3.4,0) [left] {};
	\node (opO2) at (3.4,0) [right] {};
	\node at (-3.4,0) [below] {$\cO_1$};
	\node at (3.4,0) [below] {$\cO_2$};
	\node (shad) at (0,0.8) [twopt] {};
	\node (splitL) at (-1,0.8) [inner sep=0pt] {};
	\node (splitR) at (1,0.8) [inner sep=0pt] {};
	\node at (0,0.85) [above] {$\f'$};
	\node (shad2) at (0,-0.8) [twopt] {};
	\node (splitL2) at (-1,-0.8) [inner sep=0pt] {};
	\node (splitR2) at (1,-0.8) [inner sep=0pt] {};
	\node at (0,-0.85) [below] {$\f''$};
	\draw [spinning] (propL)-- (opO1);
	\draw [spinning] (propR)-- (opO2);
	\draw [spinning bulk] (propL)-- (vertL);
	\draw [spinning bulk] (propR)-- (vertR);
	\draw [scalar bulk] (splitL)-- (vertL);
	\draw [scalar bulk] (splitR)-- (vertR);
	\draw [scalar bulk] (splitL2)-- (vertL);
	\draw [scalar bulk] (splitR2)-- (vertR);
	\draw [scalar bulk] (shad)-- (splitL);
	\draw [scalar bulk] (shad)-- (splitR);
	\draw [scalar bulk] (shad2)-- (splitL2);
	\draw [scalar bulk] (shad2)-- (splitR2);
\end{tikzpicture}}\,.
\ee
A similar simplification exists for the three-point one-loop diagram. One way to see this is to start with \eqref{eq:3pt-1loop-diagram-2} and move two contracted weight shifting operators into the loop using the special crossing equation \eqref{eq:6jdefinition_bulk_single_ts}. After this operation all legs in the loop are still scalars and the bubble inside the loop can be removed because in a triangle the operators are only separated by one vertex. This is then repeated for the remaining weight shifting operators. For higher point diagrams like the four point box diagram this trick does not reduce all representations in the loop to scalars, since weight shifting operators 
in the loop may be separated by more than one vertex. In such a case there is no obvious way to remove the bubble without introducing a spinning leg.

\subsection{From loops to infinite dimensional $6j$ symbols}
\label{sec:loops-to-6j}
The computation of loop diagrams such as the ones appearing in \eqref{eq:2pt-1loop-diagram-5} or \eqref{eq:3pt-1loop-diagram-2} is in principle equivalent to the computation of $6j$ symbols for infinite dimensional representations\footnote{We thank Petr Kravchuk for pointing this out.}, which are defined by the crossing equation for conformal blocks
\be
\diagramEnvelope{\begin{tikzpicture}[anchor=base,baseline]
	\node (vertL) at (0,0-0.09) [threept] {$a$};
	\node (shad) at (1,0) [cross] {};
	\node (vertR) at (2,0-0.1) [threept] {$b$};
	\node (opO1) at (-0.5,-1) [below] {$\cO_1$};
	\node (opO2) at (-0.5,1) [above] {$\cO_2$};
	\node (opO3) at (2.5,1) [above] {$\cO_3$};
	\node (opO4) at (2.5,-1) [below] {$\cO_4$};	
	\node at (1,0.1) [above] {$\cO_3'$};	
	\draw [spinning] (vertL)-- (opO1);
	\draw [spinning] (vertL)-- (opO2);
	\draw [spinning] (vertL)-- (shad);
	\draw [spinning] (vertR)-- (shad);
	\draw [spinning] (vertR)-- (opO3);
	\draw [spinning] (vertR)-- (opO4);
\end{tikzpicture}}
	\quad=\quad
	\sum_{\cO_1',c,d}
	\left\{
		\begin{matrix}
		\cO_1 & \cO_2 & \cO_1' \\
		\cO_3 & \cO_4 & \cO_3'
		\end{matrix}
	\right\}^{ab}_{cd}
\diagramEnvelope{\begin{tikzpicture}[anchor=base,baseline]
	\node (vertU) at (0,0.9-0.1) [threept] {$c$};
	\node (vertD) at (0,-0.9-0.1) [threept] {$d$};
	\node (shad) at (0,0) [cross] {};
	\node (opO1) at (-1,-1.5) [below] {$\cO_1$};
	\node (opO2) at (-1,1.5) [above] {$\cO_2$};
	\node (opO3) at (1,1.5) [above] {$\cO_3$};
	\node (opO4) at (1,-1.5) [below] {$\cO_4$};	
	\node at (0.1,0) [right] {$\cO_1'$};	
	\draw [spinning] (vertD)-- (opO1);
	\draw [spinning] (vertU)-- (opO2);
	\draw [spinning] (vertU)-- (shad);
	\draw [spinning] (vertD)-- (shad);
	\draw [spinning] (vertU)-- (opO3);
	\draw [spinning] (vertD)-- (opO4);
\end{tikzpicture}}\,.
\label{eq:infinite6jdefinition}
\ee
We will demonstrate this by showing how the diagram \eqref{eq:2pt-1loop-diagram-1}, which appears in a generic two-point one-loop Witten diagram, 
can be written as a tree diagram times an infinite dimensional $6j$ symbol.
One can use the split representation \eqref{eq:spinning_omega_to_shadow} for the harmonic functions and do the AdS integrals (assuming for simplicity the simple relation between three point structures \eqref{eq:ads_integral_general_diagonal}) to write the loop diagram as a purely CFT diagram,
\be
\frac{\eqref{eq:2pt-1loop-diagram-1}}{
b(\cO_1 ,\cO_{\De(\nu),l}, \cO_{\De(\bar{\nu}),\bar{l}})\,
b(\cO_1 ,\cO_{\De(\nu),l} ,\cO_{\De(\bar{\nu}),\bar{l}})
\cT_{\De(\nu),l}
\cT_{\De(\bar{\nu}),\bar{l}}} =
\diagramEnvelope{\begin{tikzpicture}[anchor=base,baseline]
	\node (vertL) at (-1.4,-0.08) [threept] {$c$};
	\node (vertR) at (1.4,-0.12) [threept] {$d$};
	\node (opO1) at (-2.4,0) [left] {};
	\node (opO2) at (2.4,0) [right] {};
	\node at (-2.4,0) [below] {$\cO_1$};
	\node at (2.4,0) [below] {$\cO_2$};
	\node (shad) at (0,0.8) [cross] {};
	\node (splitL) at (-1,0.8) [inner sep=0pt] {};
	\node (splitR) at (1,0.8) [inner sep=0pt] {};
	\node at (0,0.9) [above] {$\cO_{\D(\nu),l}$};
	\node (shad2) at (0,-0.8) [cross] {};
	\node (splitL2) at (-1,-0.8) [inner sep=0pt] {};
	\node (splitR2) at (1,-0.8) [inner sep=0pt] {};
	\node at (0,-0.9) [below] {$\cO_{\D(\bar{\nu}),\bar{l}}$};
	\draw [spinning] (vertL)-- (opO1);
	\draw [spinning] (vertL)-- (splitL);
	\draw [spinning] (vertL)-- (splitL2);
	\draw [spinning] (vertR)-- (opO2);
	\draw [spinning] (vertR)-- (splitR);
	\draw [spinning] (vertR)-- (splitR2);
	\draw [spinning] (splitL2)-- (shad2);
	\draw [spinning] (splitR)-- (shad);
	\draw [spinning] (splitL)-- (shad);
	\draw [spinning] (splitR2)-- (shad2);
\end{tikzpicture}}.
\label{eq:2pt-1loop-diagram-cft-1}
\ee
Then using \eqref{eq:infinite6jdefinition} one finds that
\be
\eqref{eq:2pt-1loop-diagram-cft-1}
\quad=\quad
	\sum_{\cO_3,a,b}
	\left\{
		\begin{matrix}
		\cO_1 & \cO_{\D(\nu),l} & \cO_3 \\
		\cO_{\D(\nu),l} & \cO_2 & \cO_{\D(\bar{\nu}),\bar{l}}
		\end{matrix}
	\right\}^{cd}_{ab}
\diagramEnvelope{\begin{tikzpicture}[anchor=base,baseline]
	\node (vertU) at (0,0.9-0.1) [threept] {$a$};
	\node (vertD) at (0,-0.9-0.11) [threept] {$b$};
	\node (shad) at (0,0) [cross] {};
	\node (opO1) at (-1,-0.9) [left] {$\cO_1$};
	\node (opO4) at (1,-0.9) [right] {$\cO_2$};	
	\node (shad2) at (0,2) [cross] {};
	\node at (0,2.1) [above] {$\cO_{\D(\nu),l}$};
	\node at (0.1,0) [right] {$\cO_3$};	
	\draw [spinning] (vertD)-- (opO1);
	\draw [spinning] (vertU)-- (shad);
	\draw [spinning] (vertD)-- (shad);
	\draw [spinning] (vertD)-- (opO4);
	\draw [spinning] (vertU) to[out=10,in=0,looseness=1.6] (shad2);
	\draw [spinning] (vertU) to[out=170,in=180,looseness=1.6] (shad2);
\end{tikzpicture}}\,.
\label{eq:2pt-1loop-diagram-cft-2}
\ee
The tadpole diagram (a three-point function with two legs glued together) only gives a non-vanishing result if the third leg is the unit operator\footnote{In this equation there are subtleties regarding divergences. It would be interesting to understand the relation to divergences in loop Witten diagrams in detail.}
\be
\diagramEnvelope{\begin{tikzpicture}[anchor=base,baseline]
	\node (vertU) at (0,0.1-0.1) [threept] {$a$};
	\node (shad) at (0,-0.8) [below] {$\cO$};
	\node (shad2) at (0,1.2) [cross] {};
	\node at (0,1.3) [above] {$\cO'$};
	\draw [spinning] (vertU)-- (shad);
	\draw [spinning] (vertU) to[out=10,in=0,looseness=1.6] (shad2);
	\draw [spinning] (vertU) to[out=170,in=180,looseness=1.6] (shad2);
\end{tikzpicture}}
\quad\propto\quad \de_{\cO \mathbf{1}}
\,.
\ee
Hence we find that the one-loop diagram reduces to the tree level two-point function times a $6j$ symbol
\be
\eqref{eq:2pt-1loop-diagram-cft-2} \propto
	\left\{
		\begin{matrix}
		\cO_1 & \cO_{\D(\nu),l} & \mathbf{1} \\
		\cO_{\D(\nu),l} & \cO_2 & \cO_{\D(\bar{\nu}),\bar{l}}
		\end{matrix}
	\right\}^{cd}_{\uniq \uniq}
\diagramEnvelope{\begin{tikzpicture}[anchor=base,baseline]
	\node (vertD) at (0,0) [twopt] {};
	\node (opO1) at (-1,0) [left] {$\cO_1$};
	\node (opO2) at (1,0) [right] {$\cO_2$};	
	\draw [spinning] (vertD)-- (opO1);
	\draw [spinning] (vertD)-- (opO2);
\end{tikzpicture}}\,.
\ee
This technique can generally be used to reduce any loop diagram with cubic couplings to a tree diagram times $6j$ symbols, where every use of \eqref{eq:infinite6jdefinition} reduces the number of vertices in a given loop until one reaches the two-point loop discussed above.
See chapter 5 of \cite{Cvitanovic:2008zz} for a more detailed explanation.

Note however that $6j$ symbols of infinite dimensional representations are significantly more complicated than the ones involving finite dimensional representations that were used and computed in this paper.
For instance the $6j$ symbol of six scalar operators was computed in terms of a four-fold Mellin-Barnes integral in \cite{Krasnov:2005fu}.

\section{Conclusions}
\label{sec:conclusion}

In this paper we  extended the construction of CFT weight shifting operators given in \cite{Karateev:2017jgd} to the case of AdS fields. 
Our expectation is that these new operators are a new useful technical tool to study spinning fields in AdS. We showed that this is indeed true for 
Witten diagram computations. There are a number of immediate questions one could try to answer.

An obvious extension of our work would be to apply it to the spinorial formalisms for $\text{AdS}_4/\text{CFT}_3$ and $\text{AdS}_5/\text{CFT}_4$, 
which provide a uniform description of all irreducible representations of the conformal group (including fermions) in these dimensions.
The CFT weight shifting operators in these formalisms have been given in \cite{Karateev:2017jgd} and the constraints should map to the fermionic relativistic wave equations on AdS just as in the bosonic formalism discussed here. An embedding formalism for Dirac spinors in AdS was also recently developed in \cite{Nishida:2018opl}.

The expressions we wrote for the Witten diagrams are given in terms of the spectral representations of the bulk-to-bulk propagators. The lower spin components of this representation are not universal. They are off-shell terms that give rise to contact interactions. We considered the case of massive fields, whose spectral representation
(\ref{eq:SlipRepStart}) is written in terms of derivatives of harmonic functions. To convert these
harmonic functions into scalar harmonic functions we wrote such derivatives in terms of weight shifting operators and then used 
crossing relations to bring these operators to the boundary, acting on the external points as CFT weight shifting operators. 
It would be interesting to see if
a similar computation could be carried out for massless spinning fields by using the corresponding spectral representation \cite{Bekaert:2014cea,Sleight:2017cax}.
The case of loop diagrams involving spinning gauge fields also requires to add the ghost contributions. It would also be interesting to consider such contributions using the techniques of this paper.

A limitation of our work is that we did not consider quartic, or even higher, interactions in the bulk. In order to move around the weight shifting operators we would need crossing equations for these higher point couplings. 
Such crossing equations could be obtained by first studying crossing equations involving weight shifting operators for conformal structures acting on contact Witten diagrams. The corresponding crossing equations for the local couplings could then be deduced straightforwardly using the ideas of this paper. 
It would be interesting to look at this question in detail.

We have not explored the role  the new AdS weight shifting operators might have in higher spin theories. These theories are endowed with an infinite
dimensional symmetry, that relates higher spin representations and constrains interactions, thus it would be interesting to understand the role 
that the new AdS weight shifting operators have in these theories.

Another interesting problem would be to consider the action of weight shifting operators on the Mellin transform of Witten diagrams. Consider the  example of a four-point correlation function computed from a Witten diagram. For any external spins and for a single tree level exchange, we saw how to reduce those correlation functions to a sum  of all-scalar Witten diagrams. Thus we could use this result to define the Mellin transform of Witten diagrams with external spins in terms of the Mellin transform of the scalar Witten diagram. 

\section*{Acknowledgments}
The authors  benefited from discussions with Agnese Bissi and Denis Karateev.
We thank Petr Kravchuk, Jo\~ao Penedones, Charlotte Sleight, David Simmons-Duffin and Massimo Taronna for comments on the draft.
This research received funding from the grant CERN/FIS-PAR/0019/2017, the 
Simons Foundation grant 488637 (Simons collaboration on the Non-perturbative bootstrap)
and the Knut and Alice Wallenberg Foundation grant KAW 2016.0129. 
Centro de F\'\i sica do Porto is partially funded by the Foundation for Science and Technology of Portugal (FCT).

\appendix

\section{Weight shifting operators for  mixed-symmetry tensor projectors}
\label{sec:projectors}

In this appendix we spell out the weight shifting operators that relate projectors to traceless mixed-symmetry tensors, more specifically the 
projectors to irreps of $SO(d)$ labeled by Young diagrams with two rows of lengths $(l_1,l_2)$.
These projectors can be encoded in a polynomial depending on four vectors
\bea
\pi_{(l_1,l_2)}(z_1,z_2,\bar z_1, \bar z_2)
=z_1^{a_1}  \ldots  z_1^{a_{l_1}}
z_2^{b_{1}} \ldots z_2^{b_{l_2}}
\pi_{(l_1,l_2)} {\,\!}_{a_1 \ldots a_{l_1} b_1 \ldots b_{l_2} ,c_1 \ldots c_{l_1} d_1 \ldots d_{l_2}}
\barz_1^{c_1}  \ldots  \barz_1^{c_{l_1}}
\barz_2^{d_{1}} \ldots \barz_2^{d_{l_2}}\,.
\eea{eq:gen_poly_def}
We showed in \cite{Costa:2016hju} that, up to normalization, these functions can be constructed by imposing
the conditions of mixed symmetry and tracelessness. The normalization is fixed by requiring idempotence.
The mixed symmetry or Young symmetrization of a tensor in the symmetric representation of the irrep $(l_1,l_2)$ amounts to the condition
\beq
\left( z_1 \cdot \frac{\partial}{\partial z_2}\right)  \pi_{(l_1,l_2)}(z_1,z_2,\bar z_1, \bar z_2) = 0\,.
\label{eq:Young_symmetry}
\eeq
The tracelessness conditions can be written as
\bea
\partial_{z_1}^2  \pi_{(l_1,l_2)}(z_1,z_2,\bar z_1, \bar z_2) =
\partial_{z_1} \cdot \partial_{z_2}  \pi_{(l_1,l_2)}(z_1,z_2,\bar z_1, \bar z_2) =
\partial_{z_2}^2   \pi_{(l_1,l_2)}(z_1,z_2,\bar z_1, \bar z_2) = 0 \,.
\eea{eq:laplac}
Analogous conditions hold for the variables $\barz_1, \barz_2$.
These four conditions are essentially the same as the conditions (\ref{eq:dZ1squaredOnOmega}-\ref{eq:dX1dZ1OnOmega}) for AdS harmonic functions. As a consequence, the projectors are related by the same weight shifting operators after some relabelling.

\subsection{Weight shifting operators for projectors}

The relabelling we have to perform on the operators \eqref{eq:ws_operators_inverted}
to obtain operators relating projectors is
\beq
\D \to  -l_1\,, \qquad J \to l_2\,,\qquad
\delta_\D \to - \delta_1\,, \qquad
\delta_J \to \delta_2\,, \qquad 
d \to d-2\,.
\eeq
In this way one finds the operators\footnote{These operators appeared previously in the higher spin literature as cell operators (see for instance \cite{Rahman:2015pzl}).}
\bea
\calL_{0-}^a(z_1,z_2) ={}& \partial_{z_2}^a \,,
\\
\calL_{-0}^a(z_1,z_2) ={}& \partial_{z_1}^b \Big( (l_1 - l_2 + 1) \delta^a_b + z_{2 b} \partial_{z_2}^a \Big)\,,
\\
\calL_{+0}^a(z_1,z_2) ={}& z_1^c\Big( (d-2+2 l_1)\delta^b_c -z_{1 c} \partial_{z_1}^b \Big)
\Big( (-l_2 -d +3 - l_1) \delta_b^a + z_{2 b} \partial_{z_2}^a \Big) \,,
\\
\calL_{0+}^a(z_1,z_2) ={}& z_2^d \Big( c_3' \de_d^c - \partial_{z_1 d} z_1^c  \Big)
\Big( c_2' \de_c^b + z_{1 c} \partial_{z_1}^b  \Big)
\Big( c_1' \de_b^a + z_{2 b} \partial_{z_2}^a  \Big)\,,
\eea{eq:L_operators_projectors}
where
\bea
c_1' = 4 - d -2l_2\,, \qquad
c_2' = 3 - d + l_1 +l_2\,, \qquad
c_3' = l_1-l_2+1 \,. 
\eea{eq:c_i_projectors}
These operators can be used to increase or decrease $l_1$ or $l_2$ of a projector
\bea
\pi_{(l_1+\delta_1,l_2+\delta_2)} (z_1,z_2,\bar z_1, \bar z_2)
&\propto
\delta_{ab}
\calL_{\delta_1 \delta_2}^a(z_1,z_2)
\calL_{\delta_1 \delta_2}^b(\barz_1,\barz_2)
\pi_{(l_1,l_2)} (z_1,z_2,\bar z_1, \bar z_2)
\,.
\eea{eq:L_on_pi}
This was checked for the projectors up to $l_2=4$, which we calculated in \cite{Costa:2016hju} using a different method.
The action on projectors  also satisfies a simple crossing equation,
\beq
\calL_{\delta_1 \delta_2}^a(z_1,z_2)
\pi_{(l_1,l_2)} (z_1,z_2,\bar z_1, \bar z_2)
\propto
\calL_{-\delta_1 -\delta_2}^a(\barz_1,\barz_2)
\pi_{(l_1+\delta_1,l_2+\delta_2)} (z_1,z_2,\bar z_1, \bar z_2)\,,
\eeq
where on the right hand side we replace $l_i \to l_i + \delta_i$ according to the homogeneities of the function
the operator acts on.
Analogous weight shifting operators relating Young diagrams with three rows can be obtained from \eqref{eq:ms_ws_L_operators} by relabelling
\beq
\D \to  -l_1\,, \quad J_1 \to l_2\,,\quad J_2 \to l_3\,,\quad
\delta_\D \to - \delta_1\,, \quad
\delta_{J_1} \to \delta_2\,, \quad 
\delta_{J_2} \to \delta_3\,, \quad 
d \to d-2\,.
\eeq

\subsection{Normalization of AdS harmonic functions}
Let us also compare the normalization of the spinning harmonic functions with that of the projectors to mixed-symmetry tensors.
The normalization of the projector $\pi_{(l_1,l_2)}(z_1,z_2,\bar z_1, \bar z_2)$ was derived in \cite{Costa:2016hju}
by looking at the term $(z_1 \cdot \bar z_1)^{l_1} (z_2 \cdot \bar z_2)^{l_2}$, that is
\beq
\pi_{(l_1,l_2)}(z_1,z_2,\bar z_1, \bar z_2) =
\frac{l_1-l_2+1}{l_1+1}(z_1 \cdot \bar z_1)^{l_1} (z_2 \cdot \bar z_2)^{l_2} + \text{other terms}\,.
\eeq
In order to compare normalizations, we would like to extract the analogous term from the harmonic function on AdS,
which can be written in terms of the bulk-to-bulk propagator \cite{Costa:2014kfa}
\begin{align}
&\Omega_{\D,J}(X_1,X_2;W_1,W_2)=\frac{\D-\frac{d}{2}}{2\pi} \Big(\Pi_{\D,J}(X_1,X_2;W_1,W_2)-{\tilde \Pi}_{\D,J}(X_1,X_2;W_1,W_2) \Big)\,.
\label{integrationboundary}
\end{align}
We have to extract the term containing $(X_1 \cdot X_2)^{-\D} (W_1 \cdot W_2)^{J}$ out of $\Pi_{\D,J}$.
This can be achieved by setting $X_i^2=-1$ and then scaling the $X_i$ to infinity,
which is the same prescription that was used to compute the bulk-to-boundary propagator \eqref{eq:eq propagador boundary}. So the normalization of the term in the harmonic function that we want to use to compare to the projector is just the normalization constant of the bulk-to-boundary propagator
\beq
\O_{\D,J}(X_1,X_2;W_1,W_2) =
 \frac{\D-\tfrac{d}{2}}{2\pi}  \,{\cal C}_{\Delta,J} (-2)^{-\D} (X_1 \cdot X_2)^{-\D} (W_1 \cdot W_2)^{J}  + \text{other terms}\,.
\eeq
Hence the relative normalization does not depend on $J$ and is the same as for the scalar harmonic function
\beq
\frac{\O_{\D,J}(X_1,X_2;W_1,W_2)}{\pi_{(-\D,J)}^{d+2} (X_1,X_2;W_1,W_2)} = 
\frac{(-2)^{-\D-3}(d-2\D)\Gamma(\D)}{\pi^{\frac{d+2}{2}} \Gamma(1-\frac{d}{2}+\D)}\,.
\eeq

\section{Explicit computations for section 4}
\label{app:crossing}

In this appendix we perform explicit computations with the weight shifting operators in the vector representation (\ref{eq:ws_operators}, \ref{eq:ws_operators_inverted}) that were introduced in general terms in section \ref{sec:crossing}.

\subsection{Crossing for two-point functions}
\label{app:2pt-6j-symbols}

In the case of CFT weight shifting operators in the vector representation \eqref{eq:twoptcrossing} reads \cite{Karateev:2017jgd} 
\begin{align}
{}&\calD_{\delta_\D, \delta_J}^a(P_2,Z_2)
\big\< \cO_{\D,J} (P_1, Z_1) \cO_{\D,J} (P_2, Z_2) \big\>=
\label{eq:crossing_CFT}
\\
&
\left\{ \begin{matrix}
\cO_{\D,J}\\ \cO_{\D+\delta_\D,J+\delta_J}
\end{matrix} \right\}^{(\delta_\D, \delta_J)}_{(-\delta_\D, -\delta_J)}
\calD_{-\delta_\D, -\delta_J}^a(P_1,Z_1)
\big\< \cO_{\D+\delta_\D,J+\delta_J} (P_1, Z_1) \cO_{\D+\delta_\D,J+\delta_J} (P_2, Z_2) \big\> \,,
\nonumber
\end{align}
with $6j$ symbols
\begin{align}
\left\{ \begin{matrix}
\cO_{\D,J}\\ \cO_{\D,J+1}
\end{matrix} \right\}^{(0+)}_{(0-)}
&=
\left(
\left\{ \begin{matrix}
\cO_{\D,J+1}\\ \cO_{\D,J}
\end{matrix} \right\}^{(0-)}_{(0+)}
\right)^{-1}
= \frac{J+\D}{(1 + J) (d + 2 J - 2) (\D + 1 - d - J)}\,,
\label{eq:6j_CFT}
\\
\left\{ \begin{matrix}
\cO_{\D,J}\\ \cO_{\D+1,J}
\end{matrix} \right\}^{(+0)}_{(-0)}
&=
\left(
\left\{ \begin{matrix}
\cO_{\D+1,J}\\ \cO_{\D,J} 
\end{matrix} \right\}^{(-0)}_{(+0)}
\right)^{-1}
= 2 (d +  J - \D -2) (\D-1) (J + \D) \big(d - 2 (1 + \D)\big) \,.
\nonumber
\end{align}
Equation \eqref{eq:twoptcrossing_bulk} for the AdS weight shifting operators in the vector representation becomes
\bea
{}&\calL_{\delta_\D, \delta_J}^a(X_2,W_2)
\,\Omega_{\D,J} (X_1, X_2; W_1, W_2)=
\\
&
\left\{ \begin{matrix}
\f_{\D,J}\\ \f_{\D+\delta_\D,J+\delta_J}
\end{matrix} \right\}^{(\delta_\D, \delta_J)}_{(-\delta_\D, -\delta_J)}
\calL_{-\delta_\D, -\delta_J}^a(X_1,W_1)
\,\Omega_{\D+\delta_\D,J+\delta_J} (X_1,  X_2;W_1, W_2)\,,
\eea{eq:crossing_bulk}
with the bulk $6j$ symbols
\bea
\left\{ \begin{matrix}
\f_{\D,J}\\ \f_{\D,J+1} 
\end{matrix} \right\}^{(0+)}_{(0-)}
&=
\left(
\left\{ \begin{matrix}
\f_{\D,J+1}\\ \f_{\D,J} 
\end{matrix} \right\}^{(0-)}_{(0+)}
\right)^{-1}
= \frac{(1-\D-J)(d-2+2J)(d-1-\D+J)}{J+1}\,,
\\
\left\{ \begin{matrix}
\f_{\D,J}\\  \f_{\D+1,J} 
\end{matrix} \right\}^{(+0)}_{(-0)}
&=
\left(
\left\{ \begin{matrix}
\f_{\D+1,J}\\ \f_{\D,J} 
\end{matrix} \right\}^{(-0)}_{(+0)}
\right)^{-1}
= \frac{(1-\D-J)(\frac{d}{2}-\D)}{(d-2-\D+J)(\frac{d}{2}-\D-1)}
\,.
\eea{eq:6j_bulk-bulk}
Finally, the crossing equation for the bulk-to-boundary propagators \eqref{eq:twoptcrossing_bulk_boundary} becomes
\bea
{}&\calL_{\delta_\D, \delta_J}^a(X_2,W_2)
\,\Pi_{\D,J} (P_1, X_2; Z_1, W_2) = \\
&
\left\{ \begin{matrix}
\cO_{\D,J}\\ \f_{\D+\delta_\D,J+\delta_J}
\end{matrix} \right\}^{(\delta_\D, \delta_J)}_{(-\delta_\D, -\delta_J)}
\calD_{-\delta_\D, -\delta_J}^a(P_1,Z_1)
\,\Pi_{\D+\delta_\D,J+\delta_J} (P_1, X_2; Z_1, W_2)\,,
\eea{eq:crossing_bulk_boundary}
with the bulk-boundary $6j$ symbols
\bea
\left\{ \begin{matrix}
\cO_{\D,J}\\ \f_{\D,J+1}
\end{matrix} \right\}^{(0+)}_{(0-)}
&=
\left(
\left\{ \begin{matrix}
\cO_{\D,J+1}\\ \f_{\D,J}
\end{matrix} \right\}^{(0-)}_{(0+)}
\right)^{-1}
= \frac{\D+J-1}{J+1}\,,
\\
\left\{ \begin{matrix}
\cO_{\D,J}\\ \f_{\D+1,J}
\end{matrix} \right\}^{(+0)}_{(-0)}
&=
\left(
\left\{ \begin{matrix}
\cO_{\D+1,J}\\ \f_{\D,J}
\end{matrix} \right\}^{(-0)}_{(+0)}
\right)^{-1}
= (\D+J-1)(d-2\D-2)
\,.
\eea{eq:6j_bulk-boundary}

Weight shifts by higher integers can be realized using multiple copies of the same weight shifting operators. To commute $n$ copies of the same weight shifting operator past a two-point function one has to use the crossing equations \eqref{eq:crossing_CFT}, \eqref{eq:crossing_bulk} or \eqref{eq:crossing_bulk_boundary} $n$ times, so the $6j$ symbols for this operation are products of the ones stated above, for example
\bea
\left\{ \begin{matrix}
\cO_{\D,J}\\ \cO_{\D,J+n}
\end{matrix} \right\}^{(0,n)}_{(0,-n)}
&=
\left(
\left\{ \begin{matrix}
\cO_{\D,J+n}\\ \cO_{\D,J}
\end{matrix} \right\}^{(0,-n)}_{(0,n)}
\right)^{-1}
= \prod_{k=1}^{n} \left\{ \begin{matrix}
\cO_{\D,J+k-1}\\ \cO_{\D,J+k}
\end{matrix} \right\}^{(0+)}_{(0-)} \,,
\\
\left\{ \begin{matrix}
\cO_{\D,J}\\ \cO_{\D+n,J}
\end{matrix} \right\}^{(n,0)}_{(-n,0)}
&=
\left(
\left\{ \begin{matrix}
\cO_{\D+n,J}\\ \cO_{\D,J} 
\end{matrix} \right\}^{(-n,0)}_{(n,0)}
\right)^{-1}
= \prod_{k=1}^{n}
\left\{ \begin{matrix}
\cO_{\D+k-1,J}\\ \cO_{\D+k,J}
\end{matrix} \right\}^{(+0)}_{(-0)} \,,
\eea{eq:6j_CFT_n}
and similar for the generalizations of \eqref{eq:6j_bulk-bulk} and \eqref{eq:6j_bulk-boundary}.

\subsection{Bubbles}
\label{app:bubbles}

The bubble coefficients defined by \eqref{eq:bubble} and \eqref{eq:bubble_bulk} 
can be computed for our choice of weight shifting operators by acting with the equations on
a CFT two-point function
\bea
{}&\eta_{ab}
\calD_{-\delta_\D, -\delta_J}^a(P_1, Z_1)
\calD_{\delta_\D, \delta_J}^b(P_1, Z_1)
\big\< \cO_{\D,J} (P_1, Z_1) \cO_{\D,J} (P_2, Z_2) \big\>=
\\
&
\begin{pmatrix}
\cO_{\D,J}\\
\cO_{\D+\delta_\D,J+\delta_J} \calV
\end{pmatrix}^{(-\delta_\D, -\delta_J)(\delta_\D, \delta_J)}
\< \cO_{\D,J} (P_1, Z_1) \cO_{\D,J} (P_2, Z_2) \>
\,,
\eea{eq:D_on_2pt}
or a bulk harmonic function
\bea
{}&\eta_{ab}
\calL_{-\delta_\D, -\delta_J}^a(X_1, W_1)
\calL_{\delta_\D, \delta_J}^b(X_1, W_1)
\,\Omega_{\Delta,J} (X_1,X_2; W_1,W_2) =\\
&
\begin{pmatrix}
\f_{\D,J}\\
\f_{\D+\delta_\D,J+\delta_J} \calV
\end{pmatrix}^{(-\delta_\D, -\delta_J)(\delta_\D, \delta_J)}
\Omega_{\Delta,J} (X_1,X_2; W_1,W_2)
\,.
\eea{eq:L_on_Omega}
One finds that the coefficients are the same in both cases
\begin{align}
\begin{pmatrix}
\cO_{\D,J}\\
\cO_{\D+1,J}\ \calV
\end{pmatrix}^{\!\!(-0)(+0)}
\hspace{-10pt}
=
\begin{pmatrix}
\f_{\D,J}\\
\f_{\D+1,J}\ \calV
\end{pmatrix}^{\!\!(-0)(+0)}
\hspace{-10pt}
&= (1-\D)(\D+J)(d+J-\D-2)(d-2\D-2)\,,
\nonumber
\\
\begin{pmatrix}
\cO_{\D,J}\\
\cO_{\D-1,J}\ \calV
\end{pmatrix}^{\!\!(+0)(-0)}
\hspace{-10pt}
=
\begin{pmatrix}
\f_{\D,J}\\
\f_{\D-1,J}\ \calV
\end{pmatrix}^{\!\!(+0)(-0)}
\hspace{-10pt}
&= (2-\D-J)(\D-J-d)(d-\D-1)(d+2-2\D)\,,
\nonumber
\\
\begin{pmatrix}
\cO_{\D,J}\\
\cO_{\D,J+1}\ \calV
\end{pmatrix}^{\!\!(0-)(0+)}
\hspace{-10pt}
=
\begin{pmatrix}
\f_{\D,J}\\
\f_{\D,J+1}\ \calV
\end{pmatrix}^{\!\!(0-)(0+)}
\hspace{-10pt}
&= (\D+J)(d+2J)(d+J-\D)(2-d-J)\,,
\label{eq:bubble_symbols_bulk}
\\
\begin{pmatrix}
\cO_{\D,J}\\
\cO_{\D,J-1}\ \calV
\end{pmatrix}^{\!\!(0+)(0-)}
\hspace{-10pt}
=
\begin{pmatrix}
\f_{\D,J}\\
\f_{\D,J-1}\ \calV
\end{pmatrix}^{\!\!(0+)(0-)}
\hspace{-10pt}
&=J(d-4+2J)(2-J-\D)(d-2+J-\D)\,.
\nonumber
\end{align}
For larger weight shifts, implemented by $n$ copies of the same weight shifting operators, the coefficients are
\begin{align}
\begin{pmatrix}
\cO_{\D,J}\\
\cO_{\D\pm n,J}\ \cW
\end{pmatrix}^{\!\!(\mp n,0)(\pm n,0)}
\hspace{-10pt}
=
\begin{pmatrix}
\f_{\D,J}\\
\f_{\D\pm n,J}\ \cW
\end{pmatrix}^{\!\!(\mp n,0)(\pm n,0)}
\hspace{-10pt}
&= \prod_{k=1}^{n}
\begin{pmatrix}
\cO_{\D\pm(k-1),J}\\
\cO_{\D\pm k,J}\ \cW
\end{pmatrix}^{\!\!(\mp 0)(\pm 0)}\,,
\nonumber
\\
\begin{pmatrix}
\cO_{\D,J}\\
\cO_{\D,J\pm n}\ \cW
\end{pmatrix}^{\!\!(0,\mp n)(0,\pm n)}
\hspace{-10pt}
=
\begin{pmatrix}
\f_{\D,J}\\
\f_{\D,J\pm n}\ \cW
\end{pmatrix}^{\!\!(0,\mp n)(0,\pm n)}
\hspace{-10pt}
&= \prod_{k=1}^{n}
\begin{pmatrix}
\cO_{\D,J\pm (k-1)}\\
\cO_{\D,J \pm k}\ \cW
\end{pmatrix}^{\!\!(0 \mp)(0 \pm)}\,.
\label{eq:bubble_symbols_bulk_n}
\end{align}

\subsection{Crossing for three-point functions}
\label{app:3pt-6j-symbols}

In this appendix the $6j$ symbols for conformal three-point correlators that appear in the main text are computed. 
We consider cases where the three-point functions have a unique tensor structure. 
For example, if $\cO_1$ and $\cO_2$ in the crossing equation \eqref{eq:6jdefinition} are scalars, it simplifies to
\be
\diagramEnvelope{\begin{tikzpicture}[anchor=base,baseline]
	\node (vertL) at (0,0.07) [twopt] {};
	\node (vertR) at (2,-0.05) [threept] {$m$};
	\node (opO1) at (-0.5,-1) [below] {$\cO_1$};
	\node (opO2) at (-0.5,1) [above] {$\cO_2$};
	\node (opO3) at (2.5,1) [above] {$\cO_3$};
	\node (opW) at (2.5,-1) [below] {$\cW$};	
	\node at (1,0.1) [above] {$\cO_3'$};	
	\draw [scalar] (vertL)-- (opO1);
	\draw [scalar] (vertL)-- (opO2);
	\draw [spinning] (vertL)-- (vertR);
	\draw [spinning] (vertR)-- (opO3);
	\draw [finite with arrow] (vertR)-- (opW);
\end{tikzpicture}}
	\quad=\quad
	\sum_{\cO_1',n}
	\left\{
		\begin{matrix}
		\cO_1 & \cO_2 & \cO_1' \\
		\cO_3 & \cW & \cO_3'
		\end{matrix}
	\right\}^{\uniq m}_{b n}
\diagramEnvelope{\begin{tikzpicture}[anchor=base,baseline]
	\node (vertU) at (0,0.7) [threept] {$b$};
	\node (vertD) at (0,-0.7) [threept] {$n$};
	\node (opO1) at (-1,-1.5) [below] {$\cO_1$};
	\node (opO2) at (-1,1.5) [above] {$\cO_2$};
	\node (opO3) at (1,1.5) [above] {$\cO_3$};
	\node (opW) at (1,-1.5) [below] {$\cW$};	
	\node at (0.1,0) [right] {$\cO_1'$};	
	\draw [scalar] (vertD)-- (opO1);
	\draw [scalar] (vertU)-- (opO2);
	\draw [spinning] (vertU)-- (vertD);
	\draw [spinning] (vertU)-- (opO3);
	\draw [finite with arrow] (vertD)-- (opW);
\end{tikzpicture}}.
\label{eq:6j_single_tensor_structure_O1_scalar}
\ee
The three-point function on the right hand side is unique unless $n=(0-)$ in which case $\cO_1'$ has spin 1. 
Except from this case the three-point structures are given by \eqref{eq:3pt_correlator}.
The $6j$ symbols with $n=p=(\pm 0)$ can be extracted by acting on both sides of the equation with another weight shifting operator $\cL_{\bar{p}}^{a}$ on $\cO_1$ and contracting with the weight shifting operators already present,
\be
\diagramEnvelope{\begin{tikzpicture}[anchor=base,baseline]
	\node (vertL) at (0,0.07) [twopt] {};
	\node (vertR) at (2,-0.05) [threept] {$m$};
	\node (vertD) at (-0.5,-1) [threept] {$\bar{p}$};
	\node (opO1) at (-1.2,-2) [below] {$\cO_1'$};
	\node (opO2) at (-0.5,1) [above] {$\cO_2$};
	\node (opO3) at (2.5,1) [above] {$\cO_3$};
	\node at (1,-1.1) [below] {$\cW$};	
	\node at (1,0.1) [above] {$\cO_3'$};	
	\node at (-0.2,-0.2) [left] {$\cO_1$};
	\draw [scalar] (vertL)-- (vertD);
	\draw [scalar] (vertD)-- (opO1);
	\draw [scalar] (vertL)-- (opO2);
	\draw [spinning] (vertL)-- (vertR);
	\draw [spinning] (vertR)-- (opO3);
	\draw [finite with arrow] (vertR) to[out=-90,in=0]  (vertD);
\end{tikzpicture}}
	\quad=\quad
	\left\{
		\begin{matrix}
		\cO_1 & \cO_2 & \cO_1' \\
		\cO_3 & \cW & \cO_3'
		\end{matrix}
	\right\}^{\uniq m}_{\uniq p}
\diagramEnvelope{\begin{tikzpicture}[anchor=base,baseline]
	\node (vertU) at (0,0.7) [twopt] {};
	\node (vertD) at (0,-0.7) [threept] {$p$};
	\node (vertD2) at (0,-2.1) [threept] {$\bar{p}$};
	\node (opO1) at (-1,-2.9) [below] {$\cO_1'$};
	\node (opO2) at (-1,1.5) [above] {$\cO_2$};
	\node (opO3) at (1,1.5) [above] {$\cO_3$};
	\node at (0.7,-1.3) [right] {$\cW$};	
	\node at (-0.1,0) [left] {$\cO_1'$};	
	\node at (-0.1,-1.3) [left] {$\cO_1$};	
	\draw [scalar] (vertD2)-- (opO1);
	\draw [scalar] (vertU)-- (opO2);
	\draw [scalar] (vertU)-- (vertD);
	\draw [scalar] (vertD)-- (vertD2);
	\draw [spinning] (vertU)-- (opO3);
	\draw [finite with arrow] (vertD) to[out=0,in=0] (vertD2);
\end{tikzpicture}}.
\label{eq:6j_single_tensor_structure_extra_D}
\ee
In this way we can compute the following $6j$ symbols
\begin{align}
\left\{\begin{matrix}
\calO_{\D_1} & \cO_{\D_2} & \cO_{\D_1-1}\\
\cO_{\D,J} & \calV & \cO_{\D-1,J}
\end{matrix} \right\}^{\uniq (+0)}_{\uniq (+0)}
={}& 
\frac{(\Delta -2) (d-\Delta -1) \left(\Delta -\Delta _1+\Delta _2+J-1\right) }{2
   \left(\Delta _1-2\right) \left(\Delta _1-1\right) \left(d-2 \Delta _1\right) \left(d-\Delta _1-1\right)}\nn\\
&\times \left(d-\Delta +\Delta _1-\Delta _2+J-1\right)
\,,\nn\\
\left\{\begin{matrix}
\calO_{\D_1} & \cO_{\D_2} & \cO_{\D_1-1}\\
\cO_{\D,J} & \calV & \cO_{\D+1,J}
\end{matrix} \right\}^{\uniq (-0)}_{\uniq (+0)}
={}& 
\frac{1}{2 \left(\Delta _1-2\right) \left(\Delta _1-1\right) \left(-d+\Delta _1+1\right) \left(2 \Delta _1-d\right)}
 \,,\nn\\
\left\{\begin{matrix}
\calO_{\D_1} & \cO_{\D_2} & \cO_{\D_1-1}\\
\cO_{\D,J} & \calV & \cO_{\D,J-1}
\end{matrix} \right\}^{\uniq (0+)}_{\uniq (+0)}
={}& 
\frac{-\Delta +\Delta _1-\Delta _2-J+1}{2 \left(\Delta _1-2\right) \left(\Delta _1-1\right) \left(-d+\Delta _1+1\right) \left(2
   \Delta _1-d\right)}
\,,\nn\\
\left\{\begin{matrix}
\calO_{\D_1} & \cO_{\D_2} & \cO_{\D_1-1}\\
\cO_{\D,J} & \calV & \cO_{\D,J+1}
\end{matrix} \right\}^{\uniq (0-)}_{\uniq (+0)}
={}& 
\frac{(J+1) (d+J-2) \left(d-\Delta +\Delta _1-\Delta _2+J-1\right)}{2 \left(\Delta _1-2\right) \left(\Delta _1-1\right)
   \left(d-2 \Delta _1\right) \left(d-\Delta _1-1\right)}
 \,,\nn\\
\left\{\begin{matrix}
\calO_{\D_1} & \cO_{\D_2} & \cO_{\D_1+1}\\
\cO_{\D,J} & \calV & \cO_{\D-1,J}
\end{matrix} \right\}^{\uniq (+0)}_{\uniq (-0)}
={}& 
\frac{(\Delta -2) (d-\Delta -1) \left(\Delta +\Delta _1-\Delta _2+J-1\right) }{2 \left(d-2 \Delta
   _1\right) \left(d-\Delta _1-1\right)}
\nn\\
&\times
\left(d-\Delta -\Delta _1-\Delta _2-J+1\right)
   \left(2 d-\Delta -\Delta _1-\Delta _2+J-1\right) 
 \,,\nn\\
&\times
\left(d-\Delta -\Delta _1+\Delta _2+J-1\right)
 \,,\nn\\
\left\{\begin{matrix}
\calO_{\D_1} & \cO_{\D_2} & \cO_{\D_1+1}\\
\cO_{\D,J} & \calV & \cO_{\D+1,J}
\end{matrix} \right\}^{\uniq (-0)}_{\uniq (-0)}
={}& 
-\frac{\left(-\Delta +\Delta _1+\Delta _2+J-1\right) \left(d+\Delta -\Delta _1-\Delta _2+J-1\right)}{2 \left(d-2 \Delta
   _1\right) \left(d-\Delta _1-1\right)}
 \,,\nn
\end{align}
\begin{align}
\left\{\begin{matrix}
\calO_{\D_1} & \cO_{\D_2} & \cO_{\D_1+1}\\
\cO_{\D,J} & \calV & \cO_{\D,J-1}
\end{matrix} \right\}^{\uniq (0+)}_{\uniq (-0)}
={}& 
-\frac{\left(\Delta +\Delta _1-\Delta _2+J-1\right) \left(-\Delta +\Delta _1+\Delta _2+J-1\right) }{2 \left(d-2 \Delta _1\right) \left(d-\Delta _1-1\right)}\nn\\
&\times \left(d-\Delta -\Delta
   _1-\Delta _2-J+1\right)
 \,,\nn\\
\left\{\begin{matrix}
\calO_{\D_1} & \cO_{\D_2} & \cO_{\D_1+1}\\
\cO_{\D,J} & \calV & \cO_{\D,J+1}
\end{matrix} \right\}^{\uniq (0-)}_{\uniq (-0)}
={}& 
-\frac{(J+1) (d+J-2) \left(2 d-\Delta -\Delta _1-\Delta _2+J-1\right) }{2 \left(d-2 \Delta _1\right) \left(d-\Delta _1-1\right)}\label{eq:6j_3pt}\\
&\times \left(d+\Delta -\Delta _1-\Delta _2+J-1\right)
   \left(d-\Delta -\Delta _1+\Delta _2+J-1\right)
 \,.\nn
\end{align}
Another example with unique tensor structures is when $\cO_2$, $\cO_3$ and $\cO_3'$ in \eqref{eq:6jdefinition} are scalars
\be
\diagramEnvelope{\begin{tikzpicture}[anchor=base,baseline]
	\node (vertL) at (0,0.07) [twopt] {};
	\node (vertR) at (2,-0.05) [threept] {$m$};
	\node (opO1) at (-0.5,-1) [below] {$\cO_1$};
	\node (opO2) at (-0.5,1) [above] {$\cO_2$};
	\node (opO3) at (2.5,1) [above] {$\cO_3$};
	\node (opW) at (2.5,-1) [below] {$\cW$};	
	\node at (1,0.1) [above] {$\cO_3'$};	
	\draw [spinning] (vertL)-- (opO1);
	\draw [scalar] (vertL)-- (opO2);
	\draw [scalar] (vertL)-- (vertR);
	\draw [scalar] (vertR)-- (opO3);
	\draw [finite with arrow] (vertR)-- (opW);
\end{tikzpicture}}
	\quad=\quad
	\sum_{\cO_1',n}
	\left\{
		\begin{matrix}
		\cO_1 & \cO_2 & \cO_1' \\
		\cO_3 & \cW & \cO_3'
		\end{matrix}
	\right\}^{\uniq m}_{\uniq n}
\diagramEnvelope{\begin{tikzpicture}[anchor=base,baseline]
	\node (vertU) at (0,0.7) [twopt] {};
	\node (vertD) at (0,-0.7) [threept] {$n$};
	\node (opO1) at (-1,-1.5) [below] {$\cO_1$};
	\node (opO2) at (-1,1.5) [above] {$\cO_2$};
	\node (opO3) at (1,1.5) [above] {$\cO_3$};
	\node (opW) at (1,-1.5) [below] {$\cW$};	
	\node at (0.1,0) [right] {$\cO_1'$};	
	\draw [spinning] (vertD)-- (opO1);
	\draw [scalar] (vertU)-- (opO2);
	\draw [spinning] (vertU)-- (vertD);
	\draw [scalar] (vertU)-- (opO3);
	\draw [finite with arrow] (vertD)-- (opW);
\end{tikzpicture}}.
\label{eq:6j_single_tensor_structure_O3_scalar}
\ee
The possible $6j$ symbols for this case are
\be
\left\{\begin{matrix}
\calO_{\D,J} & \cO_{\D_2} & \cO_{\D-1,J}\\
\cO_{\D_3} & \calV & \cO_{\D_3-1}
\end{matrix} \right\}^{\uniq (+0)}_{\uniq (+0)}
={}& 
-\frac{\left(\Delta _3-2\right) \left(d-\Delta _3-1\right) \left(\Delta -\Delta _2-\Delta _3-J+1\right) }{2 (\Delta -2) (2 \Delta -d) (\Delta +J-1) (-d+\Delta -J+1)}\nn\\
&\times
\left(d+\Delta -\Delta
   _2-\Delta _3+J-1\right)
\,,\nn\\
\left\{\begin{matrix}
\calO_{\D,J} & \cO_{\D_2} & \cO_{\D+1,J}\\
\cO_{\D_3} & \calV & \cO_{\D_3-1}
\end{matrix} \right\}^{\uniq (+0)}_{\uniq (-0)}
={}& 
\frac{(\Delta -1) \left(\Delta _3-2\right) \left(d-\Delta _3-1\right) \left(\Delta -\Delta _2+\Delta _3+J-1\right)
   }{2 (d-2 \Delta ) (\Delta +J-1) (d-\Delta +J-1)}\nn\\
&\times
\left(d-\Delta -\Delta _2-\Delta _3-J+1\right) \left(2 d-\Delta -\Delta _2-\Delta _3+J-1\right) \nn\\
&\times \left(d-\Delta +\Delta
   _2-\Delta _3+J-1\right)
\,,\nn\\
\left\{\begin{matrix}
\calO_{\D,J} & \cO_{\D_2} & \cO_{\D,J-1}\\
\cO_{\D_3} & \calV & \cO_{\D_3-1}
\end{matrix} \right\}^{\uniq (+0)}_{\uniq (0+)}
={}& 
-\frac{\left(\Delta _3-2\right) J \left(d-\Delta _3-1\right) \left(2 d-\Delta -\Delta _2-\Delta _3+J-1\right) }{2 (d+2 J-2) (\Delta +J-1) (d-\Delta +J-1)}\nn\\
&\times
\left(d+\Delta
   -\Delta _2-\Delta _3+J-1\right) \left(d-\Delta +\Delta _2-\Delta _3+J-1\right)
\,,\nn\\
\left\{\begin{matrix}
\calO_{\D,J} & \cO_{\D_2} & \cO_{\D,J+1}\\
\cO_{\D_3} & \calV & \cO_{\D_3-1}
\end{matrix} \right\}^{\uniq (+0)}_{\uniq (0-)}
={}& 
\frac{\left(\Delta _3-2\right) \left(d-\Delta _3-1\right) \left(\Delta -\Delta _2+\Delta _3+J-1\right) }{2 (J+1) (d+2 J-2) (\Delta +J-1) (d-\Delta +J-1)}
\nn\\
&\times
\left(-\Delta +\Delta
   _2+\Delta _3+J-1\right) \left(-d+\Delta +\Delta _2+\Delta _3+J-1\right)
\,,\nn\\
\left\{\begin{matrix}
\calO_{\D,J} & \cO_{\D_2} & \cO_{\D-1,J}\\
\cO_{\D_3} & \calV & \cO_{\D_3+1}
\end{matrix} \right\}^{\uniq (-0)}_{\uniq (+0)}
={}& 
\frac{1}{2 (\Delta -2) (2 \Delta -d) (\Delta +J-1) (-d+\Delta -J+1)}
\,,\nn
\ee
\be
\left\{\begin{matrix}
\calO_{\D,J} & \cO_{\D_2} & \cO_{\D+1,J}\\
\cO_{\D_3} & \calV & \cO_{\D_3+1}
\end{matrix} \right\}^{\uniq (-0)}_{\uniq (-0)}
={}& 
\frac{(1-\Delta) \left(\Delta +\Delta _2-\Delta _3+J-1\right) \left(d-\Delta -\Delta _2+\Delta _3+J-1\right)}{2 (d-2 \Delta )
   (\Delta +J-1) (d-\Delta +J-1)}
\,,\nn\\
\left\{\begin{matrix}
\calO_{\D,J} & \cO_{\D_2} & \cO_{\D,J-1}\\
\cO_{\D_3} & \calV & \cO_{\D_3+1}
\end{matrix} \right\}^{\uniq (-0)}_{\uniq (0+)}
={}& 
-\frac{J \left(d-\Delta -\Delta _2+\Delta _3+J-1\right)}{2 (d+2 J-2) (\Delta +J-1) (d-\Delta +J-1)}
\,,\nn\\
\left\{\begin{matrix}
\calO_{\D,J} & \cO_{\D_2} & \cO_{\D,J+1}\\
\cO_{\D_3} & \calV & \cO_{\D_3+1}
\end{matrix} \right\}^{\uniq (-0)}_{\uniq (0-)}
={}& 
\frac{\Delta +\Delta _2-\Delta _3+J-1}{2 (J+1) (d+2 J-2) (\Delta +J-1) (d-\Delta +J-1)}
\,.\label{eq:6j_3pt_2}
\ee

Next we consider $6j$ symbols for shifts greater than one. These are particularly simple if they can be implemented by $n$ copies of the same weight shifting operators on either side of the crossing equation. We will explain this for a specific example, namely the $6j$ symbol
\beq
\left\{\begin{matrix}
\calO_{\D_1+n} & \cO_{\D_2} & \cO_{\D_1}\\
\cO_{\D,n} & \cW & \cO_{\D}
\end{matrix} \right\}^{\uniq (0,n)}_{\uniq (n,0)}\,,
\eeq
which appears in \eqref{eq:unique_coupling5}.
Consider the case $n=2$ and use \eqref{eq:6j_single_tensor_structure_O1_scalar} twice
\be
{}&\diagramEnvelope{\begin{tikzpicture}[anchor=base,baseline]
	\node (vertL) at (-1.2,0) [twopt] {};
	\node (vertR) at (1.2,-0.05) [threept,inner sep=1pt] {\tiny{$0+$}};
	\node (vertM) at (0,-0.05) [threept,inner sep=1pt] {\tiny{$0+$}};
	\node (opO1) at (-1.7,-1) [below] {$\cO_{\De_1+2}$};
	\node (opO2) at (-1.7,1) [above] {$\cO_{\De_2}$};
	\node (opO3) at (1.7,1) [above] {$\cO_{\De,2}$};
	\node (opWp) at (0.5,-1) [below] {$\calV'$};	
	\node (opW) at (1.7,-1) [below] {$\calV$};	
	\node at (-0.6,0.1) [above] {$\cO_{\De}$};	
	\node at (0.6,0.1) [above] {$\cO_{\De,1}$};	
	\draw [scalar] (vertL)-- (opO1);
	\draw [scalar] (vertL)-- (opO2);
	\draw [spinning] (vertL)-- (vertM);
	\draw [spinning] (vertM)-- (vertR);
	\draw [spinning] (vertR)-- (opO3);
	\draw [finite with arrow] (vertM)-- (opWp);
	\draw [finite with arrow] (vertR)-- (opW);
\end{tikzpicture}}
	\quad=\quad
\left\{\begin{matrix}
\calO_{\D_1+2} & \cO_{\D_2} & \cO_{\D_1+1}\\
\cO_{\D,1} & \calV & \cO_{\D}
\end{matrix} \right\}^{\uniq (0+)}_{\uniq (+0)}
\diagramEnvelope{\begin{tikzpicture}[anchor=base,baseline]
	\node (vertL) at (0,0.6) [twopt] {};
	\node (vertR) at (1.2,0.55) [threept,inner sep=1pt] {\tiny{$0+$}};
	\node (vertD) at (0,-0.65) [threept,inner sep=1pt] {\tiny{$+0$}};
	\node (opO1) at (-1,-1.6) [below] {$\cO_{\De_1+2}$};
	\node (opO2) at (-0.8,1.4) [above] {$\cO_{\De_2}$};
	\node (opO3) at (2.1,1.6) [above] {$\cO_{\De,2}$};
	\node (opWp) at (1,-1.6) [below] {$\calV'$};	
	\node (opW) at (2.1,-0.4) [below] {$\calV$};	
	\node at (-0.1,0) [left] {$\cO_{\De_1+1}$};	
	\node at (0.6,0.7) [above] {$\cO_{\De,1}$};	
	\draw [scalar] (vertD)-- (opO1);
	\draw [scalar] (vertL)-- (opO2);
	\draw [spinning] (vertL)-- (vertR);
	\draw [scalar] (vertL)-- (vertD);
	\draw [spinning] (vertR)-- (opO3);
	\draw [finite with arrow] (vertD)-- (opWp);
	\draw [finite with arrow] (vertR)-- (opW);
\end{tikzpicture}} + \ldots
\label{eq:6j_two_wso}
\\
&=\quad
\left\{\begin{matrix}
\calO_{\D_1+2} & \cO_{\D_2} & \cO_{\D_1+1}\\
\cO_{\D,1} & \calV & \cO_{\D}
\end{matrix} \right\}^{\uniq (0+)}_{\uniq (+0)}
\left\{\begin{matrix}
\calO_{\D_1+1} & \cO_{\D_2} & \cO_{\D_1}\\
\cO_{\D,2} & \calV & \cO_{\D,1}
\end{matrix} \right\}^{\uniq (0+)}_{\uniq (+0)}
\diagramEnvelope{\begin{tikzpicture}[anchor=base,baseline]
	\node (vertU) at (0,1.2) [twopt] {};
	\node (vertD) at (0,-1.2) [threept,inner sep=1pt] {\tiny{$+0$}};
	\node (vertM) at (0,0) [threept,inner sep=1pt] {\tiny{$+0$}};
	\node (opO1) at (-1,-2) [below] {$\cO_{\De_1+2}$};
	\node (opO2) at (-1,2) [above] {$\cO_{\De_2}$};
	\node (opO3) at (1,2) [above] {$\cO_{\De,2}$};
	\node (opWp) at (1,-2) [below] {$\calV'$};	
	\node (opW) at (1,-0.8) [below] {$\calV$};	
	\node at (-0.1,-0.6) [left] {$\cO_{\De_1+1}$};	
	\node at (-0.1,0.6) [left] {$\cO_{\De_1}$};	
	\draw [scalar] (vertD)-- (opO1);
	\draw [scalar] (vertU)-- (opO2);
	\draw [scalar] (vertU)-- (vertM);
	\draw [scalar] (vertM)-- (vertD);
	\draw [spinning] (vertU)-- (opO3);
	\draw [finite with arrow] (vertM)-- (opW);
	\draw [finite with arrow] (vertD)-- (opWp);
\end{tikzpicture}}  + \ldots\,.
\nonumber
\ee
The dots indicate terms with other weight shifting operators in which we are not interested here.
The steps performed above generalize straightforwardly to the case of $n$ weight shifting operators and the resulting $6j$ symbol is
\beq
\left\{\begin{matrix}
\calO_{\D_1+n} & \cO_{\D_2} & \cO_{\D_1}\\
\cO_{\D,J+n} & \cW & \cO_{\D,J}
\end{matrix} \right\}^{\uniq (0,n)}_{\uniq (n,0)}
=
\prod_{k=1}^n
\left\{\begin{matrix}
\calO_{\D_1+n-k+1} & \cO_{\D_2} & \cO_{\D_1+n-k}\\
\cO_{\D,J+k} & \calV & \cO_{\D,J+k-1}
\end{matrix} \right\}^{\uniq (0+)}_{\uniq (+0)}\,.
\eeq
All the other $6j$ symbols in \eqref{eq:6j_3pt} and \eqref{eq:6j_3pt_2} generalize to higher shifts in a similar way.

Now consider the special crossing equation introduced in \eqref{eq:6jdefinition_single_ts},
\be
\diagramEnvelope{\begin{tikzpicture}[anchor=base,baseline]
	\node (vertL) at (0,0) [twopt] {};
	\node (vertR) at (1.5,-0.12) [threept] {$m$};
	\node (opO1) at (-0.5,-1) [below] {$\cO_1$};
	\node (opO2) at (-0.5,1) [above] {$\cO_2$};
	\node (opO3) at (2,1) [above] {$\cO_3$};
	\node (opW) at (2,-1) [below] {$\cW$};	
	\node at (0.75,0.1) [above] {$\cO_3'$};	
	\draw [scalar] (vertL)-- (opO1);
	\draw [scalar] (vertL)-- (opO2);
	\draw [spinning] (vertL)-- (vertR);
	\draw [spinning] (vertR)-- (opO3);
	\draw [finite with arrow] (vertR)-- (opW);
\end{tikzpicture}}
	\hspace{-0.3cm}=
	\sum_{\cO_1',n}
	\left[
		\begin{matrix}
		\cO_1 & \cO_2 & \cO_1' \\
		\cO_3 & \cW & \cO_3'
		\end{matrix}
	\right]^{m}_{n}
\hspace{-0.8cm}
\diagramEnvelope{\begin{tikzpicture}[anchor=base,baseline]
	\node (vertU) at (0,0.7) [twopt] {};
	\node (vertD) at (0,-0.7) [threept] {$n$};
	\node (opO1) at (-1,-1.5) [below] {$\cO_1$};
	\node (opO2) at (-1,1.5) [above] {$\cO_2$};
	\node (opO3) at (1,1.5) [above] {$\cO_3$};
	\node (opW) at (1,-1.5) [below] {$\cW$};	
	\node at (0.1,0) [right] {$\cO_1'$};	
	\draw [scalar] (vertD)-- (opO1);
	\draw [scalar] (vertU)-- (opO2);
	\draw [scalar] (vertU)-- (vertD);
	\draw [spinning] (vertU)-- (opO3);
	\draw [finite with arrow] (vertD)-- (opW);
\end{tikzpicture}}
\hspace{-0.3cm}+  (-1)^{J_3-J_3'}
	\sum_{\cO_2',n}
	\left[
		\begin{matrix}
		\cO_2 & \cO_1 & \cO_2' \\
		\cO_3 & \cW & \cO_3'
		\end{matrix}
	\right]^{m}_{n}
\hspace{-0.8cm}
\diagramEnvelope{\begin{tikzpicture}[anchor=base,baseline]
	\node (vertU) at (0,-0.7) [twopt] {};
	\node (vertD) at (0,0.7) [threept] {$n$};
	\node (opO1) at (-1,1.5) [above] {$\cO_2$};
	\node (opO2) at (-1,-1.5) [below] {$\cO_1$};
	\node (opO3) at (1,-1.5) [below] {$\cO_3$};
	\node (opW) at (1,1.5) [above] {$\cW$};	
	\node at (0.1,0) [right] {$\cO_2'$};	
	\draw [scalar] (vertD)-- (opO1);
	\draw [scalar] (vertU)-- (opO2);
	\draw [scalar] (vertU)-- (vertD);
	\draw [spinning] (vertU)-- (opO3);
	\draw [finite with arrow] (vertD)-- (opW);
\end{tikzpicture}}
\hspace{-0.3cm}.
\label{eq:6jdefinition_single_ts_appendix}
\ee
Explicitly the equation reads
\be
\label{eq:CFT_crossing_example}
&\cD^{a}_{\de_\De,\de_J}(P,Z)
\< \cO_{\D_1}(P_1) \cO_{\D_2}(P_2) \cO_{\D-\de_\De,J-\de_J}(P, Z)  \>\\
&=
\sum_{\delta_{\D 1} = \pm 1} 
\left[
\begin{matrix}
\cO_{\D_1} & \cO_{\D_2} & \cO_{\D_1-\delta_{\D 1}} \\
\cO_{\D,J} & \cW & \cO_{\D-\de_\De,J-\de_J}
\end{matrix}
\right]^{(\de_\De,\de_J)}_{(\delta_{\D 1}, 0)}
 \cD^{a}_{\delta_{\D 1}, 0}(P_1)\< \cO_{\D_1-\delta_{\D 1}}(P_1) \cO_{\D_2}(P_2) \cO_{\D,J}(P, Z)  \>\nn\\
&+ (-1)^{\de_J}
\sum_{\delta_{\D 2} = \pm 1}
\left[
\begin{matrix}
\cO_{\D_2} & \cO_{\D_1} & \cO_{\D_2-\delta_{\D 2}} \\
\cO_{\D,J} & \cW & \cO_{\D-\de_\De,J-\de_J}
\end{matrix}
\right]^{(\de_\De,\de_J)}_{(\delta_{\D 2}, 0)}
 \cD^{a}_{\delta_{\D 2}, 0}(P_2)\< \cO_{\D_1}(P_1) \cO_{\D_2-\delta_{\D 2}}(P_2) \cO_{\D,J}(P, Z)  \>\,,
\nn
\ee
We find the coefficients
\be
\left[
\begin{matrix}
\cO_{\D_1} & \cO_{\D_2} & \cO_{\D_1-1} \\
\cO_{\D,J} & \cW & \cO_{\D-1,J}
\end{matrix}
\right]^{(+0)}_{(+0)} 
={}&
\frac{(\Delta -2) \left(\Delta -\Delta _1+\Delta _2+J-1\right) \left(d-\Delta +\Delta _1-\Delta _2+J-1\right)}{2 \left(\Delta
   _1-2\right) \left(d-2 \Delta _1\right) \left(d-\Delta _1-1\right)}
\nn \,, \\
\left[
\begin{matrix}
\cO_{\D_1} & \cO_{\D_2} & \cO_{\D_1+1} \\
\cO_{\D,J} & \cW & \cO_{\D-1,J}
\end{matrix}
\right]^{(+0)}_{(-0)} 
={}& 
-\frac{(\Delta -2) \left(\Delta +\Delta _1-\Delta _2+J-1\right) \left(-d+\Delta +\Delta _1-\Delta _2-J+1\right) }{2 \left(d-2 \Delta _1\right)}\nn \\
&\times
\left(-2
   d+\Delta +\Delta _1+\Delta _2-J+1\right) \left(-d+\Delta +\Delta _1+\Delta _2+J-1\right)
\nn \,, \\
\left[
\begin{matrix}
\cO_{\D_1} & \cO_{\D_2} & \cO_{\D_1-1} \\
\cO_{\D,J} & \cW & \cO_{\D+1,J}
\end{matrix}
\right]^{(-0)}_{(+0)} 
={}&
\frac{1}{2 (\Delta -1) \left(\Delta _1-2\right) \left(d-2 \Delta _1\right) \left(d-\Delta _1-1\right)}
\nn \,, \\
\left[
\begin{matrix}
\cO_{\D_1} & \cO_{\D_2} & \cO_{\D_1+1} \\
\cO_{\D,J} & \cW & \cO_{\D+1,J}
\end{matrix}
\right]^{(-0)}_{(-0)} 
={}& 
-\frac{\left(-\Delta +\Delta _1+\Delta _2+J-1\right) \left(d+\Delta -\Delta _1-\Delta _2+J-1\right)}{2 (\Delta -1) \left(d-2
   \Delta _1\right)}
\nn \,,
\ee
\be
\left[
\begin{matrix}
\cO_{\D_1} & \cO_{\D_2} & \cO_{\D_1-1} \\
\cO_{\D,J} & \cW & \cO_{\D,J-1}
\end{matrix}
\right]^{(0+)}_{(+0)} 
={}&
\frac{\Delta -\Delta _1+\Delta _2+J-1}{2 \left(\Delta _1-2\right) J \left(d-2 \Delta _1\right) \left(d-\Delta _1-1\right)}
\nn \,, \\
\left[
\begin{matrix}
\cO_{\D_1} & \cO_{\D_2} & \cO_{\D_1+1} \\
\cO_{\D,J} & \cW & \cO_{\D,J-1}
\end{matrix}
\right]^{(0+)}_{(-0)} 
={}& 
-\frac{\left(\Delta +\Delta _1-\Delta _2+J-1\right) \left(-\Delta +\Delta _1+\Delta _2+J-1\right) }{2 J \left(d-2 \Delta _1\right)}\nn\\
&\times
\left(-d+\Delta +\Delta
   _1+\Delta _2+J-1\right)
\nn \,, \\
\left[
\begin{matrix}
\cO_{\D_1} & \cO_{\D_2} & \cO_{\D_1-1} \\
\cO_{\D,J} & \cW & \cO_{\D,J+1}
\end{matrix}
\right]^{(0-)}_{(+0)} 
={}&
\frac{(J+1) \left(d-\Delta +\Delta _1-\Delta _2+J-1\right)}{2 \left(\Delta _1-2\right) \left(d-2 \Delta _1\right)
   \left(d-\Delta _1-1\right)}
\nn \,, \\
\left[
\begin{matrix}
\cO_{\D_1} & \cO_{\D_2} & \cO_{\D_1+1} \\
\cO_{\D,J} & \cW & \cO_{\D,J+1}
\end{matrix}
\right]^{(0-)}_{(-0)} 
={}& 
-\frac{ \left(2 d-\Delta -\Delta _1-\Delta _2+J-1\right) \left(d+\Delta -\Delta _1-\Delta _2+J-1\right) }{2 \left(d-2 \Delta _1\right)}\nn\\
&\times
(J+1)\left(d-\Delta
   -\Delta _1+\Delta _2+J-1\right)
\label{eq:special_6j_symbols} \,.
\ee

\subsection{Local AdS couplings}
\label{app:local_ads_couplings}

\subsubsection{Non-unique couplings}
\label{app:non-unique_couplings}

In this appendix we give an example how AdS couplings that admit multiple tensor structures are related to CFT three-point functions, as stated in \eqref{eq:ads_integral_general}. We consider the simplest example with multiple tensor structures, two massive spin 1 fields $\f_1$, $\f_2$ and a massive scalar $\f_3$. Consider the couplings
\be
\diagramEnvelope{\begin{tikzpicture}[anchor=base,baseline]
	\node (vert) at (0,-0.1) [threept] {$1$};
	\node (opO1) at (-0.5,-1) [below] {$\f_1$};
	\node (opO2) at (-0.5,1) [above] {$\f_2$};
	\node (opO3) at (1,0) [right] {$\f_3$};	
	\draw [spinning bulk] (opO1)-- (vert);
	\draw [spinning bulk] (opO2)-- (vert);
	\draw [scalar bulk] (opO3)-- (vert);
\end{tikzpicture}}
\quad&=\quad \int_{\rm AdS} dX \ \partial_{W_1}^{\{a\}} \< \f_1 (X,W_1)| \, \partial_{W_2 \{a\}} \< \f_2 (X,W_2)| \, \< \f_3 (X)| \,, \nonumber\\
\diagramEnvelope{\begin{tikzpicture}[anchor=base,baseline]
	\node (vert) at (0,-0.1) [threept] {$2$};
	\node (opO1) at (-0.5,-1) [below] {$\f_1$};
	\node (opO2) at (-0.5,1) [above] {$\f_2$};
	\node (opO3) at (1,0) [right] {$\f_3$};	
	\draw [spinning bulk] (opO1)-- (vert);
	\draw [spinning bulk] (opO2)-- (vert);
	\draw [scalar bulk] (opO3)-- (vert);
\end{tikzpicture}}
\quad&=\quad \int_{\rm AdS} dX \ \partial_{W_1}^{\{a\}} \< \f_1 (X,W_1)| \, \partial_{W_2}^{\{b\}} \< \f_2 (X,W_2)| \, \nabla_{\{a\}} \nabla_{\{b\}}\< \f_3 (X)| \,.
\label{eq:non-unique_coupling}
\ee
In order to relate these couplings to CFT three-point functions, we dress them with bulk-to-boundary propagators and compute the corresponding Witten diagrams
\be
\int_{\rm AdS} dX \ P_a^b \partial_{W_1}^{a} \Pi_{\De_1,1} (X,P_1,W_1,Z_1) \,
\partial_{W_2 b} \Pi_{\De_2,1} (X,P_2,W_2,Z_2)
\Pi_{\De_3} (X,P_3)\,,
\label{eq:non-unique_WD1}
\ee
and 
\be
\int_{\rm AdS} dX \ \partial_{W_1}^{a} \Pi_{\De_1,1} (X,P_1,W_1,Z_1) \,
\partial_{W_2}^{b} \Pi_{\De_2,1} (X,P_2,W_2,Z_2)
P_a^c P_b^d \partial_{X c} P_d^e  \partial_{X e}
\Pi_{\De_3} (X,P_3)\,.
\label{eq:non-unique_WD2}
\ee
The derivatives can be expressed in terms of bulk weight shifting operators
\be
\eqref{eq:non-unique_WD1}\quad=\quad
\diagramEnvelope{\begin{tikzpicture}[anchor=base,baseline]
	\node (vert) at (-0.1,-0.1) [threept] {$1$};
	\node (prop1) at (-0.4,-0.8) [twopt] {};
	\node (prop2) at (-0.4,0.8) [twopt] {};
	\node (prop3) at (0.8,0) [twopt] {};	
	\node (opO1) at (-0.8,-1.6) [below] {$\cO_1$};
	\node (opO2) at (-0.8,1.6) [above] {$\cO_2$};
	\node (opO3) at (1.6,0) [right] {$\cO_3$};	
	\draw [spinning bulk] (prop1)-- (vert);
	\draw [spinning bulk] (prop2)-- (vert);
	\draw [scalar bulk] (prop3)-- (vert);
	\draw [spinning] (prop1)-- (opO1);
	\draw [spinning] (prop2)-- (opO2);
	\draw [scalar] (prop3)-- (opO3);
\end{tikzpicture}}
\quad=\quad  
\diagramEnvelope{\begin{tikzpicture}[anchor=base,baseline]
	\node (vert) at (-0.1,0) [twopt] {};
	\node (prop1) at (-1,-1.8) [twopt] {};
	\node (prop2) at (-1,1.8) [twopt] {};
	\node (prop3) at (0.8,0) [twopt] {};	
	\node (ws1) at (-0.5,-0.9-0.1) [threept] {\tiny{$0-$}};
	\node (ws2) at (-0.5,0.9-0.1) [threept] {\tiny{$0-$}};
	\node (opO1) at (-1.5,-2.4) [below] {$\cO_1$};
	\node (opO2) at (-1.5,2.4) [above] {$\cO_2$};
	\node (opO3) at (1.6,0) [right] {$\cO_3$};	
	\draw [scalar bulk] (ws1)-- (vert);
	\draw [scalar bulk] (ws2)-- (vert);
	\draw [spinning bulk] (prop1)-- (ws1);
	\draw [spinning bulk] (prop2)-- (ws2);
	\draw [scalar bulk] (prop3)-- (vert);
	\draw [spinning] (prop1)-- (opO1);
	\draw [spinning] (prop2)-- (opO2);
	\draw [scalar] (prop3)-- (opO3);
	\draw [finite with arrow] (ws2) to[out=210,in=150]  (ws1);
\end{tikzpicture}}\quad .
\ee
Similarly \eqref{eq:non-unique_WD2} is given by
\be
\diagramEnvelope{\begin{tikzpicture}[anchor=base,baseline]
	\node (vert) at (-0.1,-0.1) [threept] {$2$};
	\node (prop1) at (-0.4,-0.8) [twopt] {};
	\node (prop2) at (-0.4,0.8) [twopt] {};
	\node (prop3) at (0.8,0) [twopt] {};	
	\node (opO1) at (-0.8,-1.6) [below] {$\cO_1$};
	\node (opO2) at (-0.8,1.6) [above] {$\cO_2$};
	\node (opO3) at (1.6,0) [right] {};	
	\node at (1.6,0) [above] {$\cO_3$};	
	\draw [spinning bulk] (prop1)-- (vert);
	\draw [spinning bulk] (prop2)-- (vert);
	\draw [scalar bulk] (prop3)-- (vert);
	\draw [spinning] (prop1)-- (opO1);
	\draw [spinning] (prop2)-- (opO2);
	\draw [scalar] (prop3)-- (opO3);
\end{tikzpicture}}
=
\frac{1}{\De_3(\De_3-1)} \hspace{-1cm}
\diagramEnvelope{\begin{tikzpicture}[anchor=base,baseline]
	\node (vert) at (-0.1,0) [twopt] {};
	\node (prop1) at (-1,-1.8) [twopt] {};
	\node (prop2) at (-1,1.8) [twopt] {};
	\node (prop3) at (2.8,0) [twopt] {};	
	\node (ws1) at (-0.5,-0.9-0.1) [threept] {\tiny{$0-$}};
	\node (ws2) at (-0.5,0.9-0.1) [threept] {\tiny{$0-$}};
	\node (ws31) at (0.8,-0.1) [threept] {\tiny{$+0$}};	
	\node (ws32) at (1.8,-0.1) [threept] {\tiny{$+0$}};	
	\node (opO1) at (-1.5,-2.4) [below] {$\cO_1$};
	\node (opO2) at (-1.5,2.4) [above] {$\cO_2$};
	\node (opO3) at (3.7,0) [right] {};
	\node at (3.7,0) [above] {$\cO_3$};
	\draw [scalar bulk] (ws1)-- (vert);
	\draw [scalar bulk] (ws2)-- (vert);
	\draw [spinning bulk] (prop1)-- (ws1);
	\draw [spinning bulk] (prop2)-- (ws2);
	\draw [scalar bulk] (prop3)-- (ws32);
	\draw [scalar bulk] (ws32)-- (ws31);
	\draw [scalar bulk] (ws31)-- (vert);
	\draw [spinning] (prop1)-- (opO1);
	\draw [spinning] (prop2)-- (opO2);
	\draw [scalar] (prop3)-- (opO3);
	\draw [finite with arrow] (ws2) to[out=45,in=90]  (ws32);
	\draw [finite with arrow] (ws1) to[out=330,in=270]  (ws31);
\end{tikzpicture}}
- \ \De_3 \hspace{-.5cm}
\diagramEnvelope{\begin{tikzpicture}[anchor=base,baseline]
	\node (vert) at (-0.1,0) [twopt] {};
	\node (prop1) at (-1,-1.8) [twopt] {};
	\node (prop2) at (-1,1.8) [twopt] {};
	\node (prop3) at (0.8,0) [twopt] {};	
	\node (ws1) at (-0.5,-0.9-0.1) [threept] {\tiny{$0-$}};
	\node (ws2) at (-0.5,0.9-0.1) [threept] {\tiny{$0-$}};
	\node (opO1) at (-1.5,-2.4) [below] {$\cO_1$};
	\node (opO2) at (-1.5,2.4) [above] {$\cO_2$};
	\node (opO3) at (1.6,0) [right] {};	
	\node at (1.6,0) [above] {$\cO_3$};	
	\draw [scalar bulk] (ws1)-- (vert);
	\draw [scalar bulk] (ws2)-- (vert);
	\draw [spinning bulk] (prop1)-- (ws1);
	\draw [spinning bulk] (prop2)-- (ws2);
	\draw [scalar bulk] (prop3)-- (vert);
	\draw [spinning] (prop1)-- (opO1);
	\draw [spinning] (prop2)-- (opO2);
	\draw [scalar] (prop3)-- (opO3);
	\draw [finite with arrow] (ws2) to[out=210,in=150]  (ws1);
\end{tikzpicture}}\,.
\ee
Using \eqref{eq:twoptcrossing_bulk_boundary} and \eqref{eq:ads_integral} the two remaining diagrams can easily be expressed in the differential CFT three-point function basis
\beq
\diagramEnvelope{\begin{tikzpicture}[anchor=base,baseline]
	\node (vert) at (-0.1,0) [twopt] {};
	\node (ws1) at (-0.5,-0.9-0.1) [threept] {\tiny{$0+$}};
	\node (ws2) at (-0.5,0.9-0.1) [threept] {\tiny{$0+$}};
	\node (opO1) at (-1,-1.7) [below] {$\cO_1$};
	\node (opO2) at (-1,1.6) [above] {$\cO_2$};
	\node (opO3) at (1,0) [right] {$\cO_3$};	
	\draw [scalar] (vert)-- (ws1);
	\draw [scalar] (vert)-- (ws2);
	\draw [spinning] (ws1)-- (opO1);
	\draw [spinning] (ws2)-- (opO2);
	\draw [scalar] (vert)-- (opO3);
	\draw [finite with arrow] (ws2) to[out=210,in=150]  (ws1);
\end{tikzpicture}}
\quad , \quad
\diagramEnvelope{\begin{tikzpicture}[anchor=base,baseline]
	\node (vert) at (-0.1,0) [twopt] {};
	\node (ws1) at (-0.5,-0.9-0.1) [threept] {\tiny{$0+$}};
	\node (ws2) at (-0.5,0.9-0.1) [threept] {\tiny{$0+$}};
	\node (ws31) at (0.8,-0.1) [threept] {\tiny{$-0$}};	
	\node (ws32) at (1.8,-0.1) [threept] {\tiny{$-0$}};	
	\node (opO1) at (-1,-1.7) [below] {$\cO_1$};
	\node (opO2) at (-1,1.6) [above] {$\cO_2$};
	\node (opO3) at (2.9,0) [right] {$\cO_3$};
	\draw [scalar] (vert)-- (ws1);
	\draw [scalar] (vert)-- (ws2);
	\draw [spinning] (ws1)-- (opO1);
	\draw [spinning] (ws2)-- (opO2);
	\draw [scalar] (ws31)-- (ws32);
	\draw [scalar] (vert)-- (ws31);
	\draw [scalar] (ws32)-- (opO3);
	\draw [finite with arrow] (ws2) to[out=45,in=90]  (ws32);
	\draw [finite with arrow] (ws1) to[out=330,in=270]  (ws31);
\end{tikzpicture}}
\quad ,
\eeq
which in turn can be related to the two tensor structures
\bea
&\frac{\big( 2(Z_1 \cdot P_2)( Z_2 \cdot P_1) - 2 (Z_1 \cdot Z_2)( P_1 \cdot P_2) \big)}{P_{12}^{\frac{\D_1+\D_2-\D+2}{2}} P_{23}^{\frac{\D_2+\D-\D_1}{2}} P_{31}^{\frac{\D+\D_1-\D_2}{2}}} \ , \\
&\frac{\big((Z_1 \cdot P_3)P_{21} - (Z_1 \cdot P_2)P_{31} \big)\big((Z_2 \cdot P_3)P_{12} - (Z_2 \cdot P_1)P_{32} \big)}{P_{12}^{\frac{\D_1+\D_2-\D+2}{2}} P_{23}^{\frac{\D_2+\D-\D_1+2}{2}} P_{31}^{\frac{\D+\D_1-\D_2+2}{2}}} \ .
\eea{eq:non-unique-ts}
\subsubsection{Computing bulk $6j$ symbols}
\label{app:computing_bulk_6j}

We demonstrate how to compute bulk $6j$ symbols as defined by \eqref{eq:Bulk6jdefinition}
from the usual $6j$ symbols by performing the steps for the special case \eqref{eq:6j_single_tensor_structure_O3_scalar}, where the three-point structures are unique.
We would like to derive the coefficients in the relation
\be
\diagramEnvelope{\begin{tikzpicture}[anchor=base,baseline]
	\node (vertL) at (0,0.07) [twopt] {};
	\node (vertR) at (2,-0.05) [threept] {$m$};
	\node (opO1) at (-0.5,-1) [below] {$\f_1$};
	\node (opO2) at (-0.5,1) [above] {$\f_2$};
	\node (opO3) at (2.5,1) [above] {$\f_3$};
	\node (opW) at (2.5,-1) [below] {$\cW$};	
	\node at (1,0.1) [above] {$\f_3'$};	
	\draw [spinning bulk] (opO1)-- (vertL);
	\draw [scalar bulk] (opO2)-- (vertL);
	\draw [scalar bulk] (vertR)-- (vertL);
	\draw [scalar bulk] (opO3)-- (vertR);
	\draw [finite with arrow] (vertR)-- (opW);
\end{tikzpicture}}
	\quad=\quad
	\sum_{\f_1',n}
	\left\{
		\begin{matrix}
		\f_1' & \f_2 & \f_1 \\
		\f_3' & \cW & \f_3
		\end{matrix}
	\right\}^{\uniq m}_{\uniq n}
\diagramEnvelope{\begin{tikzpicture}[anchor=base,baseline]
	\node (vertU) at (0,0.7) [twopt] {};
	\node (vertD) at (0,-0.7) [threept] {$n$};
	\node (opO1) at (-1,-1.5) [below] {$\f_1$};
	\node (opO2) at (-1,1.5) [above] {$\f_2$};
	\node (opO3) at (1,1.5) [above] {$\f_3$};
	\node (opW) at (1,-1.5) [below] {$\cW$};	
	\node at (0.1,0) [right] {$\f_1'$};	
	\draw [spinning bulk] (opO1)-- (vertD);
	\draw [scalar bulk] (opO2)-- (vertU);
	\draw [spinning bulk] (vertD)-- (vertU);
	\draw [scalar bulk] (opO3)-- (vertU);
	\draw [finite with arrow] (vertD)-- (opW);
\end{tikzpicture}},
\label{eq:6j_bulk_single_tensor_structure}
\ee
where $\f_2$, $\f_3$ and $\f_3'$ are scalars in order to make all tensor structures unique. We start by inserting \eqref{eq:ads_integral_unique} into \eqref{eq:6j_single_tensor_structure_O3_scalar},
\be
\diagramEnvelope{\begin{tikzpicture}[anchor=base,baseline]
	\node (vertL) at (-1.1,0) [twopt] {};
	\node (vertR) at (1.1,-0.08) [threept] {$m$};
	\node (opO1) at (-1.8,-1.6) [below] {$\cO_1$};
	\node (opO2) at (-1.8,1.6) [above] {$\cO_2$};
	\node (W) at (1.8,-1.6) [below] {$\cW$};
	\node (opO3) at (1.8,1.6) [above] {$\cO_3$};
	\node (prop1) at (-1.4,-0.8) [twopt] {};
	\node (prop2) at (-1.4,0.8) [twopt] {};
	\node (prop) at (0,0) [twopt] {};
	\node at (0.55,-0.1) [below] {$\cO_3'$};	
	\draw [spinning] (prop1)-- (opO1);
	\draw [scalar] (prop2)-- (opO2);
	\draw [spinning bulk] (prop1)-- (vertL);
	\draw [scalar bulk] (prop2)-- (vertL);
	\draw [scalar bulk] (prop)-- (vertL);
	\draw [scalar] (prop)-- (vertR);
	\draw [finite with arrow] (vertR)-- (W);
	\draw [scalar] (vertR)-- (opO3);
\end{tikzpicture}}
	\quad=\quad
	\sum_{\cO_1',n}
	\left\{
		\begin{matrix}
		\cO_1 & \cO_2 & \cO_1' \\
		\cO_3 & \cW & \cO_3'
		\end{matrix}
	\right\}^{\uniq m}_{\uniq n}
\frac{b(\cO_2, \cO_3' ,\cO_1)}{b(\cO_2, \cO_3 ,\cO_1')}
\diagramEnvelope{\begin{tikzpicture}[anchor=base,baseline]
	\node (vertU) at (0,0.8) [twopt] {};
	\node (vertD) at (0,-0.8) [threept] {$n$};
	\node (opO1) at (-1.4,-1.6) [below] {$\cO_1$};
	\node (opO2) at (-1.4,1.6) [above] {$\cO_2$};
	\node (opO3) at (1.4,1.6) [above] {$\cO_3$};
	\node (opW) at (1.4,-1.6) [below] {$\cW$};	
	\node (prop2) at (-0.53,1.2) [twopt] {};
	\node (prop3) at (0.53,1.2) [twopt] {};
	\node (prop) at (0,0.1) [twopt] {};
	\node at (0,-0.3) [right] {$\cO_1'$};	
	\draw [spinning] (vertD)-- (opO1);
	\draw [scalar] (prop2)-- (opO2);
	\draw [spinning] (prop)-- (vertD);
	\draw [spinning bulk] (prop)-- (vertU);
	\draw [scalar bulk] (prop2)-- (vertU);
	\draw [scalar bulk] (prop3)-- (vertU);
	\draw [scalar] (prop3)-- (opO3);
	\draw [finite with arrow] (vertD)-- (opW);
\end{tikzpicture}}.
\ee
Next the weight shifting operators are moved towards the local AdS couplings by commuting with the bulk-to-boundary propagators
\be
\diagramEnvelope{\begin{tikzpicture}[anchor=base,baseline]
	\node (vertL) at (-0.7,0) [twopt] {};
	\node (vertR) at (0.7,-0.08) [threept] {$\bar{m}$};
	\node (opO1) at (-1.4,-1.6) [below] {$\cO_1$};
	\node (opO2) at (-1.4,1.6) [above] {$\cO_2$};
	\node (W) at (1.4,-1.6) [below] {$\cW$};
	\node (opO3) at (1.4,1.6) [above] {$\cO_3$};
	\node (prop1) at (-1,-0.8) [twopt] {};
	\node (prop2) at (-1,0.8) [twopt] {};
	\node (prop3) at (1.05,0.95) [twopt] {};
	\node at (0,-0.1) [below] {$\f_3'$};	
	\draw [spinning] (prop1)-- (opO1);
	\draw [scalar] (prop2)-- (opO2);
	\draw [spinning bulk] (prop1)-- (vertL);
	\draw [scalar bulk] (prop2)-- (vertL);
	\draw [scalar bulk] (prop3)-- (vertR);
	\draw [scalar bulk] (vertR)-- (vertL);
	\draw [finite with arrow] (vertR)-- (W);
	\draw [scalar] (prop3)-- (opO3);
\end{tikzpicture}}
	=\quad
	\sum_{\f_1',n}
	\left\{
		\begin{matrix}
		\cO_1 & \cO_2 & \cO_1' \\
		\cO_3 & \cW & \cO_3'
		\end{matrix}
	\right\}^{\uniq m}_{\uniq n}
\frac{b(\cO_2 ,\cO_3', \cO_1)\left\{ \begin{matrix}
\cO_3\\ \f_3'
\end{matrix} \right\}^{\bar{m}}_{m} 
}{
b(\cO_2, \cO_3 ,\cO_1')\left\{ \begin{matrix}
\cO_1\\ \f_1'
\end{matrix} \right\}^{\bar{n}}_{n} }
\diagramEnvelope{\begin{tikzpicture}[anchor=base,baseline]
	\node (vertU) at (0,0.8) [twopt] {};
	\node (vertD) at (0,-0.7) [threept] {$\bar{n}$};
	\node (opO1) at (-1.4,-1.6) [below] {$\cO_1$};
	\node (opO2) at (-1.4,1.6) [above] {$\cO_2$};
	\node (opO3) at (1.4,1.6) [above] {$\cO_3$};
	\node (opW) at (1.4,-1.6) [below] {$\cW$};	
	\node (prop1) at (-0.53,-1.12) [twopt] {};
	\node (prop2) at (-0.53,1.2) [twopt] {};
	\node (prop3) at (0.53,1.2) [twopt] {};
	\node at (0.1,0) [right] {$\f_1'$};	
	\draw [spinning] (prop1)-- (opO1);
	\draw [scalar] (prop2)-- (opO2);
	\draw [spinning bulk] (vertD)-- (vertU);
	\draw [spinning bulk] (prop1)-- (vertD);
	\draw [scalar bulk] (prop2)-- (vertU);
	\draw [scalar bulk] (prop3)-- (vertU);
	\draw [scalar] (prop3)-- (opO3);
	\draw [finite with arrow] (vertD)-- (opW);
\end{tikzpicture}}.
\ee
We can bring this into the form \eqref{eq:6j_bulk_single_tensor_structure} by removing the propagators from the
external lines and relabelling $\bar{m} \to m$, $\bar{n} \to n$. Hence the bulk $6j$ symbols are related to the ones for conformal structures by
\beq
	\left\{
		\begin{matrix}
		\f_1' & \f_2 & \f_1 \\
		\f_3' & \cW & \f_3
		\end{matrix}
	\right\}^{\uniq m}_{\uniq n}
=
	\left\{
		\begin{matrix}
		\cO_1 & \cO_2 & \cO_1' \\
		\cO_3 & \cW & \cO_3'
		\end{matrix}
	\right\}^{\uniq \bar{m}}_{\uniq \bar{n}}
\frac{b(\cO_2 ,\cO_3' ,\cO_1)\left\{ \begin{matrix}
\cO_3\\ \f_3'
\end{matrix} \right\}^{m}_{\bar{m}} 
}{
b(\cO_2 ,\cO_3 ,\cO_1')\left\{ \begin{matrix}
\cO_1\\ \f_1'
\end{matrix} \right\}^{n}_{\bar{n}} }\,.
\eeq
For this derivation it was assumed that the local couplings are dressed with bulk-to-boundary propagators. Due to the split representation of the harmonic functions \eqref{eq:spinning_omega_to_shadow} it is clear that the same crossing relation also holds if some or all of the lines are attached to harmonic functions.

\subsection{AdS $6j$ symbols with identity field}
\label{app:compatibility}

In this appendix we show equation (\ref{eq:compatibility}). We start by taking $\f_2$ in \eqref{eq:Bulk6jdefinition} to be the identity
\be
\diagramEnvelope{\begin{tikzpicture}[anchor=base,baseline]
	\node (vertU) at (0,0) [twopt] {};
	\node (vertD) at (1,-0.12) [threept] {$\bar{m}$};
	\node (opO1) at (-1,0) [left] {$\f_1$};
	\node (opO3) at (2,0) [right] {$\f_3$};
	\node (opW) at (1,-1) [below] {$\cW$};	
	\node at (0.6,0.1) [above] {$\f_3'$};
	\draw [spinning bulk] (opO1)-- (vertU);
	\draw [spinning bulk] (vertD)-- (vertU);
	\draw [spinning bulk] (opO3)-- (vertD);
	\draw [finite with arrow] (vertD)-- (opW);
\end{tikzpicture}}
	\quad=\quad
	\sum_{\f_1',n}
	\left\{
		\begin{matrix}
		\f_1' & \mathbf{1} & \f_1 \\
		\f_3' & \cW & \f_3
		\end{matrix}
	\right\}^{\uniq \bar{m}}_{\uniq n}
\diagramEnvelope{\begin{tikzpicture}[anchor=base,baseline]
	\node (vertL) at (1,0) [twopt] {};
	\node (vertR) at (0,-0.08) [threept] {$n$};
	\node (opO1) at (-1,0) [left] {$\f_1$};
	\node (opO3) at (2,0) [right] {$\f_3$};
	\node (opW) at (0,-1) [below] {$\cW$};
	\node at (0.4,0.1) [above] {$\f_1'$};
	\draw [spinning bulk] (opO1)-- (vertR);
	\draw [spinning bulk] (vertR)-- (vertL);
	\draw [spinning bulk] (opO3)-- (vertL);
	\draw [finite with arrow] (vertR)-- (opW);
\end{tikzpicture}},
\label{eq:bulk_6j_2pt_vs_3pt_1}
\ee
where the dot with two incoming arrows indicates a local two-point coupling, i.e.\ indices have to be contracted and AdS coordinates integrated. This implies that $\f_1$ and $\f_3'$ (and $\f_1'$, $\f_3$) have the same spin.
Next we attach harmonic functions to both sides of the equation and use \eqref{eq:twoptcrossing_bulk} to move each weight shifting operator past one of them
\be
{}&\left\{ \begin{matrix}
\f_3\\ \f_3'
\end{matrix} \right\}^{\bar{m}}_{m} 
\diagramEnvelope{\begin{tikzpicture}[anchor=base,baseline]
	\node (propL) at (-1,0) [twopt] {};
	\node (vertL) at (0,0) [twopt] {};
	\node (propR) at (1,0) [twopt] {};
	\node (vertR) at (2,-0.11) [threept] {$m$};
	\node (opO1) at (-2,0) [left] {$\f_1$};
	\node (opO3) at (3,0) [right] {$\f_3$};
	\node (opW) at (2,-1) [below] {$\cW$};	
	\node at (1,0.1) [above] {$\f_3'$};
	\draw [spinning bulk] (propL)-- (opO1);
	\draw [spinning bulk] (propL)-- (vertL);
	\draw [spinning bulk] (propR)-- (vertL);
	\draw [spinning bulk] (propR)-- (vertR);
	\draw [spinning bulk] (vertR)-- (opO3);
	\draw [finite with arrow] (vertR)-- (opW);
\end{tikzpicture}}\\
={}&\quad
	\sum_{\f_1',n}
	\left\{
		\begin{matrix}
		\f_1' & \mathbf{1} & \f_1 \\
		\f_3' & \cW & \f_3
		\end{matrix}
	\right\}^{\uniq \bar{m}}_{\uniq n}
\left\{ \begin{matrix}
\f_1\\ \f_1'
\end{matrix} \right\}^{n}_{\bar{n}} 
\diagramEnvelope{\begin{tikzpicture}[anchor=base,baseline]
	\node (opO1) at (-2,0) [left] {$\f_1$};
	\node (vertL) at (-1,-0.11) [threept] {$\bar{n}$};
	\node (propL) at (0,0) [twopt] {};
	\node (vertR) at (1,0) [twopt] {};
	\node (propR) at (2,0) [twopt] {};
	\node (opO3) at (3,0) [right] {$\f_3$};
	\node (opW) at (-1,-1) [below] {$\cW$};
	\node at (0,0.1) [above] {$\f_1'$};
	\draw [spinning bulk] (vertL)-- (opO1);
	\draw [spinning bulk] (propL)-- (vertL);
	\draw [spinning bulk] (propL)-- (vertR);
	\draw [spinning bulk] (propR)-- (vertR);
	\draw [spinning bulk] (propR)-- (opO3);
	\draw [finite with arrow] (vertL)-- (opW);
\end{tikzpicture}}.
\label{eq:bulk_6j_2pt_vs_3pt_2}
\ee
Now we use the orthogonality relation for harmonic functions \cite{Costa:2014kfa}
\beq
\diagramEnvelope{\begin{tikzpicture}[anchor=base,baseline]
	\node (propL) at (-1,0) [twopt] {};
	\node (vertL) at (0,0) [twopt] {};
	\node (propR) at (1,0) [twopt] {};
	\node (opO1) at (-2,0) [left] {};
	\node (opO3) at (2,0) [right] {};
	\node at (-2,0) [below] {$\f_{\De(\nu),J}$};
	\node at (2,0) [below] {$\f_{\De(\bar{\nu}),J}$};
	\draw [spinning bulk] (propL)-- (opO1);
	\draw [spinning bulk] (propL)-- (vertL);
	\draw [spinning bulk] (propR)-- (vertL);
	\draw [spinning bulk] (propR)-- (opO3);
\end{tikzpicture}}
= \quad
\frac{1}{2}\big( \de(\nu+\bar{\nu}) + \de(\nu-\bar{\nu}) \big)
\diagramEnvelope{\begin{tikzpicture}[anchor=base,baseline]
	\node (prop) at (0,0) [twopt] {};
	\node (opO1) at (-1,0) [left] {};
	\node (opO3) at (1,0) [right] {};
	\node at (-1,0) [below] {$\f_{\De(\nu),J}$};
	\node at (1,0) [below] {$\f_{\De(\nu),J}$};
	\draw [spinning bulk] (prop)-- (opO1);
	\draw [spinning bulk] (prop)-- (opO3);
\end{tikzpicture}}.
\eeq
By inserting it on both sides of \eqref{eq:bulk_6j_2pt_vs_3pt_2} we conclude
\be
\left\{ \begin{matrix}
\f_3\\ \f_1
\end{matrix} \right\}^{\bar{m}}_{m} 
\diagramEnvelope{\begin{tikzpicture}[anchor=base,baseline]
	\node (propR) at (1,0) [twopt] {};
	\node (vertR) at (2,-0.10) [threept] {$m$};
	\node (opO1) at (0,0) [left] {$\f_1$};
	\node (opO3) at (3,0) [right] {$\f_3$};
	\node (opW) at (2,-1) [below] {$\cW$};	
	\draw [spinning bulk] (propR)-- (opO1);
	\draw [spinning bulk] (propR)-- (vertR);
	\draw [spinning bulk] (vertR)-- (opO3);
	\draw [finite with arrow] (vertR)-- (opW);
\end{tikzpicture}}
=	\left\{
		\begin{matrix}
		\f_3 & \mathbf{1} & \f_1 \\
		\f_1 & \cW & \f_3
		\end{matrix}
	\right\}^{\uniq \bar{m}}_{\uniq m}
\left\{ \begin{matrix}
\f_1\\ \f_3
\end{matrix} \right\}^{m}_{\bar{m}} 
\diagramEnvelope{\begin{tikzpicture}[anchor=base,baseline]
	\node (opO1) at (-2,0) [left] {$\f_1$};
	\node (vertL) at (-1,-0.12) [threept] {$\bar{m}$};
	\node (propL) at (0,0) [twopt] {};
	\node (opO3) at (1,0) [right] {$\f_3$};
	\node (opW) at (-1,-1) [below] {$\cW$};
	\draw [spinning bulk] (vertL)-- (opO1);
	\draw [spinning bulk] (propL)-- (vertL);
	\draw [spinning bulk] (propL)-- (opO3);
	\draw [finite with arrow] (vertL)-- (opW);
\end{tikzpicture}}.
\label{eq:bulk_6j_2pt_vs_3pt_3}
\ee
Comparing with \eqref{eq:twoptcrossing_bulk} one sees that (\ref{eq:compatibility}) holds,
in complete analogy to the relation \eqref{eq:shorthand_6j_2pt} for conformal structures.

\subsection{Coupling of bulk-to-bulk propagator in diagrams}
\label{app:bulk-to-bulk-diagram}

In this appendix we examine how the derivatives in the combination of a local coupling and a bulk-to-bulk propagator can be replaced by weight shifting operators.
Consider the coupling \eqref{eq:unique_coupling} for two scalar fields and a spin $J$ field,
\beq
\int_{\rm AdS} dX \ \< \f_2 (X)| \ \partial_{X \{a_1} \ldots \partial_{X a_J\}}  \< \f_1 (X)| \ \partial_{W}^{\{a_1} \ldots \partial_{W}^{a_J\}} \< \f_3 (X,W)| 
\,.
\eeq
We assume that $\< \f_1 (X)|$ belongs to a bulk-to-boundary propagator $\Pi_{\Delta_1}(X,P_1)$
and $\< \f_3 (X,W)|$ to a spin $J$ bulk-to-bulk propagator.
The spin $l$ contribution to the spectral representation of the spin $J$ bulk-to-bulk propagator \eqref{eq:SlipRepStart} has the form
\beq
\partial_{X}^{\{a_1} \ldots \partial_{X}^{a_{J-l}}
\partial_{W}^{a_{J-l+1}} \ldots \partial_{W}^{a_{J}\}}
\partial_{X_2}^{\{b_1}\ldots \partial_{X_2}^{b_{J-l}}
\partial_{W_2}^{b_{J-l+1}} \ldots \partial_{W_2}^{b_{J}\}}
 \, \Omega_{\D,l}(X,X_2;W,W_2)\,.
\eeq
Hence we need to study the expression
\bea
(*) \equiv{}& 
\frac{1}{l!}
\left[\partial_{X \{a_1} \ldots \partial_{X a_J\}}
\Pi_{\Delta_1}(X,P_1)\right]
\partial_{X}^{\{a_1} \ldots \partial_{X}^{a_{J-l}}
\partial_{W}^{a_{J-l+1}} \ldots \partial_{W}^{a_{J}\}}
\Omega_{\Delta,l}(X,X_2;W,W_2)\,.
\eea{eq:3pt_cont_manipulation}
Next the differential operators can be replaced by weight shifting operators. The $\partial_{W}^{a}$ is equal to $\cL_{0-}^{a}(X,W)$ and the $\partial_{X}^{a}$ acts on expressions without any dependence on $W$ so they can be replaced by $\cL_{+0}^{a}(X,W)$, thus
\begin{align}
(*) 
&={}
\frac{1}{l!(2-\D_1-J)_J (2-\D-J+l)_{J-l}}
\left[\cL_{+0 \, \{a_1}(X,W) \ldots \cL_{+0 \, a_J\}}(X,W)
\Pi_{\Delta_1}(X,P_1)\right]
\nonumber
\\
&\times \cL_{+0}(X,W)^{\{a_1} \ldots \cL_{+0}(X,W)^{a_{J-l}}
\cL_{0-}(X,W)^{a_{J-l+1}} \ldots \cL_{0-}(X,W)^{a_{J}\}}
\Omega_{\Delta,l}(X,X_2;W,W_2)\,.
\label{eq:3pt_cont_manipulation3}
\end{align}
We are not quite done yet, because the traceless and transverse contraction depends on $X_1$.
In order to use the crossing equation \eqref{eq:twoptcrossing_bulk_boundary} and relate everything to a Witten diagram for the exchange of a scalar we need to get rid of this dependence.
So one has to take apart the transverse and traceless contraction.
This can be done by looking at the explicit formula in equation (242) of \cite{Costa:2014kfa}.
The result is
\bea
(*) 
=
\sum_{m=0}^{J-l} \beta_{J,l,m}^{\D_1 \D} \Big(
&
\left[\prod_{i=1}^{J-m} \cL_{+0 \, a_i}(X,W)
\Pi_{\Delta_1}(X,P_1)\right]\\
&\prod_{j=1}^{J-l-m} \cL^{a_j}_{+0}(X,W)
\prod_{j=J-l-m+1}^{J-m} \cL^{a_j}_{0-}(X,W)
\Omega_{\Delta,l}(X,X_2;W,W_2)
\Big)\,,
\eea{eq:3pt_with_Ls}
with the coefficients
\bea
\beta_{J,l,m}^{\D_1 \D} ={}& \frac{(\D_1+J-m)_{m} (\D+J-m)_{m}}{l!(2-\D_1-J+m)_{J-m} (2-\D-J+l+m)_{J-l-m}}\\
&\times \frac{(J-l)! \Gamma(1-h-J+m)\Gamma(\frac{3-2h-2J}{2})}{\Gamma(1-h-J+\frac{m}{2})\Gamma(\frac{3-2h-2J+m}{2}) m! (J-l-m)!}\,.
\eea{eq:b_coeff}

\bibliographystyle{JHEP}

\bibliography{wso_for_witten_diagrams}

\end{document}